%% file: paper.tex
\documentclass[aps,prd,final,groupedaddress,showpacs,showkeys,floatfix,preprintnumbers,nofootinbib,letterpaper]{revtex4}

\usepackage{graphicx}
\usepackage{float}

\usepackage[latin1]{inputenc}                    
\usepackage{latexsym}                            
\usepackage{amsfonts}                            
\usepackage{amssymb}                             
\usepackage{amsmath}                             
\usepackage[mathscr]{eucal}                      
\usepackage{dcolumn}                             
\usepackage{subfigure}
\usepackage[english]{babel}
\usepackage{theorem}                            
\usepackage{hyperref}
\usepackage{slashed}

\input{text/setup}

\begin{document}

\input{text/titlepage}

\section{\label{sect:Introduction}Introduction}
  \input{text/introduction}

\section{\label{sect:lattice_methodology}Lattice methodology}

  \subsection{Dynamical domain wall ensembles}
    \input{text/methodology/dwf_ensemble}

  \subsection{\label{sect:pion_obsv}Pion mass, decay constant and setting the fine lattice scale}
    \input{text/methodology/pion_obsv}

  \subsection{Extraction of nucleon matrix elements}
    \input{text/methodology/ratio_method}

  \subsection{Overdetermined analysis of form factors}
    \input{text/methodology/overdet_ff_analysis}

    \section{\label{sect:isovector}Isovector form factors}
    \input{text/lattice_res/isovector_ff}

    \subsection{\label{sect:renorm}Vector current renormalization}
    \input{text/lattice_res/current_renorm}

    \subsection{\label{sect:q2_dependence}$Q^2$ dependence}
    \input{text/lattice_res/q2_dependence}

    \subsection{Chiral extrapolations}
    \input{text/lattice_res/chiral_extrap}

    \section{\label{sect:isoscalar}Isoscalar form factors}
    \input{text/lattice_res/isoscalar_ff}

\section{\label{sect:sys_errs}Systematic errors}
  \subsection{\label{sect:excited_states_systematics}Effect of the excited states}
    \input{text/syst_errors/me_plateau_study}

  \subsection{\label{sect:finite_vol}Finite volume dependence}

\input{text/syst_errors/finite_vol}

\section{\label{sect:comparison}Comparison with previous calculations}
  \input{text/comparison}

\section{\label{sect:summary-conclusions}Summary and conclusions}
  \input{text/conclusions}

\section*{Acknowledgments}
This work is supported in part by the U.~S.~Department of Energy under Grants DE-FG02-94ER40818,  
DE-FG02-05ER25681, and DE-FG02-96ER40965.
Ph.~H.~acknowledges support by the Emmy-Noether program and the cluster of excellence 
``Origin and Structure of the Universe'' of the DFG, 
M.~P.~acknowledges support by a Feodor Lynen Fellowship from the Alexander von Humboldt Foundation, 
and T.~R.~H.~is supported by DFG via SFB/TR 55.
W.~S.~wishes to thank the Institute of Physics at Academia Sinica for their kind hospitality and
support as well as  Jiunn-Wei~Chen at National Taiwan University and Hsiang-nan~Li
at Academia Sinica for their hospitality and for valuable physics discussions and suggestions.
Computations for this work were carried out using the Argonne Leadership Computing Facility at
Argonne National Laboratory, which is supported by the Office of Science 
of the U.~S.~Department of Energy under contract DE-AC02-06CH11357; 
using facilities of the USQCD Collaboration, which are funded by the Office of Science 
of the U.~S.~Department of Energy; 
and using resources provided by the New Mexico Computing Applications Center (NMCAC) on Encanto. 
The authors also wish to acknowledge use of  dynamical domain wall configurations 
and universal propagators calculated by the RBC and LHPC collaborations and the use of 
Chroma~\cite{Edwards:2004sx} SciDAC software.

\appendix
\clearpage
\section{\label{sect:appendix1}Tables}
  \input{text/app_tables/ff_tables}

\clearpage

\section{\label{sect:appendix2}Smeared nucleon sources for domain wall fermions}
  \input{text/app_smearing/smearing}

\bibliography{paper}

\printtables

\printfigures

\end{document}

%% file: text/setup.tex
\newcommand{\bea}{\begin{eqnarray}}
\newcommand{\eea}{\end{eqnarray}}

\usepackage{color}

\theoremstyle{plain}
\theorembodyfont{\slshape}
\theoremheaderfont{\scshape}

%% file: text/titlepage.tex
\preprint{ MIT-CTP-4032} 
\title[DW_FF]{Nucleon Electromagnetic Form Factors from Lattice QCD using 
  2+1 Flavor Domain Wall Fermions on Fine Lattices and Chiral Perturbation Theory}

\author{ S.~N.~Syritsyn} 
  \affiliation{Center for Theoretical Physics, Massachusetts Institute of Technology, 
               Cambridge, MA 02139}

\author{ J.~D.~Bratt, M.~F.~Lin, H.~B.~Meyer, J.~W.~Negele, A.~V.~Pochinsky, M.~Procura} 
  \affiliation{Center for Theoretical Physics, Massachusetts Institute of Technology, 
               Cambridge, MA 02139}
\author{M.~Engelhardt}
  \affiliation{Physics Department, New Mexico State University, Las Cruces, NM 88003-8001}
 
\author{Ph.~H{\"a}gler}
  \affiliation{Institut f\"ur Theoretische Physik T39, Physik-Department der TU M\"unchen, 
               James-Franck-Stra\ss{}e, D-85747 Garching, Germany}

\author{T.~R.~Hemmert}
  \affiliation{Theoretische Physik, Universit\"at Regensburg, D-93040 Regensburg, Germany}

\author{W.~Schroers}
  \affiliation{Institute of Physics, Academia Sinica, Taipei 115, Taiwan, R.O.C.}
   
\author{(LHPC Collaboration)}
\noaffiliation

\begin{abstract}

We present a high-statistics calculation of nucleon electromagnetic form factors in $N_f=2+1$ lattice
QCD using  domain  wall quarks on fine lattices, to attain a new level of precision in
systematic and statistical errors.  
Our calculations use $32^3 \times 64$ lattices  with lattice
spacing $a=0.084\text{ fm}$ for pion masses of  297, 355, and 403 MeV,  and we perform an
overdetermined analysis using on the order of 3600 to 7000 measurements to calculate nucleon electric and
magnetic form factors up to $Q^2 \approx$ 1.05 GeV$^2$. 
Results are shown to be consistent with
those obtained using valence domain wall quarks with improved staggered sea quarks, and
using coarse domain wall lattices.     
We determine the isovector Dirac radius $r_1^v$, Pauli radius $r_2^v$ 
and anomalous magnetic moment $\kappa_v$.
We also determine connected contributions to the corresponding isoscalar observables.
We extrapolate these observables to the physical pion  mass using two different
formulations of two-flavor chiral effective field theory at one loop: the heavy
baryon Small Scale Expansion (SSE) and covariant baryon chiral perturbation
theory.
The isovector results and the connected contributions to
the isoscalar results are compared with experiment, and the need for calculations at smaller
pion masses is discussed. 

\end{abstract}

\pacs{ 12.38.Gc,13.40.Gp }

\keywords{electromagnetic form factors, lattice QCD,
  hadron structure}

\maketitle

%% file: text/introduction.tex
Electromagnetic form factors characterize fundamental aspects of the structure of protons and
neutrons, in particular they specify the spatial distribution of charge and magnetization.
For non-relativistic systems the electric and magnetic form factors would just be Fourier
transforms of the charge and current densities.  
At each $Q^2$, the Sachs form factors $G_E(Q^2)$ and $G_M(Q^2)$ may be 
regarded as three dimensional Fourier transforms of charge and magnetization densities 
 defined in the corresponding Breit frame.
A probabilistic interpretation of the Dirac and Pauli form factors $F_1(Q^2)$ and $F_2(Q^2)$ 
can be obtained from a two dimensional Fourier transformation to impact parameter space
in the infinite momentum frame~\cite{Burkardt:2000za,Burkardt:2002hr}.
At high momentum transfer, the elastic form factor specifies the amplitude for a single quark 
in the nucleon to absorb a very large momentum kick and share it with the other constituents 
in such a way that the nucleon remains in its ground state instead of being excited.  
It thus describes the onset of scaling and the scale at which quark counting rules become 
applicable, which is an unresolved theoretical question in nonperturbative QCD.  
The combination of precision experimental measurements and crisp theoretical interpretation 
renders elastic nucleon form factors particularly significant. 
Given the constantly improving experimental measurements of form
factors and their fundamental significance, it is an important challenge for lattice QCD to
calculate them accurately from first principles.

The nucleon Dirac and Pauli form factors,  $F_1(Q^2)$ and $F_2(Q^2)$ respectively,  
are defined as follows for each quark flavor $(f)$:
 \begin{equation}
\langle P^\prime, S^\prime|V^\mu_{(f)}|P,S \rangle = \bar U(P^\prime, S^\prime)
  \left[\gamma^\mu F_1^{(f)}(Q^2) +i \sigma^{\mu \nu} \frac{q_\nu}{2 M_N} F_2^{(f)}(Q^2) \right] 
U(P,S)\,,
\quad V_{(f)}^\mu = \bar\psi_{(f)} \gamma^\mu \psi_{(f)}\,,
\end{equation}
where $P$, $P^\prime$ are the initial and final nucleon momenta, 
$S$, $S^\prime$ are the corresponding spin vectors, 
the momentum transfer is $q = P^\prime - P$ with $Q^2 = -q^2 \ge0$, and $M_N$ is the nucleon mass. 
The Sachs form factors  $G_E(Q^2)$ and $G_M(Q^2)$ are defined by:
\begin{eqnarray}
G_E(Q^2) & = & F_1(Q^2) - \frac{Q^2}{(2 M_N)^2} F_2(Q^2) \\
G_M(Q^2) & = & F_1(Q^2) +  F_2(Q^2)\, .
\end{eqnarray}
Finally, it is useful to define  isoscalar and isovector form factors as the sum 
and difference of proton and neutron form factors as follows:
\begin{align}
\label{eqn:def_ff_isovector}
F_{1,2}^v(Q^2) &= F_{1,2}^p(Q^2) - F_{1,2}^n(Q^2) = 
  F_{1,2}^u(Q^2) - F_{1,2}^d(Q^2) \equiv F_{1,2}^{u-d}(Q^2), \\
\label{eqn:def_ff_isoscalar}
F_{1,2}^s(Q^2) &= F_{1,2}^p(Q^2) + F_{1,2}^n(Q^2) = 
  \frac13\left(F_{1,2}^u(Q^2) + F_{1,2}^d(Q^2)\right) \equiv  \frac13 F_{1,2}^{u+d}(Q^2),
\end{align}
where $F_i^{p,n}$ are the form factors of the electromagnetic current in a proton and a neutron, 
respectively:
\begin{equation}
\label{eqn:def_nucleon_emff}
V^\mu_{\text{em},p} = \frac23\bar{u}\gamma^\mu u - \frac13\bar{d}\gamma^\mu d\,,
\quad\quad
V^\mu_{\text{em},n} = -\frac13\bar{u}\gamma^\mu u + \frac23\bar{d}\gamma^\mu d\,.
\end{equation}
Although proton and neutron form factors contain both connected diagrams, calculated in this work, and disconnected diagrams, which are currently omitted, the disconnected diagrams do not contribute to the isovector form factors $F_i^v$. Hence, we will devote particular attention in this work to the isovector form factors.

Precise experimental measurements of the set of  all four nucleon form factors remains
challenging, and the field is marked both by significant recent developments and open questions.
Although the most straightforward measurement is $F_1(Q^2)$ for the proton, the slope at very
small values of $Q^2$ remains controversial. Phenomenological fits to experimental form 
factors~\cite{Friedrich:2003iz,Arrington:2007ux} appear to be inconsistent with analyses based on 
dispersion theory~\cite{Hohler:1976ax,Mergell:1995bf,Belushkin:2006qa}, with phenomenological
fits yielding larger Dirac radii. 
Hence, a new generation of
precision measurements of form factors at low momentum transfer is currently being undertaken at 
Mainz~\cite{Bernauer:2008zz}. 
Spin polarization experiments~\cite{Milbrath:1997de,Pospischil:2001pp,Gayou:2001qd,Gayou:2001qt,Punjabi:2005wq}
yielded results for $F_2(Q^2)$ significantly different from traditional measurements based on Rosenbluth separation, and there is a consensus that two-photon exchange processes contribute much more strongly to the backward cross section used in Rosenbluth separation than to polarization transfer~\cite{Arrington:2007ux}.
However, there are not yet precise theoretical calculations of two photon exchange that fully resolve the discrepancy between the two experimental methods, and hence  experiments using positron scattering, for which the relative contribution of the two-photon term changes sign, are being prepared~\cite{Arrington:2004hk,olympus}.  Neutron form factors are more uncertain than proton form factors because of the need to know the nuclear wave function to
go from experimental scattering results from deuterium or $^3\mathrm{He}$ to a statement about the neutron form factor. Over the years, nuclear models and theoretical calculations have been refined, but it is still a challenge to provide  a definitive estimate of the uncertainty in the claimed neutron form factors extracted from nuclear targets. Given the level of precision to which we aspire in lattice calculations, systematic uncertainties in isovector and isoscalar form factors are not necessarily negligible.  In the future when lattice calculations reliably include precise calculations of disconnected contributions, it may well be that lattice calculations play a role in
guiding the resolution of some of these experimental questions.

Electromagnetic form factors have now been calculated in lattice QCD using a variety of actions.  Quenched calculations of form factors have used both Wilson~\cite{Gockeler:2003ay,Alexandrou:2006ru} and domain wall~\cite{Sasaki:2007gw}  fermion actions, and additional quenched calculations have addressed magnetic moments and root-mean-squared (rms) radii~\cite{Tang:2003jh,Boinepalli:2006xd}. Dynamical calculations with two flavors have used  Wilson~\cite{Alexandrou:2006ru}, clover improved Wilson~\cite{Gockeler:2007hj}, twisted 
mass~\cite{Alexandrou:2009xk,Alexandrou:2008rp}
and domain wall~\cite{Lin:2008uz} actions. Extensive 2+1 flavor calculations have been performed
with a mixed action, which combines domain wall valence quarks and improved staggered sea
quarks~\cite{Hagler:2007xi,Bratt:2008uf,LHPC_mixedaction_nucleonstr_2008}, using the same methodology as in the present work, and comparisons will be made to assess the consistency of the full domain wall and  mixed action results.  
Dynamical domain wall results with 2+1 flavors on coarse lattices with $a = 0.114\text{ fm}$ 
have recently been reported~\cite{Ohta:2008kd,Bratt:2008uf,Yamazaki:2009zq}, 
and initial results from the present work on fine lattices with $a=0.084\text{ fm}$ were presented in Ref.~\cite{Syritsyn:2009np}.

The goal of this work is to achieve a new level of precision in calculating form factors from
first principles in lattice QCD.  
Hence, we have done everything feasible within the constraints
of our computational resources to reduce both statistical and systematic errors.  
Since this involves a number of new developments, we describe our methodology, innovations, 
and tests in detail. 
Because the signal to noise for baryon observables degrades with increasing Euclidean time $t$
as $e^{-(M_n -3/2m_\pi)t}$,
we have obtained high statistics using from 3688 to 7064 measurements of operators at a given
mass by performing 8 measurements per lattice and have verified their statistical independence.
The  source-sink separation distance is a crucial issue, since an excessively large distance
degrades the statistical accuracy whereas too small a distance introduces systematic errors from
the contributions of excited states. 
We present a quantitative analysis of the contributions of
excited states, and using this analysis, provide  compelling numerical evidence that with our
choice, which has been questioned in the literature~\cite{Yamazaki:2009zq}, excited state contributions are
negligible in our present work. 
Our overdetermined analysis of form factors provides a general framework for optimizing 
the precision of our lattice calculations by combining measurements of as many distinct  
nucleon matrix elements involving the form factors at the same $Q^2$ value as practical.
We also describe how we choose which contributions to include, and treat error correlations.
We compare domain wall calculations on fine lattices at three
masses  with a calculation on a coarse  lattice at one mass, and present evidence that the ${\cal
O}(a^2)$ corrections are indeed small.  
We also compare our results with mixed action results, showing essential consistency 
between mixed and domain wall actions 
and emphasizing the small size of finite volume corrections to calculations on a 2.5 fm lattice at 
$m_\pi$ = 350 MeV that have been calculated
to high precision with the mixed action.
We perform chiral extrapolations of the Dirac
and Pauli mean squared radii, $\left(r_1^{v,s}\right)^2$ and $\left(r_2^{v,s}\right)^2$, 
respectively, and of the anomalous magnetic moments $\kappa_{v,s}$. 
We use two different formulations of $SU(2)$ chiral effective field theory: the heavy
baryon Small Scale Expansion (SSE) which includes explicit $\Delta\,(1232)$ degrees 
of freedom~\cite{Hemmert:1997ye} and covariant baryon chiral perturbation theory (CBChPT) 
without  an explicit $\Delta\,(1232)$ in the $\overline{IR}$-scheme~\cite{Dorati:2007bk,Gail:2007phd}, 
which represents a variant of infrared regularization~\cite{Becher:1999he}\footnote{
  For recent work on chiral extrapolations of nucleon magnetic form factors and octet-baryon
  charge radii in heavy baryon ChPT with finite range regularization, we refer the reader
  to Refs.~\cite{Wang:2007iw, Wang:2008vb}.
}. 
We explore the degree to which the relevant low-energy constants can be determined in the range 
of masses we consider and the variation of the extrapolated results in both schemes.
We conclude that with the new level of precision we achieve, it is necessary to extend 
the lattice calculations to substantially lower masses to make contact with the regime
of applicability of chiral effective theory and possibly reach agreement with experiment.

The remainder of this paper is organized as follows.  
In Sect.~\ref{sect:lattice_methodology}, we present a detailed description of our methodology,
including setting the scale, 
computation of nucleon matrix elements and coherent sink technique, optimization of sources,
treatment of error correlations and constraints in the overdetermined analysis,
and a check of the independence of multiple measurements per configuration.
Sect.~\ref{sect:isovector} presents the results of our lattice calculations for isovector form factors, including
phenomenological fits to the momentum transfer dependence and determination of the Dirac
radius $\left(r_1^{v}\right)^2$, the Pauli radius $\left(r_2^{v}\right)^2$,  
and the anomalous magnetic moment $\kappa_{v}$.  
Comparisons are made with domain wall calculations on a coarse lattice
and with mixed-action calculations using valence domain wall valence quarks and improved
staggered sea quarks.   
We also present the chirally extrapolated values of $\left(r_1^{v}\right)^2$, 
$\left(r_2^{v}\right)^2$,  and $\kappa_{v}$
to the physical pion mass using the SSE and covariant chiral effective field theories
and compare them with experiment.  Corresponding results for isoscalar form factors are presented in 
Sect.~\ref{sect:isoscalar}.  
Systematic errors are discussed in Sect.~\ref{sect:sys_errs},  results are compared with other work in 
Sect.~\ref{sect:comparison},   
and conclusions and opportunities for further
understanding of nucleon form factors are discussed in the final
Sect.~\ref{sect:summary-conclusions}. 
Selected numerical results are tabulated in Appendix~\ref{sect:appendix1} and the optimized sources are described in Appendix~\ref{sect:appendix2}.

%% file: text/methodology/dwf_ensemble.tex
In our calculations, we analyze gauge configurations generated by the RBC and UKQCD collaborations 
\cite{Allton:2008pn} with the  Iwasaki gauge action and 
$N_f=2+1$ flavors of dynamical domain wall fermions.
The gauge configuration ensembles are summarized in Tab.~\ref{tab:gauge_config}.
We obtain the relevant physical results from three \textit{fine} lattice ensembles 
with lattice spacing $a=0.084\text{ fm}$.
We use one \textit{coarse} 
lattice ensemble with known lattice spacing $a=0.114\,\mathrm{fm}$ \cite{Allton:2008pn}
to set the scale on the \textrm{fine} lattices and control the systematic errors
due to discretization.

In our analysis, we use only a unitary fermion action, where the sea and valence 
fermion actions and masses are exactly the same.
The extent of the fifth dimension is chosen to be $L_s=16$, 
which keeps the residual mass $m_\text{res}$ smaller than the bare quark masses 
for all ensembles.

In order to maximize the signal to noise ratio and suppress excited state contamination, we 
carefully optimize the quark propagator sources.
We use Wuppertal smearing of quark sources combined with APE smearing of the source gauge fields to reach the maximum 
overlap of the lattice nucleon operators with the nucleon ground state and reduce its
fluctuation.
The details of optimization and the source parameters we use are given in
Appendix~\ref{sect:appendix2}.

To increase statistics, we perform eight measurements of nucleon correlation functions 
on each gauge configuration.
To do so, we compute four forward quark propagators and 
construct nucleon and antinucleon correlators advancing in the positive and negative 
time directions, respectively.
The data for antinucleons are transformed according to the reflection symmetry
and combined with the data for nucleons into a single data set.
We save computing time by using the ``coherent'' backward propagator technique,
in which we compute only a sum of four backward propagators for 
four separate sequential sources with the same hadron type, flavor and sink momentum.
To check for possible systematic effects, we recalculate the nucleon three-point
functions using independent backward propagators and larger source-sink separation on a subset of 
our lightest pion ensemble, and the extracted form factors 
(see Fig.~\ref{fig:comp-pltx-av-fits-indep-dt14}) show no significant deviation from 
the method we use.
Since lattice data may be autocorrelated, we block all the
measurements on the two consecutive gauge configurations, 
and also check that the measurements we get are indeed independent by increasing the block size
to include eight consecutive configurations (see Fig.~\ref{fig:solve_od_different_analyses}).

\begin{table}
\centering
\caption{\label{tab:gauge_config} Gauge configuration ensembles used for our analysis, 
  with one \textit{coarse} and three \textit{fine} lattice spacings. 
  These configurations were generated by the RBC and UKQCD \cite{Allton:2008pn} collaborations. 
  The coarse lattice spacing was determined in \cite{Allton:2008pn}, 
  and the fine lattice spacing is determined in Sect.~\ref{sect:pion_obsv}.
  Measurement count includes a factor of $8$ for each gauge configuration.
  Note that for $m_\pi$, $F_\pi$, $m_\text{res}^\prime$ the measurement count is the number of
  configurations multiplied by $4$ instead of $8$.
}
\begin{tabular}{ccr|r|cr@{.}l|cc|cc|cc}
\hline
\hline
$L_s^3\times L_t $ & $a\text{ [fm]}$ & $T$ & \# & $am_l/am_h$ & 
  \multicolumn{2}{c}{$am_\text{res}^\prime\times 10^3$} & 
  $am_\pi$ & $m_\pi\text{ [MeV]}$ & 
  $aF_\pi$ & $F_\pi\text{ [MeV]}$ & 
  $aM_N$ & $M_N\text{ [MeV]}$ 
 \\
\hline
$24^3\times64$  & $0.114$ & $9$ &
  3208 & $0.005/0.04$ & $3$&$15(1)$ & 
  $0.1901(3)$ &  $329(5)$ &
  $0.06100(11)$ &  $105.5(1.7)$ &
  $0.657(4)$  &  $1136(20)$
\\
\hline
$32^3\times64$  & $0.084$ & $12$ &
4928  & $0.004/0.03$ & $0$&$665(3)$ & 
  $0.1268(3)$ &  $297(5)$ &
  $0.04400(15)$ &  $102.9(1.8)$ &
  $0.474(4)$  &  $1109(21)$
\\
$32^3\times64$  & $0.084$ & $12$ &
  7064  & $0.006/0.03$ & $0$&$663(2)$ &  
  $0.1519(3)$ &  $355(6)$  &
  $0.04571(09)$ &  $107.0(1.8)$ &
  $0.501(2)$  &  $1172(21)$
\\
$32^3\times64$  & $0.084$ & $12$ &
  4224  & $0.008/0.03$ & $0$&$668(3)$ &  
  $0.1724(3)$ &  $403(7)$ &
  $0.04755(18)$ &  $111.3(2.0)$ &
  $0.522(2)$  &  $1221(21)$
\\
\hline
\hline
\end{tabular}
\end{table}

%% file: text/methodology/pion_obsv.tex
So far, the scale has been set only for the coarse lattice ensembles \cite{Allton:2008pn}.
In order to set the scale for the fine lattice ensembles, we compare the lattice values for 
the pion decay constant $(aF_\pi)$ on coarse and fine lattices at the same value 
of the dimensionless ratio $(m_\pi/F_\pi)^2$ ignoring possible finite lattice spacing 
effects in the pion decay constant $F_\pi$.

First, we compute the pion mass, the pion decay constant and the local axial current 
renormalization constant\footnote{
  In this paper, we assume that 
  the renormalization constant $Z_{\mathcal A}$
  of the (partially) conserved domain wall axial current ${\mathcal A}_\mu$ 
  is equal to its $L_s\to\infty$ value of one, 
  and note that the finite $L_s$ deviation has been estimated in
  \cite{Sharpe:2007yd,Allton:2008pn} to give $|Z_{\mathcal A}-1|\lesssim 1\%$.
  The values of $F_\pi$ that we compute are, in fact, $F_\pi / Z_{\mathcal A}$.
}
from fits to the pseudoscalar density and axial current correlators using the PCAC relation
\cite{Blum:2000kn}:
\begin{align}
\langle A_0(t,\vec p=0) \tilde{J}_5(0) \rangle &= 
  \left(e^{-m_\pi t} - e^{-m_\pi(L_t - t)}\right) \times
  \frac{F_\pi^2 m_\pi^2}{2(m_l+m_\text{res}^\prime)} \times
  Z_A^{-1} Z_\text{sm}^{-1}, \\
\langle J_{5q}(t,\vec p=0) \tilde{J}_5(0) \rangle &= 
  \left(e^{-m_\pi t} + e^{-m_\pi(L_t - t)}\right) \times
  \frac{F_\pi^2 m_\pi^3}{4(m_l+m_\text{res}^\prime)^2} \times
  m_\text{res}^\prime Z_\text{sm}^{-1}, \\
\langle J_5(t,\vec p=0) \tilde{J}_5(0) \rangle &= 
  \left(e^{-m_\pi t} + e^{-m_\pi(L_t - t)}\right) \times
  \frac{F_\pi^2 m_\pi^3}{4(m_l+m_\text{res}^\prime)^2} \times
  Z_\text{sm}^{-1}, \\
\langle \tilde{J}_5(t,\vec p=0) \tilde{J}_5(0) \rangle &= 
  \left(e^{-m_\pi t} + e^{-m_\pi(L_t - t)}\right) \times
  \frac{F_\pi^2 m_\pi^3}{4(m_l+m_\text{res}^\prime)^2} \times
  Z_\text{sm}^{-2},
\end{align}
where $A_0$ is the local axial charge, $J_{5q}$ is the fifth dimension mid-point pseudoscalar
density and $J_5$ ($\tilde{J}_5$) is the (smeared) pseudoscalar density. 
The pion decay constant $F_\pi$ convention is such that 
\begin{equation}
\label{eqn:def_fpi_phys}
F_\pi^\text{phys}=92.4\pm0.3\text{ MeV}.
\end{equation}
We choose the range of $t$ to be $[12:52]$ to exclude any excited state contaminaions.
We define the smearing renormalization constant $Z_\text{sm}$ from the plateau 
$\langle J_5(t) \tilde{J}_5(0)\rangle / \langle \tilde{J}_5(t) \tilde{J}_5(0)\rangle$ and 
the local axial current renormalization constant $Z_A$ from the ratio of 
$\langle{\mathcal A}_0(t+1/2)\tilde{J}_5(0)\rangle$ and $\langle A_0(t)\tilde{J}_5(0)\rangle$ 
appropriately averaged to suppress $O(a)$ effects due to $a/2$ displacement of the 
conserved axial current ${\mathcal A}_0(t+1/2)$ \cite{Blum:2000kn}.
The results for $a m_\pi$, $a F_\pi$ and $am_\text{res}^\prime$ are shown in 
Tab.~\ref{tab:gauge_config}.
The error bars reflect both the statistical error and the systematic error due to different
fitting ranges.

Second, we fit $m_\pi$ and $F_\pi$ at three values of the light quark mass
using ${\mathcal O}(p^4)$ $SU(2)$ chiral perturbation theory \cite{Gasser:1983yg, Bernard:1998gv}
\begin{align}
a^2 m_\pi^2 &= 
  a^2\chi \left\{1 + \frac{2a^2\chi}{(aF)^2} l_3^r(a^{-1}) +
        \frac{a^2\chi}{32\pi^2 (aF)^2} \log\left(a^2\chi\right)\right\},
\\
a F_\pi &=
  a F \left\{1 + \frac{a^2\chi}{(aF)^2}l_4^r(a^{-1}) - 
        \frac{a^2\chi}{16\pi^2 (aF)^2} \log\left(a^2\chi\right)\right\},
\end{align}
where $a^2\chi = 2 (aB)\cdot a(m_l + m_\text{res}^\prime)$,
$l_{3,4}^r(a^{-1})$ are the next-to-leading order (NLO) low-energy constants (LECs) at the scale $\Lambda=a^{-1}$, 
and the fit variables are $(aF)$, $(aB)$ and $l_{3,4}^r$.
However, the fit is not satisfactory in terms of $\chi^2$: for two degrees of freedom, 
we get $\chi^2\approx7$, with its probability to be this or higher being $\lesssim3\%$.
This is the first of many indications that chiral perturbation theory, at the order we can use, is not accurate in the range of masses we are considering.  Hence, the LEC's are not precisely determined although, as noted below, we obtain an adequate interpolation to set the scale. 

\begin{figure}[h!]
\centering
\includegraphics[width=0.5\textwidth]{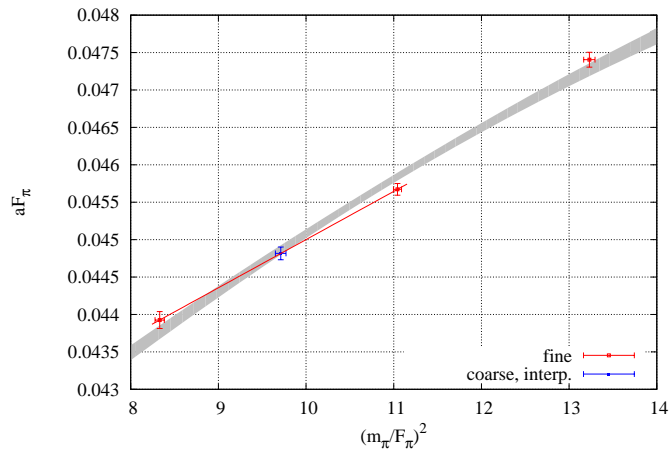}
\caption{\label{fig:mpi_fpi_chi_interpolate} 
  One-loop $SU(2)$ ChPT interpolation of the fine lattice values of $F_\pi$ and $m_\pi$. 
  The point with abscissa $9.71 = (m_\pi/F_\pi)^2_\text{coarse}$ 
  was obtained by interpolating $(aF_\pi)$ linearly in $(m_\pi/F_\pi)^2$.
}
\end{figure}

The NLO LECs $l_{3,4}^r$ from our fit can be converted to the scale-independent parameters
$\bar{l}_{3,4}$ \cite{Gasser:1983yg}. 
At the physical pion mass we obtain
\begin{equation}
\bar{l}_3 = 3.08(11),\quad \bar{l}_4 = 4.24(4).
\end{equation}
Our result for $\bar{l}_3$ is in agreement with the crude estimate 
$\bar{l}_3=2.9 \pm 2.4$ \cite{Gasser:1983yg} 
and with the lattice determination $\bar{l}_3 = 3.0(5)(1)$ \cite{DelDebbio:2006cn} 
using $N_f=2$ dynamical Wilson fermions
but disagrees with $\bar{l}_3=3.42(8)(10)$ 
(the errors are statistical and systematic due to residual lattice artifacts) 
from the ETM collaboration \cite{Dimopoulos:2008sy}.
This discrepancy could could arise from the difference between $N_f=2$ and $N_f=3$ flavors of 
dynamical fermions.
Furthermore, chiral symmetry implies that $\bar{l}_4$ determines the slope of the scalar form
factor of the pion. 
In their seminal paper, Gasser and Leutwyler obtain $\bar{l}_4=4.3 \pm 0.9$ \cite{Gasser:1983yg}. 
This estimate has been sharpened in \cite{Colangelo:2001df}: $\bar{l}_4= 4.4 \pm 0.2$,
which agrees with the value $\bar{l}_4=4.4 \pm 0.3$ obtained by Bijnens et al.
\cite{Bijnens:1998fm}. 
The ETM collaboration result \cite{Dimopoulos:2008sy} is $\bar{l}_4=4.59(4)(2)$.

The resulting interpolated functional dependence of $(aF_\pi)$ on $(m_\pi/F_\pi)^2$ is shown 
in Fig.~\ref{fig:mpi_fpi_chi_interpolate}.
For simplicity, we also estimated $(aF_\pi)|_*$ at $\left.(m_\pi/F_\pi)^2\right|_\text{coarse}$ and its error
by linear interpolation in $(m_\pi/F_\pi)^2$ between the two lightest pion masses.
The comparison in Fig.~\ref{fig:mpi_fpi_chi_interpolate} shows no difference between these two
approaches.
We also interpolated the lattice value of the nucleon mass $(aN_N)$, and obtained the ratios
\begin{equation}
\frac{(aF_\pi)|_*}{(aF_\pi)|_\text{coarse}} = 0.735(2),
\quad
\frac{(aM_N)|_*}{(aM_N)|_\text{coarse}} = 0.742(5).
\end{equation}
Although these ratios are barely consistent within errors, their discrepancy is irrelevant to
the fine scale determination as long as the fractional error in the coarse lattice scale
$a_\text{coarse} = 0.1141(18)\text{ fm}$ dominates.
We obtain the value for the fine lattice scale
\begin{equation}
a_\text{fine} = 0.0840(14)\text{ fm},
\quad
a_\text{fine}^{-1} = 2.34(4)\text{ GeV}.
\end{equation}

%% file: text/methodology/ratio_method.tex
In order to calculate nucleon matrix elements, we compute the three-point polarized nucleon
correlators involving the vector current, along with the two-point correlators
\cite{Hagler:2007xi}
:
\begin{gather}
C_\text{2pt}(t,P) = 
  \sum_x e^{-i\vec P\cdot\vec x}\sum_{\alpha\beta} \left(\Gamma_\text{pol}\right)_{\alpha\beta} 
  \langle N_{\beta}(\vec x\,,t) \bar N_{\alpha}(0\,,0) \rangle ,\\
C_\text{3pt}^{V^\mu}(\tau,T;P,P^\prime) = 
  \sum_{x,y} e^{-i\vec P^\prime\cdot\vec x 
                + i\left(\vec P^\prime - \vec P\right)\cdot\vec y}
  \sum_{\alpha\beta} \left(\Gamma_\text{pol}\right)_{\alpha\beta}
  \langle N_{\beta}(\vec x\,,T) 
          V^\mu(\vec y, \tau)
          \bar N_\alpha(0,0) \rangle
\end{gather}
where $N_\beta, \bar N_\alpha$ are the lattice nucleon operators,
$\langle\Omega\left|N_\alpha(x)\right|P,\sigma\rangle = 
  \sqrt{Z(P)} U_\alpha^{(\sigma)}(P) e^{-iPx}$, 
with $Z(P)$ parameterizing the overlap with the nucleon ground state,
$\left(\Gamma_\text{pol}\right)_{\alpha\beta} = \frac{1+\gamma_4}2\frac{1-i\gamma_3\gamma_5}2$ 
is the spin and parity projection matrix\footnote{
  In this subsection, we use Euclidean $\gamma$-matrices, 
  $(\gamma^\mu)^\dag=\gamma^\mu,\,\left\{\gamma^\mu,\gamma^\nu\right\}=2\delta^{\mu\nu}$.
},
and
$V^{a\mu} = \bar q\gamma^\mu t^a q$ is the vector current operator, 
where $t^a$ denotes an isospin generator.
In the transfer matrix formalism, these correlators take the form 
\begin{gather}
\label{eqn:trmat_nucleon_twopt}
C_\text{2pt}(t,P) = 
  \frac{Z(P) e^{-E t}}{2E}\mathrm{Tr}\left[ \Gamma_\text{pol} \left(i\slashed{P} + M_N\right) \right]
  + \mathrm{excited\ states},\\
\label{eqn:trmat_nucleon_threept}
C_\text{3pt}^{V^\mu}(\tau,T;P,P^\prime) = 
  \frac{\sqrt{Z(P^\prime)\cdot Z(P)}e^{-E^\prime (T-\tau) - E \tau}}{2E^\prime \cdot 2E}
  \mathrm{Tr}\left[ \Gamma_\text{pol} \left(i\slashed{P}^\prime + M_N\right)
    \Gamma^\mu\left(P^\prime,P\right)
    \left(i\slashed{P} + M_N\right) \right]
  + \mathrm{excited\ states},
\end{gather}
where $E$ and $E^{\prime}$ are the ground state energies of the initial and final nucleon states
and $\Gamma^\mu\left(P^\prime,P\right)$ is the electromagnetic vertex function
defined below in Eq.~(\ref{eqn:em_vertex}).
Excited state contributions have generally similar forms with different $Z$-factors, vertices and 
higher energies $E_\text{exc} > E$. 
The systematic effects related to them will be discussed in 
Sect.~\ref{sect:excited_states_systematics}.

\begin{table}[h]
\centering
\caption{\label{tab:mom_list}
  Momentum combinations used to extract the form factors 
  (only one representative of in/out momenta is given). 
  Approximate $Q^2$ values are given for the lightest $M_N=1109\mathrm{\ MeV}$.
}
\begin{tabular}{r|c|r@{.}l}
\hline
\hline
\# &  $\langle\text{out}|\text{in}\rangle$ & 
  \multicolumn{2}{c}{$Q^2\,[\text{GeV}^2]$} \\
\hline
1  &  $\langle0,0,0|0,0,0\rangle,\quad \langle{-1},0,0|{-1},0,0\rangle$ &  0&0 \\
2  &  $\langle0,0,0|1,0,0\rangle,\quad \langle{-1},0,0|0,0,0\rangle$  &  0&203 \\
3  &  $\langle{-1},0,0|{-1},0,1\rangle$ &  0&204 \\
4  &  $\langle0,0,0|1,1,0\rangle$   &  0&391 \\
5  &  $\langle{-1},0,0|{-1},1,1\rangle$ &  0&395 \\
6  &  $\langle{-1},0,0|0,0,1\rangle$  &  0&422 \\
7  &  $\langle0,0,0|1,1,1\rangle$   &  0&568 \\
8  &  $\langle{-1},0,0|0,1,1\rangle$  &  0&626 \\
9  &  $\langle{-1},0,0|1,0,0\rangle$  &  0&844 \\
10 &  $\langle{-1},0,0|1,1,0\rangle$  &  1&048 \\
\hline
\hline
\end{tabular}
\end{table}

In order to extract the combinations of matrix elements 
$\langle P^\prime,S^\prime\left|V^\mu\right|P,S\rangle = 
  \bar U(P^\prime,S^\prime) \Gamma^\mu(P^\prime,P) U(P,S)$,
we combine the lattice nucleon correlators 
(\ref{eqn:trmat_nucleon_twopt}, \ref{eqn:trmat_nucleon_threept}) into the usual ratio of  3- and 2-point correlation functions, which we find useful to write in a convenient and illuminating new form as follows.  
First, we define two ratios, a {\it normalization  ratio},  $R_N$,  and an {\it asymmetry
ratio}, $R_A$,
\begin{eqnarray}
\label{eqn:RN}
R_N & \equiv &      \frac{C_\text{3pt}^{V^\mu}(\tau,T; P, P^\prime)}{\sqrt{C_\text{2pt}(T,P)C_\text{2pt}(T,P^\prime)}} 
,\\
\label{eqn:RA}
R_A & \equiv &  \sqrt{\frac{C_\text{2pt}(T-\tau, P)C_\text{2pt}(\tau, P^\prime)}{C_\text{2pt}(T-\tau, P^\prime)C_\text{2pt}(\tau, P)}}
.
\end{eqnarray}
 The physical matrix element is then given by the product:  

\begin{gather}
\label{eqn:me_ratio}
\begin{aligned}
R^{V^\mu} \equiv  R_N R_A &= 
  \frac{C_\text{3pt}^{V^\mu}(\tau,T; P, P^\prime)}{\sqrt{C_\text{2pt}(T,P)C_\text{2pt}(T,P^\prime)}}
  \sqrt{\frac{C_\text{2pt}(T-\tau, P)C_\text{2pt}(\tau, P^\prime)}{C_\text{2pt}(T-\tau, P^\prime)C_\text{2pt}(\tau, P)}}
  \\
  &\xrightarrow{T\to\infty}
  \frac{
    \sum_{S,S^\prime}
            \left(\bar U(P,S) \Gamma_\text{pol} U(P^\prime,S^\prime)\right)
            \cdot\langle P^\prime,S^\prime\left|V_\mu\right|P,S\rangle
            }
    {\sqrt{2E(E+M_N)\cdot2E^\prime(E^\prime+M_N)}}.
\end{aligned}
\end{gather}

The normalization ratio, $R_N$, has the property that all the lattice-dependent
overlap factors $Z$ for the ground state cancel out, which motivates its name, and yields the full result in the case of forward matrix elements $P = P^\prime$.  The asymmetry ratio, $R_A$,  compensates the asymmetric exponential $\tau$ dependence of  the  three-point correlator, which motivates its name.  
In the absence of excited states, it would be equal to 
$\exp\left[-(E^\prime - E)(\tau-T/2)\right]$ and in the forward case, $P^\prime=P$, this ratio  is trivial and equal to one.
In the general case,  $P^\prime  \ne P$, this ratio is still identically one in the center of the plateau,
$\tau=T/2$, and possesses the following symmetry around the plateau center: $R_A(T-\tau) = 1/R_A(\tau)$.

The limit $T\to\infty$ should be taken to get rid of the excited state contamination. 
In practice, this requires adopting a value of source-sink separation $T$ large enough so that
the excited state contributions to Eq.~(\ref{eqn:me_ratio}) are negligible compared 
to the other sources of errors.  We will explicitly explore the contributions of excited states
to $R^{V^\mu} $in Section~\ref{sect:sys_errs}, where the decomposition into the product $ R_N R_A $ will prove extremely useful.

In order to obtain the most precise information on the form factors, we constrain the in- and
out- lattice nucleon momenta to have components $0,\pm1$.
Higher momentum components are subject to stronger finite lattice spacing effects,
i.e., discretization errors and dispersion relation deviations from the continuum expression.
There is also an indication (see Sect.~\ref{sect:excited_states_systematics}) that such states
have larger excited state contaminations.

%% file: text/methodology/overdet_ff_analysis.tex
In Minkowski space, the nucleon electromagnetic vertex $\Gamma^\mu(P^\prime,P)$ 
in Eq.~(\ref{eqn:trmat_nucleon_threept}) is parameterized with two form factors:
\begin{equation}
\label{eqn:em_vertex}
\Gamma^\mu\left(P^\prime,P\right) = 
  F_1(Q^2)\gamma^\mu + F_2(Q^2)\frac{i\sigma^{\mu\nu}q_\nu}{2M_N},
\quad q = P^\prime - P,
\quad Q^2 = -q^2.
\end{equation}
Transforming the above expression to Euclidean space and substituting it into 
Eq.~(\ref{eqn:trmat_nucleon_threept}) and then Eq.~(\ref{eqn:me_ratio})
and neglecting the excited states,
we obtain an overdetermined system of equations for the form factors $F_{1,2}(Q^2)$ 
at each fixed value of $Q^2$:
\begin{equation}
\label{eqn:ff_od_system}
A_{\alpha i} F_i(Q^2) = R_\alpha^{V_\mu},
\quad \alpha = 1,2,\dotsc
\end{equation}
where we use a summation convention over $i=1,2$ 
and $\alpha$ is a composite index specifying the current component and 
the initial and final momenta of a given matrix element (for fixed $Q^2$),
which will be discussed below.
The r.h.s. of Eq.~(\ref{eqn:ff_od_system}) is evaluated using Eq.~(\ref{eqn:me_ratio}) 
with computed lattice correlators.

We find the solution of the overdetermined system from a linear fit, 
which minimizes the functional
\begin{equation}
{\mathcal F} =
\sum_{\alpha\beta} \left(A_{\alpha i} F_i - R_\alpha\right) C^{-1}_{\alpha\beta}
  \left(A_{\beta j} F_j - R_\beta\right),
\end{equation}
where $C_{\alpha\beta}$ is the covariance matrix of $R_\alpha$ averages,
$C_{\alpha\beta} = \frac1{N-1}\left(\langle\langle R_\alpha R_\beta\rangle\rangle - 
  \langle\langle R_\alpha\rangle\rangle
  \langle\langle R_\beta\rangle\rangle\right)$,
with the double brackets denoting an ensemble average. 
Using the covariance matrix is crucial as long as the correlation functions prove to be
correlated.

\begin{figure}[h]
\centering
  \begin{minipage}{.49\textwidth}
    \includegraphics[width=\textwidth]{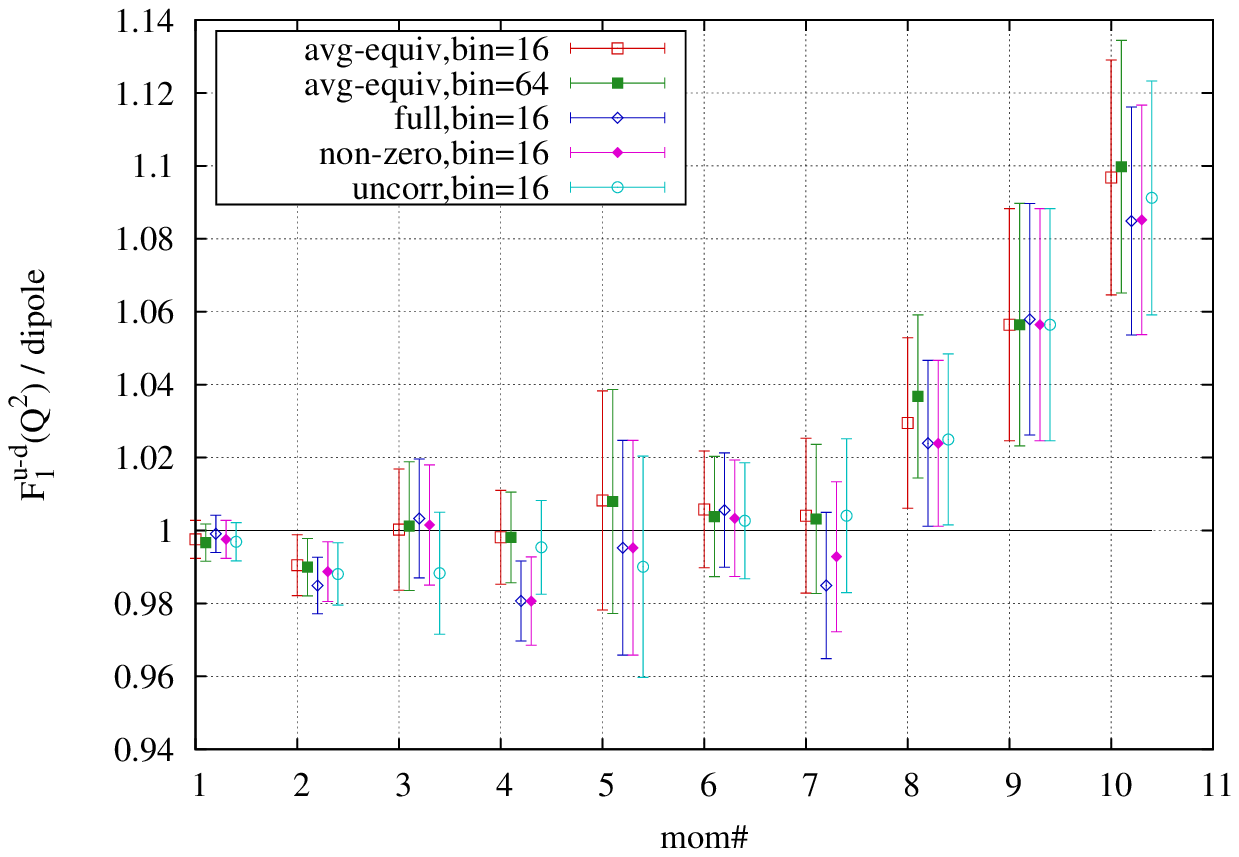}
  \end{minipage}
  \begin{minipage}{.49\textwidth}
    \includegraphics[width=\textwidth]{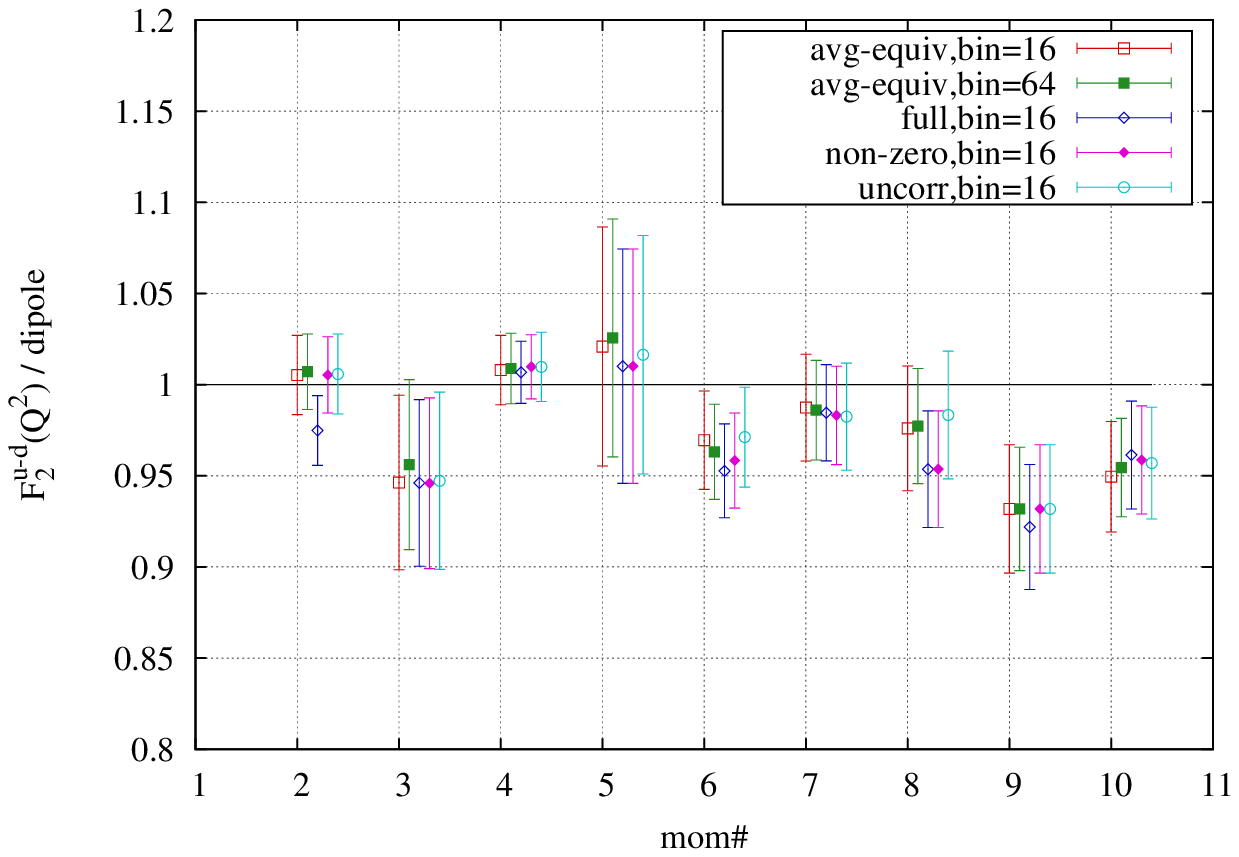}
  \end{minipage}
\caption{\label{fig:solve_od_different_analyses}
  Comparison of the nucleon form factors extracted
  from the full overdetermined system, only non-zero equations, uncorrelated fit and 
  averaged equivalence classes for $m_\pi=297\text{ MeV}$.
  Increased binning of data (eight successive configurations instead of two) 
  shows no increase in estimation of statistical errors.
  Each form factor value is divided by the central value of the dipole fit.
  Tab.~\ref{tab:mom_list} lists the momentum combinations corresponding to each index on the
  horizontal axis.
}
\end{figure}

Since the covariance matrix may be ill-determined, it can introduce
uncontrollable errors into the extracted form factors. 
In general, a covariance matrix is notoriously difficult to reliably estimate 
in a statistical analysis.
To make sure the linear fitting gives a correct result, we repeat the analysis 
with only the diagonal elements of the covariance matrix $C_{\alpha\alpha}$, 
which is equivalent to an uncorrelated linear fit. 
The comparison of these two schemes is presented in Fig.~\ref{fig:solve_od_different_analyses}.
We find that the form factors from an uncorrelated fit are consistent with the correlated fit
results.

The overdetermined system (\ref{eqn:ff_od_system}) contains a subclass of equations 
which have an exactly zero l.h.s.: $A_{\alpha i}=0,\, i=1,2$. 
The measured lattice value of a right-hand side $R_\alpha$ is
not required to be zero, and may be correlated with other matrix elements. 
In an uncorrelated fit, such equations decouple and do not contribute to the solution.
In contrast, the outcome of a correlated fit depends on such values,
thus potentially better utilizing the input from lattice calculations.
In addition, by fitting the equations with a vanishing l.h.s., we check the
symmetries of the electromagnetic vertex (\ref{eqn:em_vertex}), statistically.
Fig.~\ref{fig:solve_od_different_analyses} also shows the agreement of the full overdetermined
system solution and the system without zero l.h.s. equations, 
confirming the consistency of our analysis.

The dimension of the overdetermined system may grow large, 
especially when many momentum combinations are included. 
For example, the most precise point for $Q^2>0$ corresponds to the matrix element
$\langle0,0,0\left|V^\mu(0)\right|1,0,0\rangle$. 
All $V^\mu$ components, together with spatial rotations and reflections give 48 equations, 
only 16 of which are non-zero. 
It is useful to combine all the nucleon matrix elements for fixed $Q^2$
into equivalence classes based on spatial (rotational and reflection) symmetry.
We adopted the following heuristic equivalence criteria\footnote{
  We have not classified the matrix elements according to the hypercubic lattice symmetry
  but instead use relations derived in the continuum.
  Thus these criteria may be thought of as numerical means to improve the condition number 
  of the linear system we need to solve.} 
for three-point functions:
\begin{itemize}
  \item The momenta of the  in- and out-states  must be equivalent under the spatial symmetry.
  \item The corresponding coefficients $A_{\alpha i}$ in Eq.~(\ref{eqn:ff_od_system})
    must be equal up to an overall sign.
  \item The component of the current operator must be temporal or spatial 
    and real or imaginary for both matrix elements being compared.
\end{itemize}

Blocking the measurements in each equivalence class is advantageous for two reasons. 
First, this reduces the dimension of the system of equations~(\ref{eqn:ff_od_system}) and 
the covariance matrix we need to estimate, and we note that blocking strongly correlated values 
improves the covariance matrix condition number.
Second, as long as for the equivalent three-point functions we need spatially equivalent 
two-point functions to build the ratio in Eq.~(\ref{eqn:me_ratio}), we can block the two-point functions separately
\emph{before} computing the ratio. 
This improves the method in Eq.~(\ref{eqn:me_ratio}) by reducing the fluctuations of the two-point
functions in the denominator.

To extract the final set of the form factors, we perform a correlated fit to the reduced 
(i.e., the system with no equations whose l.h.s. is zero) 
overdetermined system with blocked equivalent equations.

%% file: text/lattice_res/isovector_ff.tex
In experiments, the proton and neutron electromagnetic form factors 
(see Eq.~(\ref{eqn:def_nucleon_emff})) are measured separately, 
and the isovector form factors~(\ref{eqn:def_ff_isovector}) can be calculated by taking 
their difference. 
In lattice calculations, the Wick contractions of the quark fields in 
Eq.~(\ref{eqn:def_nucleon_emff}) with nucleon operators 
indicate that disconnected quark loops in the current insertion would
be needed to calculate the proton and neutron form factors separately. 
The calculation of the disconnected quark loops is numerically demanding and 
has not been included in current calculations. 
However, the disconnected loop contributions cancel (in the isospin limit) in the contraction 
of the \emph{difference} of the proton and neutron 
electromagnetic currents in Eq.~(\ref{eqn:def_nucleon_emff}), 
which gives the matrix elements needed for the isovector form factors. We focus our discussion on the isovector form factors in this section. 

After presenting our lattice results for the isovector Dirac and Pauli form factors and the rms radii, 
we will compare chiral extrapolations using the SSE formulation and  covariant baryon chiral perturbation theory\footnote{
  For an analysis of nucleon electromagnetic form factors in baryon ChPT with standard infrared
  regularization, we refer to \cite{Kubis:2000zd}.
}. 
Corresponding results for the connected contributions to the isoscalar form factors will be presented in Sect.~\ref{sect:isoscalar}.

%% file: text/lattice_res/current_renorm.tex
\begin{figure}[ht]
\centering
\includegraphics[width=0.5\textwidth,clip]{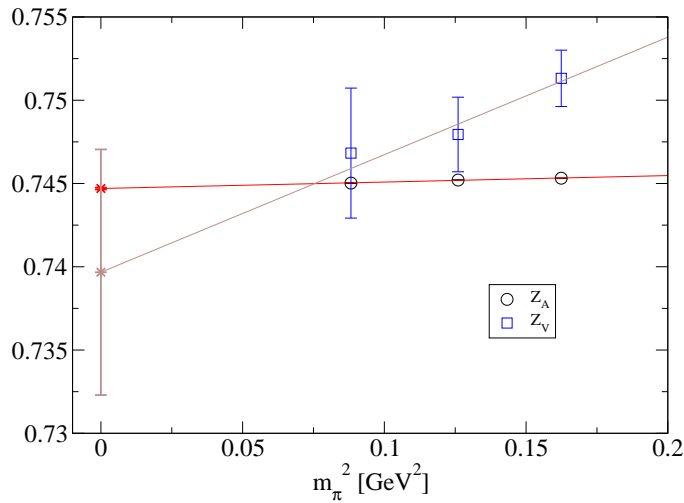}
\caption{\label{fig:Za_extrap}
  Comparison of the vector and axial vector current renormalization constants. 
  In the chiral limit, these two renormalization constants agree within errors.
  The errors on all the $Z_A$  points given in Tab.~\ref{tab:ZVZA} are too small 
  to appear on the figure.
}
\end{figure}

The isovector Dirac form factor at zero momentum transfer, $F_1^{u-d}(0)$, gives the difference of the electric charges for the proton and neutron, which is 1. Since we can measure $F_1^{u-d}(0)$ very accurately on the lattice, we use it to obtain the vector current renormalization constant, $Z_V$, by setting
\begin{equation}
Z_V  F_1^{u-d}(0) = 1.
\end{equation}
Since domain wall fermions have good chiral symmetry, in the chiral limit the vector current renormalization, $Z_V$, and the axial vector current renormalization, $Z_A$, are expected to be the same up to $O(a^2)$ corrections. $Z_A$ is measured by taking the ratio of the point-split five-dimensional conserved axial current to the local four-dimensional current (see Sect.~\ref{sect:pion_obsv} and~\cite{Blum:2000kn}). We show the results of $Z_V$ and $Z_A$ in Tab.~\ref{tab:ZVZA}. Naive linear extrapolations in $m_\pi^2$ to the chiral limit show that $Z_V$ and $Z_A$ are consistent within errors, as is clearly shown in Fig.~\ref{fig:Za_extrap}. 
\begin{table}[ht]
\centering
\caption{\label{tab:ZVZA}
  Vector and axialvector current renormalization constants. 
  The chiral limit values are obtained by linear extrapolations to $m_\pi^2 = 0$. 
}
\begin{tabular}{ccc}
\hline
\hline
$m_\pi$ [MeV] & $Z_V$ & $Z_A$ \\
\hline
297 & 0.7468(39) & 0.745025(24) \\

355 & 0.7479(22) & 0.745207(18) \\

403 & 0.7513(17) & 0.745317(20)\\
\hline
chiral limit & 0.7397(74) & 0.744700(55) \\
\hline
\hline
\end{tabular}
\end{table}

In the following analysis, we renormalize the form factors by $Z_V$ as measured on the corresponding ensemble. That is, we use a
mass-dependent renormalization condition. 
The mass dependence of the renormalization constants is very mild and consistent with the
theoretically expected form $Z_V(g_0)(1+b_v am_q)$ \cite{Bakeyev:2003ff}.

%% file: text/lattice_res/q2_dependence.tex
As will be discussed in the following section, ChPT describes the $Q^2$-dependence of the
form factors for values of $Q^2$ much less than the chiral symmetry breaking
scale (typically of the order of the nucleon mass). Lacking a model-independent
functional form applicable in the large-$Q^2$ region, we study the $Q^2$ dependence using the phenomenological dipole or tripole formula. The Dirac form factor is fixed to 1 at $Q^2 = 0$ under our renormalization scheme, and we use the following one-parameter dipole or tripole formula to describe the $Q^2$ dependence:
\begin{align}
\label{eq:one-par-dipole}
F_1(Q^2) &= \frac{1}{(1+\frac{Q^2}{{M_D}^2})^2} 
  &&\text{(one-parameter dipole)},  \\
\label{eq:one-par-tripole}
F_1(Q^2) &= \frac{1}{(1+\frac{Q^2}{{M_T}^2})^3}
  &&\text{(one-parameter tripole)}.
\end{align}
The Pauli form factor at $Q^2 = 0$, $F_2(0)$, cannot be measured on the lattice directly. We thus fit the data using the two-parameter dipole or tripole formula,
\begin{align}
\label{eq:two-par-dipole}
F_2(Q^2) &= \frac{F_2(0)}{(1+\frac{Q^2}{{M_D}^2})^2}
  &&\text{(two-parameter dipole)}, \\
\label{eq:two-par-tripole}
F_2(Q^2) &= \frac{F_2(0)}{(1+\frac{Q^2}{{M_T}^2})^3}
  &&\text{(two-parameter tripole)}.
\end{align}
We are interested in mean squared Dirac and Pauli radii, which are defined by the slope 
of the form factors at small~$Q^2$:
\begin{equation}
\label{eqn:def_rms_radius}
F_{1,2}(Q^2) = F_{1,2}(0)\left[1 - \frac16\left(r_{1,2}\right)^2 Q^2 + {\mathcal O}(Q^4)\right],
\end{equation}
and are related to the pole masses by
\begin{equation}
\left( r \right)^2 = \frac{12}{{M_D}^2},
\end{equation}
for the dipole fits, and
\begin{equation}
\left( r \right)^2 = \frac{18}{\left(M_T\right)^2},
\end{equation}
for the tripole fits. 

Note that results at different $Q^2$ from the same ensemble may be highly correlated~\cite{Bratt:2008uf}, therefore we perform \emph{correlated} least-$\chi^2$ fits to the data. We investigate the extent to which the dipole and tripole Ans\"atze describe our data and the stability of the fits by varying the maximum $Q^2$ values included in the fits.

In Tab.~\ref{tab:F1_umd_fit_comp} we show the fit results for $F_1^{u-d}(Q^2)$ using the one-parameter dipole and tripole formulae in Eqs.~(\ref{eq:one-par-dipole}) and~(\ref{eq:one-par-tripole}). Comparing the $\chi^2$/dof for the dipole and tripole fits, we see that the dipole fits are slightly preferred when larger $Q^2$ values are included in the fits. However, the Dirac radii determined from both the dipole and tripole fits agree within errors. In general, the dipole form describes the data reasonably well throughout the whole $Q^2$ range for all but one ensemble, the $m_\pi=355$ MeV ensemble, where, when $Q^2$ cutoff is larger than 0.3 GeV$^2$, $\chi^2$/dof becomes very large. This may be due to the fact that this ensemble has the most statistics, and we start to see the deviation from the phenomenological dipole formula.  For the other two ensembles, we can see the general trend that when large $Q^2$ points are included in the fits, the $\chi^2$/dof becomes slightly worse, while the fit parameters do not depend significantly on the choice of the $Q^2$ cutoff, 
indicating that the dipole fits are stable.

We do the same comparison for $F_2^{u-d}(Q^2)$ as shown in Tab.~\ref{tab:F2_umd_fit_comp}.
Judging from the $\chi^2/\text{dof}$ values, we do not see significant differences between the
dipole and tripole fits. 
Since the Pauli form factor is not constrained at $Q^2 = 0$, including larger $Q^2$ in the fits 
does not seem to affect the quality of the fits significantly.
The fit parameters $F_2(0)$ and $M_{D,T}$ prove not to be affected as well. 

\begin{figure}[htbp]
\centering
\includegraphics[width=0.9\textwidth,clip]{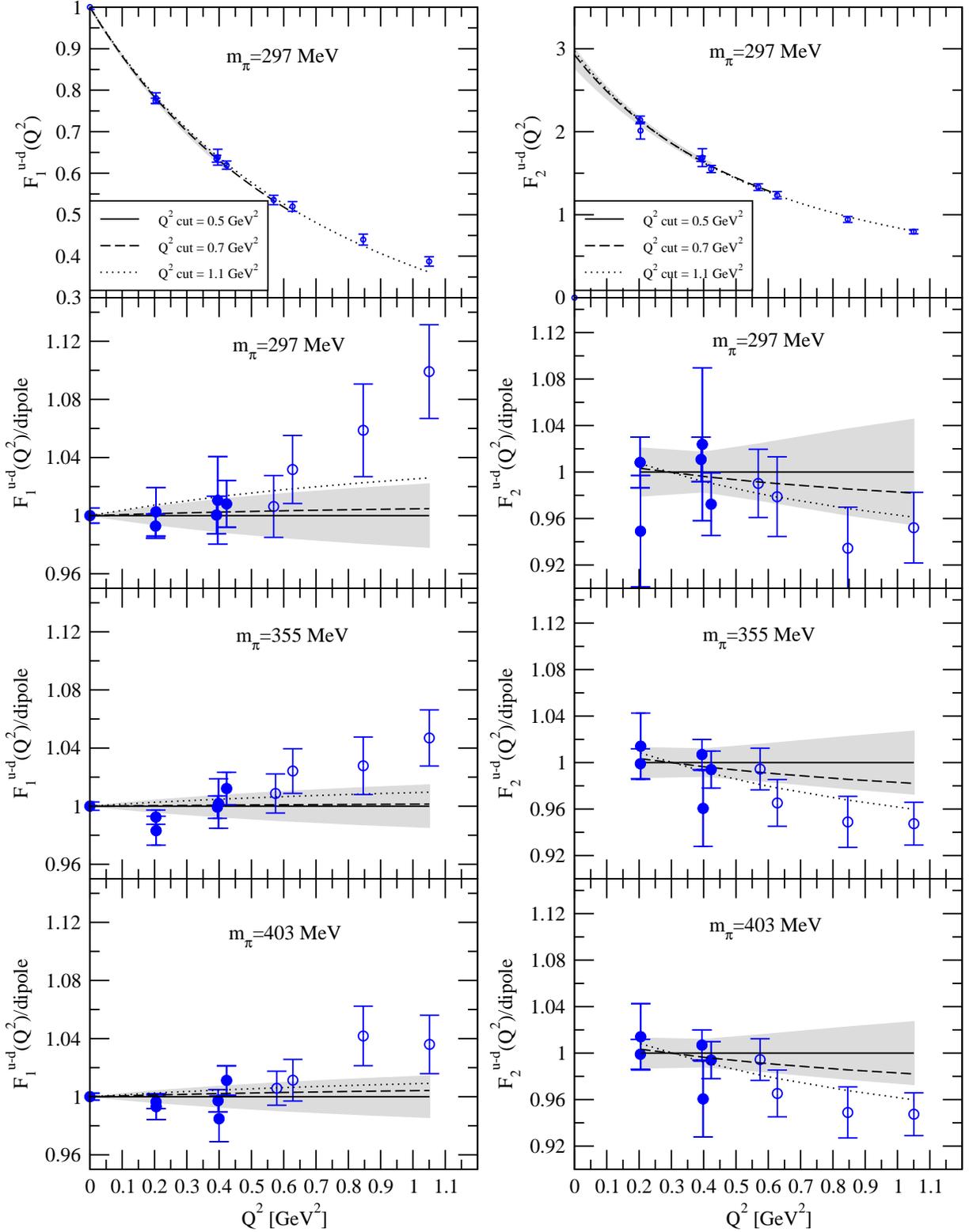}
\caption{\label{fig:data_fit_ratio}
  The top panel shows the lattice results for $F_{1,2}^{u-d}(Q^2)$ at $m_\pi = 297$ MeV along with the dipole fits with three different $Q^2$ cutoffs. The bottom left three panels show the ratios of the lattice results for $F_1^{u-d}$ to the dipole fits 
  using Eq.~(\ref{eq:one-par-dipole}), and the bottom right three panels show the ratios 
  of the lattice results for $F_2^{u-d}$ to the dipole fits using Eq.~(\ref{eq:two-par-dipole}).  
  Only the solid data points are included in the fits with cutoff   $0.5\text{ GeV}^2$, and the
  grey bands show the errors for these fits. 
  The dashed and dotted lines show the ratios of dipole fits at cutoffs $0.7\text{ GeV}^2$ 
  and $1.1\text{ GeV}^2$ relative to the fit at $0.5\text{ GeV}^2$.
  }
\end{figure}

As an example, we show the dipole fit curves with a $Q^2$ cutoff at 0.5, 0.7 and 1.1 GeV$^2$ for the $m_\pi = 297$ MeV ensemble in the top panel of  Fig.~\ref{fig:data_fit_ratio}. To show the quality of the fits more clearly, we plot the ratios of the form factor data to the dipole fit with the $Q^2$ cutoff at 0.5 GeV$^2$ in the bottom three panels of Fig.~\ref{fig:data_fit_ratio}. The error bands reflect the jackknife errors in the dipole fit parameters. 
We see that although the data included in the fits can be described reasonably well by the dipole formula  with discrepancies that are generally within 
two to three standard deviations, the clear systematic tendency indicates that the dipole Ansatz is not a good description of the data over the whole momentum transfer region. In particular, for $F_1^{u-d}$, the precisely measured points in the region of 0.2 GeV$^2$ are systematically lower than the dipole fit, whereas at high $Q^2$, the lattice data are systematically higher.  For $F_2^{u-d}$, the high $ Q^2$ lattice data are systematically lower than the dipole fit.
This is consistent with the empirical fits to the experimental data in 
Refs.~\cite{Friedrich:2003iz,Arrington:2007ux}, where the phenomenological corrections 
to the dipole form are negative in the region of $0.2\text{ GeV}^2$ 
and positive at about $0.4\text{ GeV}^2$.
For comparison, we also plot the dipole fits with $Q^2$ cutoffs at 0.7 GeV$^2$ (dashed line) and 1.1 GeV$^2$ (dotted line) relative to the 0.5 GeV$^2$ dipole fit (solid line). The differences between different $Q^2$ cutoffs are small, indicating that the fits are stable. 

It is worth noting that the Dirac and Pauli radii, $r_1^v$ and $r_2^v$, and the anomalous magnetic moment, $\kappa_v$, are defined in the $Q^2 = 0$ limit. We thus restrict the fits to the smallest $Q^2$ points possible to extract these quantities while still including enough data points to constrain the fits. For uniformity we choose to determine these quantities from the one-parameter dipole fits for $F_1^{u-d}$, and the two-parameter dipole fits for $F_2^{u-d}$, with a $Q^2$ cutoff at 0.5 GeV$^2$. 

\begin{figure}[htbp]
\centering
\includegraphics[width=0.9\textwidth,clip]{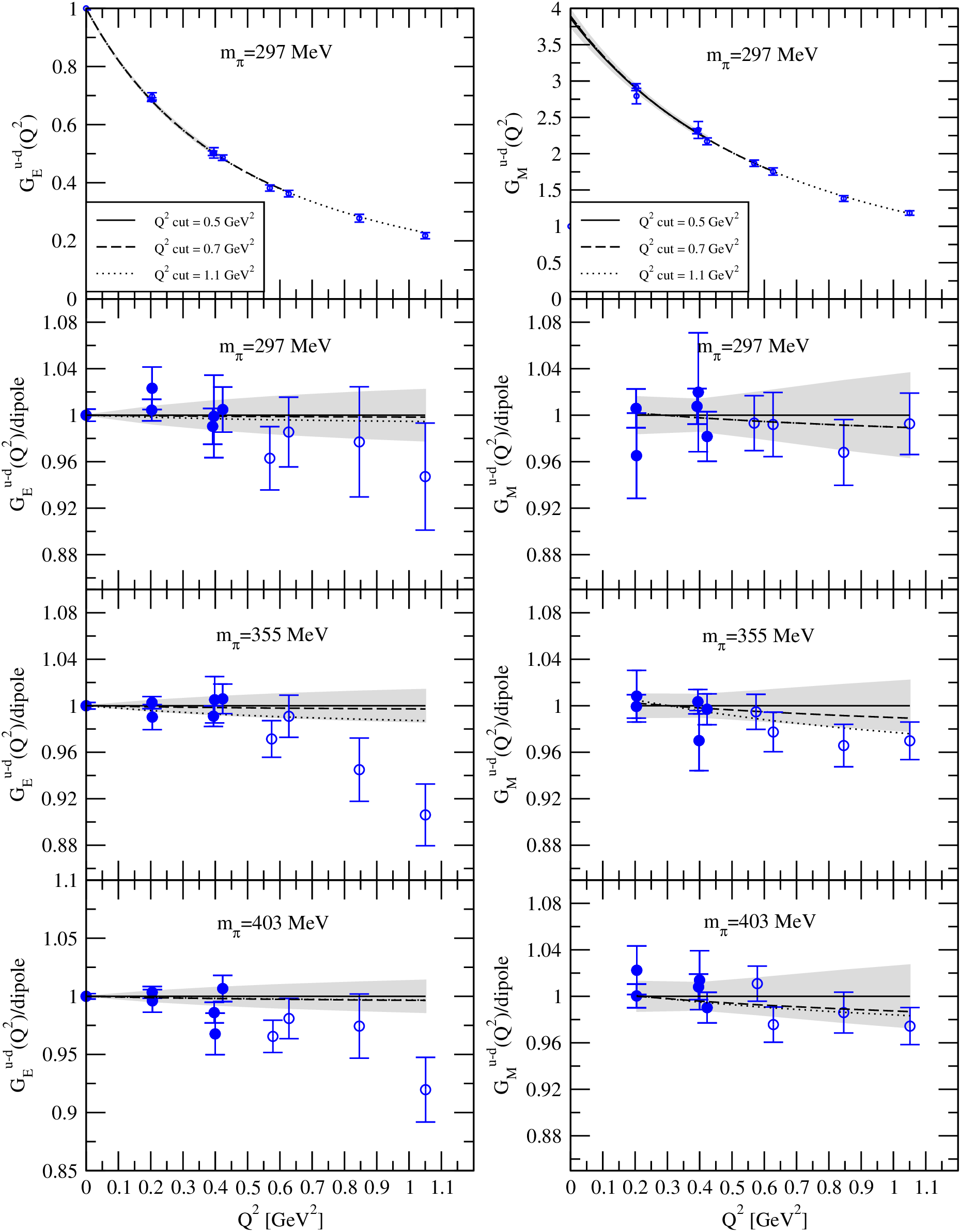}
\caption{\label{fig:Ge_Gm_data_fit_ratio}
 The top panel shows the lattice results for $G_{E,M}^{u-d}(Q^2)$ at $m_\pi = 297$ MeV along with the dipole fits with three different $Q^2$ cutoffs. The bottom left three panels show the ratios of the lattice results for $G_E^{u-d}$ to the dipole fits 
  using Eq.~(\ref{eq:one-par-dipole}), and the bottom right three panels show the ratios 
  of the lattice results for $G_M^{u-d}$ to the dipole fits using Eq.~(\ref{eq:two-par-dipole}).  
  Only the solid data points are included in the fits with cutoff   $0.5\text{ GeV}^2$, and the
  grey bands show the errors for these fits. 
  The dashed and dotted lines show the ratios of dipole fits at cutoffs $0.7\text{ GeV}^2$ 
  and $1.1\text{ GeV}^2$ relative to the fit at $0.5\text{ GeV}^2$.
  }
\end{figure}

We also perform dipole fits to $G_E(Q^2)$ and $G_M(Q^2)$ to see how well the dipole Ansatz 
describes the data. 
We find that the dipole fits to $G_E^{u-d}$ and $G_M^{u-d}$ are qualitatively similar 
to $F_1^{u-d}$ and $F_2^{u-d}$. 
However, it appears that the fits are even more stable over the whole range of $Q^2$ than Dirac 
and Pauli form factors. 
This is indicated by little change in the ratio plots in Fig.~\ref{fig:Ge_Gm_data_fit_ratio} 
with different $Q^2$ cutoffs.

\begin{figure}[htbp]
\centering
\includegraphics[width=0.5\textwidth,clip]{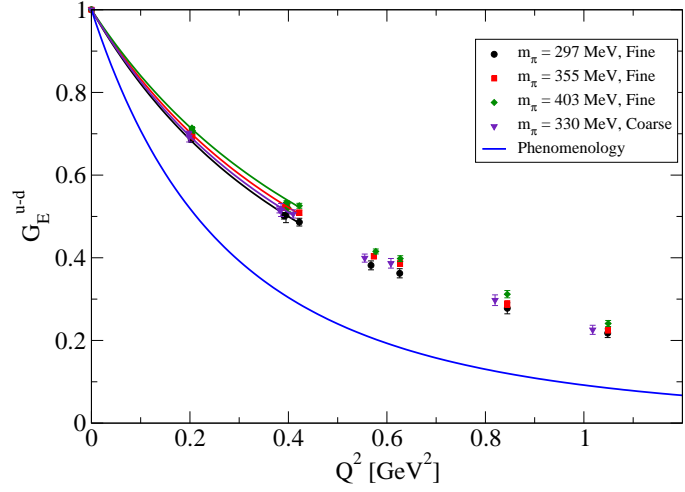}
\caption{Lattice results for $G_E^{u-d}$ at three pion masses from the fine ensembles and one pion mass from the coarse ensemble, compared with a phenomenological fit to the experimental data as parameterized in Ref.~\cite{Kelly:2004hm}. The solid curves are the dipole fits to the form factor results with a cutoff at $Q^2 = 0.5$ GeV$^2$.}
\label{fig:Geumd_all_comp_w_exp}
\end{figure}

Figure~\ref{fig:Geumd_all_comp_w_exp} shows a comparison of the lattice results for $G_E$ at
three different pion masses from the fine ensembles and one pion mass from the coarse ensemble
with  a phenomenological fit to the experimental data using the parameterization in Ref.~\cite{Kelly:2004hm} (with no indication of the  experimental errors).
The solid
curves are dipole fits to the form factor results with the $Q^2$ cutoff at 0.5 GeV$^2$. As the
pion mass decreases, the slope of the form factors at the small momentum transfer monotonically
increases.  The results from the coarse ensemble at $m_\pi = 330$ MeV is nicely surrounded
by the results from the fine ensembles at $m_\pi = 297$ and $m_\pi = 355$ MeV, 
indicating that the effect of the finite lattice spacing error should be small. 

%% file: text/lattice_res/chiral_extrap.tex
\subsubsection{\label{sec:SSE}
  Chiral extrapolations using ${\mathcal O}(\epsilon^3)$ small scale expansion}
To compare the lattice results for the nucleon form factors at 
finite momentum transfer with the experimental results, we  need to do extrapolations
for both the $m_\pi$ and $Q^2$ dependence using baryon chiral perturbation
theory. This combined dependence has been worked out both in SSE at leading one loop
accuracy and in CBChPT up to NNLO order. In particular, the 
${\mathcal O}(\epsilon^3)$ expression for the isovector Dirac form factor $F_1^{u-d}(Q^2, m_\pi)$
has been derived in Ref. \cite{Bernard:1998gv} and is given by
\begin{eqnarray} 
F_{1}^{u-d}(Q^2,m_\pi) &=&  
1 + \frac{1}{(4\pi F_\pi)^2}\left\{ -Q^2 \left(\frac{68}{81} c_A^2 
    - \frac{2}{3}g_{A}^2-2 B_{10}^{(r)}(\lambda) \right) \right. 
- Q^2 \left(\frac{40}{27} c_A^2-\frac{5}{3}g_{A}^2
  -\frac{1}{3}\right)
\log\left[\frac{m_\pi}{\lambda}\right]  
\nonumber \\ 
     & & + \int_{0}^{1}dx \left[\frac{16}{3}\Delta^2 c_A^2
          + m_{\pi}^2 \left(3 g_{A}^2+1-\frac{8}{3} c_A^2 \right) \right.
        + \left. Q^2 x(1-x)\left(5 g_{A}^2+1
        -\frac{40}{9}c_A^2\right)\right] 
        \log\left[\frac{\tilde{m}^2}{m_{\pi}^2}\right]  
\nonumber  \\
     & & + \int_{0}^{1}dx \left[ -\frac{32}{9} c_A^2 Q^2 x(1-x) 
       \frac{\Delta \log R(\tilde{m})}{\sqrt{\Delta^2-\tilde{m}^2}} \right] 
\nonumber \\  
    & & -  \left. \int_{0}^{1}dx \; \frac{32}{3}c_A^2 \Delta 
      \left[\sqrt{\Delta^2-m_{\pi}^2}\log R(m_\pi) 
      -\sqrt{\Delta^2-\tilde{m}^2}\log R(\tilde{m}) \right] \right\}, 
 \label{eqn:F1umd-SSE}
\end{eqnarray}
where
\begin{eqnarray}
R(m) & = & \frac{\Delta}{m} + \sqrt{\frac{\Delta^2}{m^2}-1}, \\
\tilde{m}^2 &=& m_\pi^2 + Q^2 x (1-x). 
\end{eqnarray}
In the above expressions, $F_\pi$ denotes the pion decay constant in the $SU(2)$ chiral limit 
with the convention in Eq.~(\ref{eqn:def_fpi_phys}).
Here $g_A$ is the nucleon axial charge in the $SU(2)$ chiral limit, 
$c_A$ is the leading-order pion-nucleon-$\Delta$ coupling\footnote{
  The coupling $c_A$ corresponds to $\dot{g}_{\pi N \Delta}$ in the notation 
  of Ref.~\cite{Hemmert:1997ye}.}, and
 $\Delta$ denotes the $\Delta\,(1232)$-nucleon mass splitting 
in the $SU(2)$ chiral limit.
For more  details on the effective Lagrangians and the definitions of the low-energy constants, 
we refer the reader to~\cite{Bernard:1998gv}.

To the same order, the expression for the isovector Pauli form factor, $F_2^{u-d}$,
is also derived in~\cite{Bernard:1998gv} and is given as
\begin{eqnarray}
 F_2^{u-d}(Q^2, m_\pi) 
 &=&  \kappa_v (m_\pi) - g_{A}^2
 \frac{4\pi M_N}{(4\pi F_\pi)^2} 
 \int_{0}^{1}dx\left[ \sqrt{\tilde{m}^2}-m_{\pi}\right]
 \nonumber\\
 & & +\frac{32c_A^2 M_N \Delta}{9 (4\pi F_\pi)^2}\int_{0}^{1}dx 
         \left[ \frac{1}{2}\log\left[\frac{\tilde{m}^2}{4\Delta^2}\right]
         -\log\left[\frac{m_\pi}{2\Delta}\right] \right.  
\nonumber \\
  & & \left. +\frac{\sqrt{\Delta^2-\tilde{m}^2}}{\Delta} \log R(\tilde{m})
              -\frac{\sqrt{\Delta^2-m_{\pi}^2}}{\Delta} \log R(m_\pi)
            \right],
\label{eqn:F2umd-SSE}
\end{eqnarray}
where, to ${\mathcal O}(\epsilon^3)$, 
\begin{eqnarray}
\kappa_v (m_\pi)&= &\kappa_v^0 - \frac{g_A^2 m_\pi M_N}{4 \pi F_\pi^2} \nonumber \\
& & + \frac{2c_A^2 \Delta M_N}{9\pi^2 F_\pi^2} \left \{ \sqrt{1-\frac{m_\pi^2}{\Delta^2}}
\log\left[R(m_\pi)\right]
 +\log\left [ \frac{m_\pi}{2\Delta} \right ] \right \} 
 + {\mathcal O}(m_\pi^2). 
\label{eqn:kappa}
\end{eqnarray}

In order to capture the most prominent ${\mathcal O}(m_\pi^2)$ corrections,
Hemmert and Weise~\cite{Hemmert:2002uh} proposed a modification of the standard SSE 
power counting to promote the leading term of the magnetic $N \to \Delta$
transition into the first order $N \Delta$ effective Lagrangian. 
This leads to the following expression for $\kappa_v(m_\pi)$:
\begin{eqnarray}
\label{eqn:kappa-SSE}
\kappa_v(m_\pi)&=&\kappa_{v}^0-\frac{g_A^2\,m_\pi M_N}{4\pi F_\pi^2}
+   \frac{2 c_A^2 \Delta M_N}{9\pi^2F_\pi^2}
\left\{\sqrt{1-\frac{m_\pi^2}{\Delta^2}}\log\left[R(m_\pi)\right]+\log\left[\frac{m_\pi}{2\Delta}\right]
\right\} \nonumber\\
& &               -   8 E_1^r(\lambda) M_Nm_\pi^2
+  \frac{4c_A c_V g_A M_N m_\pi^2}{9\pi^2F_\pi^2}\log\left[\frac{2\Delta}{\lambda}\right] \nonumber \\
& &            +  \frac{4c_A c_V g_A M_N m_\pi^3}{27\pi F_\pi^2\Delta}
-   \frac{8 c_A c_V g_A \Delta^2 M_N}{27\pi^2F_\pi^2}
\left\{\left(1-\frac{m_\pi^2}{\Delta^2}\right)^{3/2}\log\left[R(m_\pi)\right] +\left(1-\frac{3m_\pi^2}{2\Delta^2}\right)
  \log\left[\frac{m_\pi}{2\Delta}\right]
\right\},
\end{eqnarray}
where $c_V$ is the leading magnetic photon-nucleon-$\Delta$ coupling in the chiral limit and 
$\kappa_v^0$ denotes the anomalous magnetic moment in the $SU(2)$ chiral limit. 
We will use this expression in our analysis. 

Our results for the form factor $F_2$ are given in terms of a quark mass dependent 
``magneton'' (see Eq.~(\ref{eqn:em_vertex})), 
which is not accounted for in SSE at the order at which we are working. 
Therefore, in order to compare Eq.~(\ref{eqn:kappa-SSE}) with our lattice data, 
we follow Refs.\cite{Gockeler:2003ay, Alexandrou:2006ru} 
and define $\kappa^\text{norm}$ measured relative to the physical magneton:
\begin{equation}
\kappa^{\text{norm}} = 
  \frac{M_N^{\text{phys}}}{M_N^{\text{lat}}} \kappa^{\text{lat}} =
  \frac{M_N^{\text{phys}}}{M_N^{\text{lat}}} F_2(0).
\label{eq:kappa-norm}
\end{equation}
We then identify $M_N$ in the SSE expressions as the physical nucleon mass. 
In the following comparisons of our results with chiral perturbation theories, 
the normalized magnetic moment $\kappa_v^\text{norm}$ will be used throughout, 
and we drop the superscript ``norm''  unless there is an ambiguity. 

ChPT describes the $Q^2$-dependence of the form factors for values of $Q^2$ much less 
than the chiral symmetry breaking scale (typically of the order of the nucleon mass)
and $Q^2$ counts as a small quantity, of the order of $m_\pi^2$. 
In fact, we have attempted simultaneous fits to both the $m_\pi$ and $Q^2$ dependences 
of $F_1^{u-d}$ using the SSE formula in Eq.~(\ref{eqn:F1umd-SSE}), and found that the fits fail to
describe data even with $Q^2 \leq 0.4\text{ GeV}^2$ ($\chi^2/\text{dof} \approx 10$). 
This is consistent with the findings of Ref.~\cite{Bernard:1998gv}, 
where the applicability of the
${\mathcal O}(\epsilon^3)$ SSE results for the isovector nucleon form factors at physical pion
mass was found to be limited to $Q^2<0.2\text{ GeV}^2$.
Lacking a model-independent functional form applicable in the large-$Q^2$ region, 
we resort to studying the pion mass dependence of the mean squared Dirac
radius, $(r_1^v)^2$, Pauli radius, $(r_2^v)^2$, 
and the anomalous magnetic moment, $\kappa_v$, as obtained from the dipole
fits discussed in Sect.~\ref{sect:q2_dependence}. We tabulate these values in Tab.~\ref{tab:isovector_radii}.
\begin{table}[ht]
\centering
\caption{Results for the isovector Dirac and Pauli radii and anomalous magnetic moment from dipole fits with $Q^2\leq0.5$ GeV$^2$.}\label{tab:isovector_radii}
\begin{tabular}{ccccc}
\hline\hline
$m_\pi$ [MeV] & $(r_1^v)^2 $ [GeV$^{-2}$] &  $(r_2^v)^2$ [GeV$^{-2}$] & $\kappa_v^\text{norm} \cdot (r_2^v)^2$ [GeV$^{-2}$] & $\kappa_v^\text{norm}$ \\
\hline
297 & 7.83(21) & 9.82(84) & 24.1(3.0) & 2.447(99) \\
355 & 7.23(14) & 9.55(46) & 24.1(1.7) & 2.518(57) \\
403 & 6.98(13) & 9.74(41) & 24.5(1.5) & 2.508(51) \\
\hline
330 & 7.46(22) & 11.44(67)& 31.6(2.7) & 2.758(84) \\
\hline\hline
\end{tabular}
\end{table}

The ${\mathcal O}(\epsilon^3)$ SSE formulae for $(r_1^v)^2$ and $ (r_2^v)^2$ can
be derived from Eqs.~(\ref{eqn:F1umd-SSE}) and~(\ref{eqn:F2umd-SSE}),
respectively, and are given by
\begin{gather}
\label{eqn:r1umd-SSE}
\begin{aligned} 
\left(r_{1}^v\right)^2 = &
    -  \frac{1}{(4\pi F_\pi)^2}\left\{1+7 g_{A}^2 +
  \left(10 g_{A}^2 +2\right) \log\left[\frac{m_\pi}{\lambda}\right]\right\} 
\\
& - \frac{12 {B_{10}^{(r)}(\lambda)}}{(4\pi F_\pi)^2}  
     +  \frac{c_A^2}{54\pi^2 F_{\pi}^2}\Bigg\{
        26+30\log\left[\frac{m_\pi}{\lambda}\right]
\\
&         +30\frac{\Delta}{\sqrt{\Delta^2-m_{\pi}^2}}
             \log\left[\frac{\Delta}{m_\pi}
            +\sqrt{\frac{\Delta^2}{m_{\pi}^2}-1}\right] \Bigg\}
    + {\mathcal O}(m_\pi)
  \,,
\end{aligned}
\\
\label{eqn:r2umd-SSE}
\kappa_v(m_\pi)\cdot (r_{2}^v)^2 = 
  \frac{g_{A}^2 M_N}{8 \pi F_{\pi}^2 m_\pi} 
  + \frac{c_A^2 M_N}{9 \pi^2 F_{\pi}^2 \sqrt{\Delta^2-m_{\pi}^2}} 
    \log\left[\frac{\Delta}{m_\pi}+\sqrt{\frac{\Delta^2}{m_{\pi}^2}-1}\right]
    + {\mathcal O}(m_\pi^0)
  .
\end{gather}

Together with the expression for the anomalous magnetic moment in Eq.~(\ref{eqn:kappa-SSE}), 
these formulae involve six low-energy constants: $F_\pi$, $\Delta$, $c_A$,  $g_A$, $\kappa_v^0$ 
and $c_V$, as well as two counter terms: $B_{10}^r(\lambda)$ and $E_1^r(\lambda)$. 
Ideally we would like to determine all these constants from simultaneous fits to lattice results.
However, this is not feasible with the  limited number of measured observables and pion masses in the present calculation, and we thus fix some of the
low-energy constants using their phenomenological values.
We describe our choices for these values below.

In Ref.~\cite{Colangelo:2003hf}, Colangelo and D\"urr analyze numerically the NNLO expression 
for the pion mass dependence of $F_\pi$ \cite{Bijnens:1998fm}.
They use available information from phenomenology to fix all low-energy constants but the 
chiral limit value of $F_\pi$, use the physical value~(\ref{eqn:def_fpi_phys}) and obtain
\begin{equation}
F_\pi\big|_\text{chiral limit} = (86.2 \pm 0.5)\text{ MeV}.
\end{equation}

In the absence of reliable chiral extrapolations of both nucleon and $\Delta\,(1232)$ masses 
(see the discussion in Ref.~\cite{WalkerLoud:2008bp})\footnote{
  For an analysis of the quark mass dependence of nucleon and delta masses in the covariant SSE 
  at order $\epsilon^4$ we refer to \cite{Bernard:2005fy}.
}, 
we identify the $\Delta$-nucleon mass
splitting in the chiral limit with its value at the physical $m_\pi$. 
The position of the $\Delta\,(1232)$ resonance pole in the total center-of-mass energy plane
has been determined from magnetic dipole and electric quadrupole amplitudes of pion 
photoproduction.
According to the Particle Data Group average~\cite{Amsler:2008uo}, the $\Delta$-pole position
leads to $M_\Delta=(1210\pm1)\text{ MeV}$ and $\Gamma_\Delta=(100\pm2)\text{ MeV}$.
If one instead defines the $\Delta\,(1232)$ mass and width by looking at the $90^\text{o}$
$\pi N$ phase shift in the spin-3/2 isospin-3/2 channel, the PDG averages give
$M_\Delta=(1232\pm1)\text{ MeV}$ and $\Gamma_\Delta=(118\pm2)\text{ MeV}$.
With $M_N=939\text{ MeV}$, one obtains, respectively,
\begin{equation}
\Delta=(271 \pm 1)\text{ MeV},
\end{equation}
or 
\begin{equation}
\Delta=(293 \pm 1)\text{ MeV}.
\end{equation}

The $\Delta\,(1232)$ decays strongly to a nucleon and a pion with almost $100\%$
branching fraction. From the PDG values of masses and widths~\cite{Amsler:2008uo} and from
\begin{equation}
\label{eqn:DeltaNdecaywidth}
\Gamma_{\Delta \to N \pi}=\frac{c_A^2}{12 \pi\, F_\pi^2\,
M_\Delta}\,(E_\pi^2-m_\pi^2)^{3/2}\, (M_\Delta+M_N-E_\pi),
\end {equation}
where
\begin{equation}
E_\pi=\frac{M_\Delta^2-M_N^2+m_\pi^2}{2 M_\Delta}~,
\end{equation}
one obtains, respectively,
\begin{eqnarray}
|c_A|&=& 1.50 \dots 1.55
  \quad\quad\text{ if } \Gamma=(100 \pm 2)\text{ MeV}
  \text{ and } \Delta = (271 \pm 1)\text{ MeV};\\
|c_A|&=& 1.43 \dots 1.47
  \quad\quad\text{ if } \Gamma=(118 \pm 2)\text{ MeV}
  \text{ and } \Delta = (293 \pm 1)\text{ MeV}.
\end{eqnarray}
Calculating the strong decay width of $\Delta\,(1232)$ to leading order in (non-relativistic)
SSE kinematics, one obtains
\begin{equation}
\Gamma_{\Delta \to N \pi} = \frac{c_A^2}{6 \pi F_\pi^2 } (\Delta^2 - m_\pi^2)^{3/2}. 
\end{equation}
We note that this expression corresponds to the leading term in a $1/M_N$ expansion of the
result given in Eq.~(\ref{eqn:DeltaNdecaywidth}), which utilizes the full covariant kinematics.
Using the ranges of masses and decay widths mentioned above, this expression yields 
the lower values 
\begin{eqnarray}
|c_A|&=& 1.11 \dots 1.14
  \quad\quad\text{ if } \Gamma=(100 \pm 2)\text{ MeV}
  \text{ and } \Delta = (271 \pm 1)\text{ MeV};\\
|c_A|&=& 1.04 \dots 1.07
  \quad\quad\text{ if } \Gamma=(118 \pm 2)\text{ MeV}
  \text{ and }\Delta = (293 \pm 1)\text{ MeV}.
\end{eqnarray}
Furthermore, $SU(4)$ spin-flavor quark symmetry gives $c_A=3g_A /(2 \sqrt{2})=1.34$. 

Chiral extrapolations of different sets of lattice 
results~\cite{Hemmert:2003cb, Edwards:2005ym, Procura:2006gq, Khan:2006de}  
based on SSE at leading-one-loop accuracy lead to a chiral limit value for $g_A$ of about 1.2. 
From the relativistic tree-level
analysis of the process of pion photoproduction at threshold $\gamma p \to
\pi^0 p$, one obtains~\cite{Davidson:1991xz,Hemmert:1996xg} (for $g_{\pi N \Delta} =1.5$)
\begin{equation}
c_V= (-2.5 \pm 0.4)\text{ GeV}^{-1}.
\end{equation}

As specified above, at the order ${\mathcal O}(\epsilon^3)$, all the couplings in 
Eqs.~(\ref{eqn:F1umd-SSE}--\ref{eqn:r2umd-SSE}) are meant to be taken in the chiral limit. 
Replacing them with the corresponding quantities at the physical point
amounts to the inclusion of higher-order effects.
As long as the deviation between the values in the chiral limit and at the physical point  
is small, one expects such a replacement to yield little effect. 
To test this statement, in some cases we have performed the chiral fits using both 
the physical values and the chiral limit values for the low-energy constants and found 
no significant differences. 
In the following we will only present results obtained using the chiral limit values as inputs, 
which are summarized in Tab.~\ref{tab:input-SSE}.
\begin{table}[ht]
  \centering
  \caption{Input values for the low-energy constants in the fits.}
  \label{tab:input-SSE}
 \begin{tabular}{cccc}
   \hline \hline
     $g_A$ & $F_\pi$ [GeV] & $\Delta$ [GeV] \\
   
      1.2 & 0.0862 &  0.293 \\
   \hline \hline
  \end{tabular}
\end{table}

Among the low-energy constants discussed above, $c_A$ and $c_V$ are the two least known.
In addition, we have little knowledge of the counter-terms, $B_{10}^r(\lambda)$ and
$E_1^r(\lambda)$, as well as the anomalous magnetic moment in the chiral limit, $\kappa_v^0$, 
from phenomenology. 
Lattice calculations in the chiral regime have the potential to constrain these parameters 
to unprecedented accuracy.
Our attempt here is to check the consistency of our data with the predictions 
of chiral effective field theories, to estimate the range of applicability of the ChPT formulas, and to determine these low-energy constants when the formulas are applicable. 
Since $c_A$ appears in the formulas for $(r_1^v)^2$, $(r_2^v)^2$ 
and $\kappa_v$, a simultaneous fit to all these three quantities would give a better constraint 
for the value of $c_A$. 
However, we have only three data points for each of these quantities, 
and $\kappa_v$ alone has four parameters, three of which ($c_V$, $E_1^r(\lambda)$ and $\kappa_v^0$)
are not constrained by any other quantity.
Thus the quark-mass dependence of $\kappa_v$ cannot be used to constrain $c_A$.
Therefore we choose to fit simultaneously\footnote{
  We note however, that in Ref.~\cite{Gockeler:2003ay} it was already observed that the leading
  one-loop SSE formula for $(r_1^v)^2$ (Eq.~(\ref{eqn:r1umd-SSE})) is dominated by the leading 
  chiral logarithm and dropped below the level of the 
  lattice data available at that time
  for values of the pion mass as low as $m_\pi<200\text{ MeV}$. 
  This prompted the authors of Ref.~\cite{Gockeler:2003ay} to exclude the isovector Dirac
  radius from the simultaneous fit.
  Likewise, the authors of Ref.~\cite{Alexandrou:2006ru} obtained huge, unrealistic values for
  the isovector Dirac radius when trying to enforce a fit of the logarithm-dominated behavior 
  onto their data.
  Given these two negative precedents, we consider our ``fit'' to the isovector Dirac radius
  data to be of exploratory nature, testing the limits of applicability of the leading one-loop
  SSE results given in Eq.~(\ref{eqn:r1umd-SSE}).
}
only $(r_1^v)^2$ and $\kappa_v\cdot (r_2^v)^2$ 
to determine $c_A$ and $B_{10}^r(\lambda)$, and then use the resulting $c_A$  as an input 
for the fit to $\kappa_v$.  
This way the three free parameters in $\kappa_v$ are exactly specified by the three data points.

\begin{figure}[htbp]
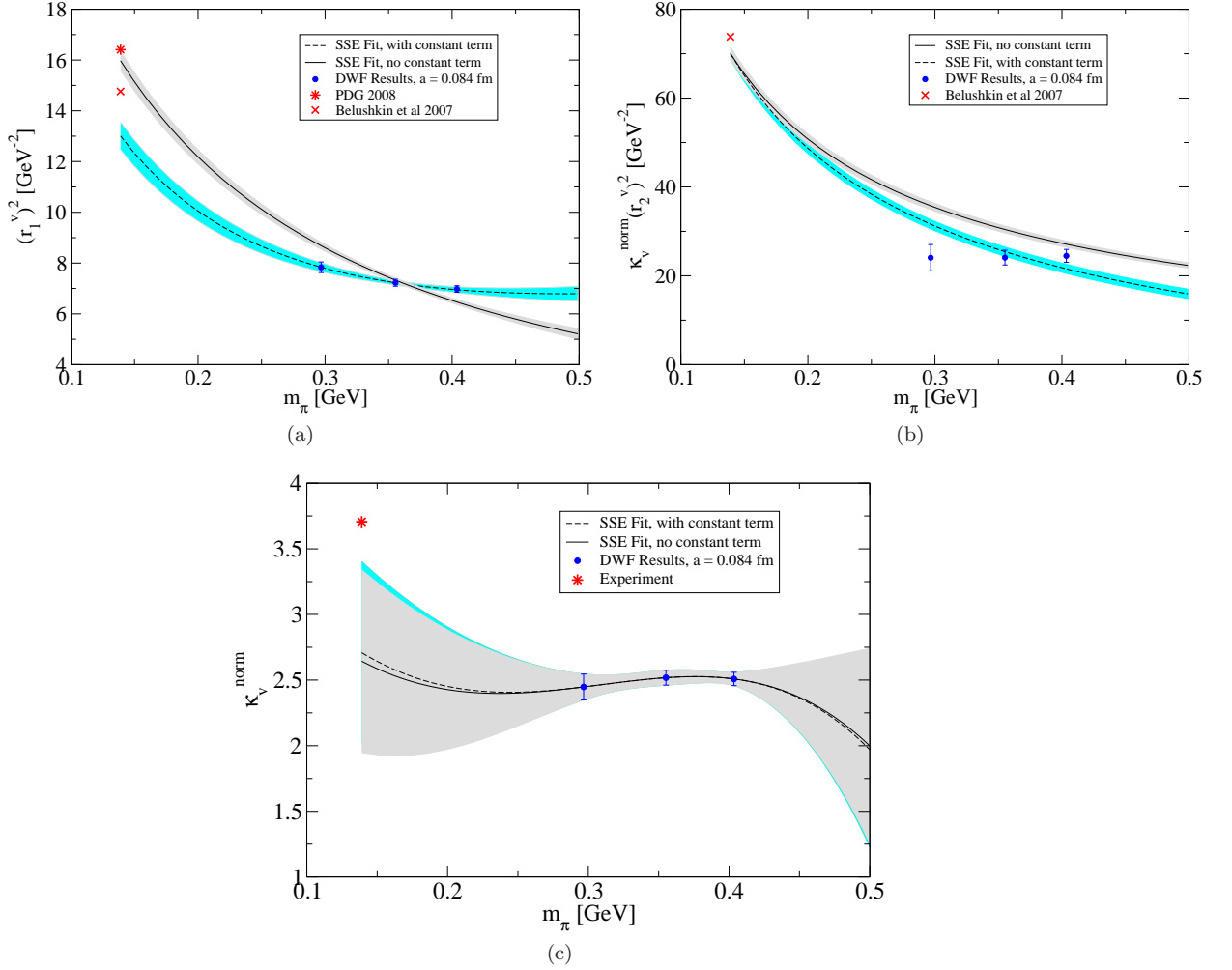

\hfill
\centering
\subfigure[]{
\includegraphics[width=0.45\textwidth,clip]{figs/chi_extrap/r1v_SSE}
\label{fig:SSE_r1v}
}
\subfigure[]{
\includegraphics[width=0.45\textwidth,clip]{figs/chi_extrap/r2v_SSE}
\label{fig:SSE_r2v}
}\\
\subfigure[]{
\includegraphics[width=0.5\textwidth,clip]{figs/chi_extrap/kappa_SSE}
\label{fig:SSE_kappa}
}
\caption{\label{fig:SSE}
  Chiral extrapolations for the isovector Dirac radius, Pauli radius and 
  the anomalous magnetic moment using the ${\mathcal O}(\epsilon^3)$ SSE formula with 
  (solid curves) or without (dashed curves) the constant term in Eq.~(\ref{eqn:r2umd-SSE-core}). 
  In both cases, $(r_1^v)^2$ and $\kappa_v\cdot(r_2^v)^2$ are fit simultaneously, 
  while $\kappa_v$ is fit separately with $c_A$ determined from the simultaneous fit.
}
\end{figure}

We present the resulting $\chi^2$/dof and fit parameters normalized at scale 
$\lambda=600\text{  MeV}$ in the first row of Tab.~\ref{tab:fit-SSE} and 
plot the fit curves as the solid lines in Fig.~\ref{fig:SSE}.  
As indicated by a $\chi^2$/dof of 17,  the simultaneous fit to $(r_1^v)^2$ 
and $\kappa_v\cdot (r_2^v)^2$ does not describe the data.  
The problem is that our results for $(r_1^v)^2$ and $\kappa_v\cdot(r_2^v)^2$ favor different values 
for $c_A$. 
In fact, an independent fit to $(r_1^v)^2$ yields $c_A = 1.98(7)$, while an independent fit 
to $\kappa_v\cdot(r_2^v)^2$ gives $c_A = 1.39(10)$. 
The tension between these two quantities results in the large $\chi^2$/dof in the simultaneous fit, 
indicating that the formulae given in Eqs.~(\ref{eqn:r1umd-SSE}) and~(\ref{eqn:r2umd-SSE}) 
do not describe our data consistently. 
As we can see from Fig.~\ref{fig:SSE_r2v}, the solid fit curve lies systematically higher 
than the data points, which then motivates us to add the ${\mathcal O}(m_\pi^0)$ correction 
to the leading one-loop result of Eq.~(\ref{eqn:r2umd-SSE}) (the so-called ``core'' 
contribution in Ref.~\cite{Gockeler:2003ay}) to $\kappa_v\cdot(r_2^v)^2$, such that

\begin{eqnarray}
\label{eqn:r2umd-SSE-core}
  \kappa_v (m_\pi) \cdot (r_{2}^v)^2 &=& 
  \frac{g_{A}^2 M_N}{8\pi F_{\pi}^2 m_\pi} 
  + \frac{c_A^2 M_N}{9\pi^2 F_{\pi}^2 \sqrt{\Delta^2-m_{\pi}^2}} 
    \log\left[\frac{\Delta}{m_\pi}+\sqrt{\frac{\Delta^2}{m_{\pi}^2}-1}\right] 
  + 24 M_N \mathcal{C}. 
\end{eqnarray}
With this modification, the simultaneous fit to $(r_1^v)^2$ and $\kappa_v\cdot(r_2^v)^2$, 
now using Eqs.~(\ref{eqn:r1umd-SSE}) and~(\ref{eqn:r2umd-SSE-core}), appears to describe 
the average value of the data much better, but still not the pion mass dependence. 
We show the results in the second row of Tab.~\ref{tab:fit-SSE}, and the fit curves 
(dashed lines) in Fig.~\ref{fig:SSE}. 
The fit describes $(r_1^v)^2$ very well, but $c_A$ turns out to be larger than the range
discussed earlier, which, not surprisingly, gives rise to a smaller extrapolated value for
$(r_1^v)^2$ than the experiments. 
Our new DWF data extend the trend of the weak pion mass dependence in $(r_2^v)^2$ observed
in Refs.~\cite{Gockeler:2003ay, Alexandrou:2006ru} now down into the range of pion masses
$\sim300\text{ MeV}$.
The appearance of such a ``plateau-like'' behavior down to such light pion masses,
which was also observed in Ref.~\cite{Yamazaki:2009zq}, is surprising.
The leading one-loop SSE formulae~(\ref{eqn:r2umd-SSE},~\ref{eqn:r2umd-SSE-core}) for this radius
cannot accomodate such a behavior, with or without the inclusion of the higher-order ``core''
term.

Using $c_A$ determined from  the above fits either with or without the constant term in Eq.~(\ref{eqn:r2umd-SSE}) to $(r_1^v)^2$ and $\kappa_v\cdot (r_2^v)^2$, we fit $\kappa_v$ to Eq.~(\ref{eqn:kappa-SSE})  with three unknown parameters, $\kappa_v^0$, $c_V$ and $E_1^r(\lambda)$. The results are shown in Tab.~\ref{tab:fit-SSE}. The value for $c_V$ from our fit turns out to have a different sign from that determined in~\cite{Davidson:1991xz,Hemmert:1996xg} mentioned earlier. This is not surprising given that we only have three data points, which have little or no pion mass dependence. We do not have the freedom to check the consistency of the fit, and we do not expect to obtain a reliable estimation for $c_V$, which, judging from Eq.~(\ref{eqn:kappa-SSE}), is very sensitive to the curvature of the data.

\begin{table}[ht]
\centering
\caption{\label{tab:fit-SSE}
  Fit parameters from the fits to the isovector Dirac radius $(r_1^{v})^2$,  
  Pauli radius $(r_2^{v})^2$ and the anomalous magnetic moment $\kappa_v$. 
  Details of the fit procedures are described in the text. 
  We have set the scale to $\lambda = 600\text{ MeV}$.
}
\begin{tabular}{c|r@{.}lr@{.}lr@{.}lr@{.}lr@{.}lr@{.}lr@{.}l}
\hline\hline
& 
  \multicolumn{2}{c}{$\chi^2$/dof} & 
  \multicolumn{2}{c}{$c_A$} &
  \multicolumn{2}{c}{$c_V\, [\text{GeV}^{-1}]$}  & 
  \multicolumn{2}{c}{$\kappa_v^0$} &
  \multicolumn{2}{c}{$B_{10}^r(\lambda)$}  & 
  \multicolumn{2}{c}{$E_1^r(\lambda)\, [\text{GeV}^{-3}]$} &
  \multicolumn{2}{c}{$\mathcal{C}\, [\text{GeV}^{-3}]$} \\
\hline
no constant term & 
  17&0(4.0) &                   
  1&54(6) &                 
  8&7(5.8)  &               
  4&13(95)  &               
  1&20(17) &                
  $-$4&67(42) &            
  \multicolumn{2}{c}{---} \\
with constant term &
  3&8(2.2) &
  1&97(7)  &
  7&5(4.5) &
  4&32(95) &
  2&58(25) &
  $-$5&58(42) &
  $-$0&51(7) \\

\hline\hline
\end{tabular}
\end{table}

To compare chiral extrapolations with experiment, we have also plotted selected experimental data in Fig.~\ref{fig:SSE}.  As noted in the introduction, there are still unresolved experimental questions, and we have indicated the range of possible values of $(r_1^v)^2$ that can be extracted from present experiments by showing two extreme results from the literature.  The highest value is from PDG 2008~\cite{Amsler:2008uo} and the lowest value is from a dispersion analysis including meson continuum contributions~\cite{Belushkin:2007dn}.  We note that none of the chiral fits simultaneously yields a good fit to the lattice data while also agreeing with experiment within statistical errors.

\begin{figure}[htbp]
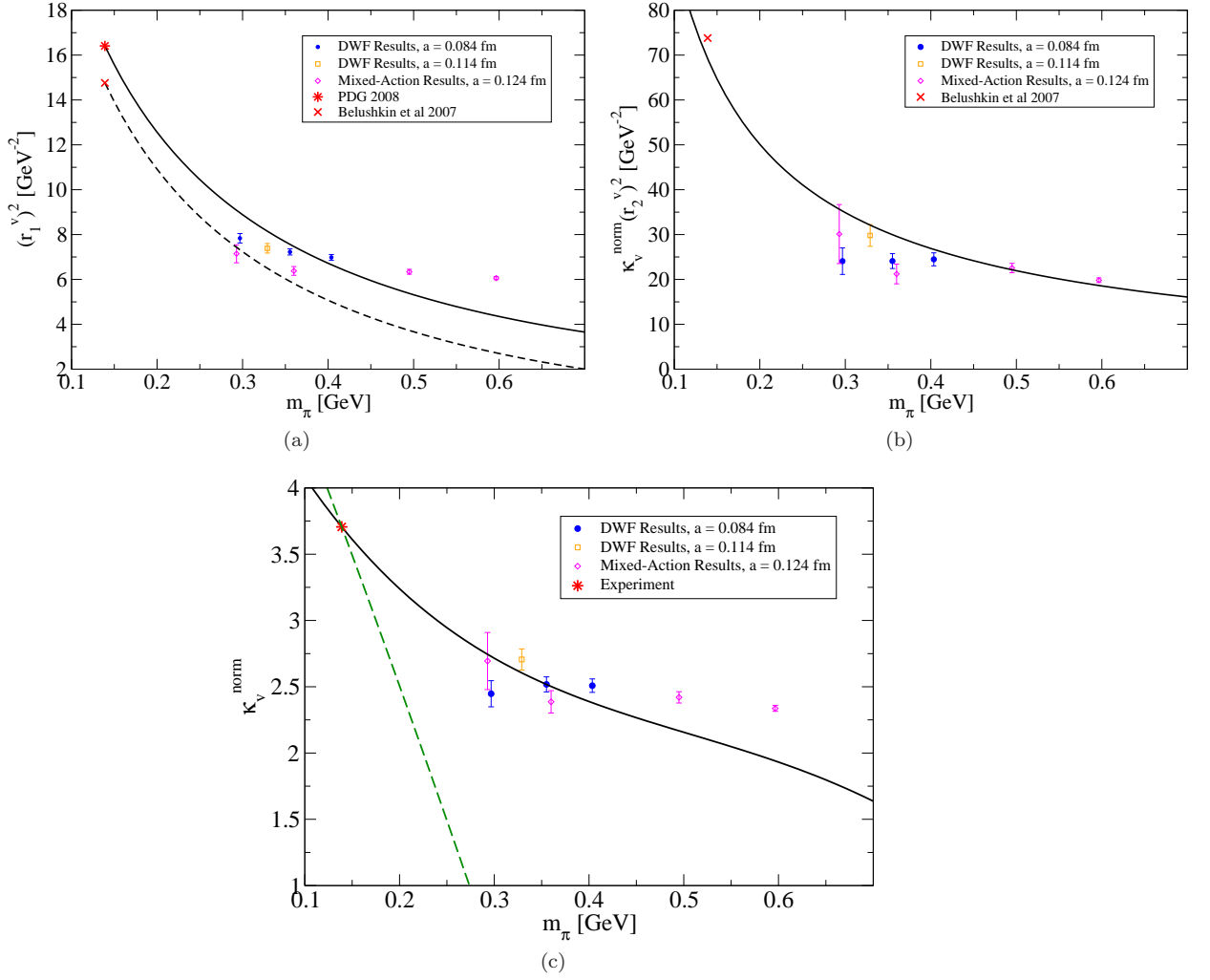

\hfill
\centering
\subfigure[]{
\includegraphics[width=0.45\textwidth,clip]{figs/chi_extrap/r1v_phenom_mixed}
\label{fig:SSE_r1v_cons}
}
\subfigure[]{
\includegraphics[width=0.45\textwidth,clip]{figs/chi_extrap/r2v_phenom_mixed}
\label{fig:SSE_r2v_cons}
}\\
\subfigure[]{
\includegraphics[width=0.5\textwidth,clip]{figs/chi_extrap/kappa_phenom_mixed}
\label{fig:SSE_kappa_cons}
}
\caption{\label{fig:SSE_cons}
  SSE chiral fits constrained to go through the physical points using the input in 
  Tab.~\ref{tab:input-SSE} as well as $c_A=1.5$ and $c_V=-2.5\text{ GeV}^{-1}$. 
  The mixed-action results at $m_\pi = 355\text{ MeV}$ are shifted slightly to the right for clarity. 
  In (a) the solid curve is constrained to go through the physical result given in PDG 2008, 
  and the dashed curve is constrained to go through the result given 
  in Ref.~\cite{Belushkin:2007dn}. 
  In (b) the curve is drawn using the input low-energy constants according to 
  Eq.~(\ref{eqn:r2umd-SSE}). 
  In (c) the solid curve is constrained to go through the physical point as well as 
  our DWF result at $m_\pi = 355\text{ MeV}$ using Eq.~(\ref{eqn:kappa-SSE}), 
  while the dashed curve is constrained to go through the physical point using 
  Eq.~(\ref{eqn:kappa}).
}
\end{figure}

To see how strongly the lattice results deviate from the SSE formulae,
we also try to determine some of the low-energy constants using experimental results 
at the physical pion mass. 
We use the values in Tab.~\ref{tab:input-SSE} as input, and also set $c_A = 1.5$ 
and $c_V = -2.5\mathrm{\ GeV}^{-1}$. Now for $(r_1^v)^2$, we have only the counter-term 
$B_{10}^r$ to determine. 
Constraining the curve to go through the higher experimental value of $(r_1^v)^2 = 0.637$ fm$^2$  
gives $B_{10}^r(\lambda=600\text{ MeV}) =1.085$,  
resulting in the solid curve shown in Fig.~\ref{fig:SSE_r1v_cons}. 
For comparison, we also plot the dashed curve that is fixed to go through the lower experimental
value~$(r_1^v)^2$.
The curve rises much more rapidly than the lattice data as the pion mass decreases.
From the slope of the leading one-loop SSE curve near the physical point and the weak pion mass
dependence displayed by our data we estimate that the applicability of Eq.~(\ref{eqn:r1umd-SSE})
for $(r_1^v)^2$ may be much less than $300\text{ MeV}$.

Without the constant term in Eq.~(\ref{eqn:r2umd-SSE-core}),  $\kappa_v\cdot(r_2^v)^2$ does not 
have any free parameters, which yields the solid curve in Fig.~\ref{fig:SSE_r2v_cons}. 
The curve undershoots the physical point by about 5\%, which may be well accounted for by 
the uncertainties in the chosen values of the low-energy constants. 
Including the higher-order term ${\mathcal C}$ of Eq.~(\ref{eqn:r2umd-SSE-core}) can of course shift
the curve up to exactly reproduce the product of physical Pauli radius and anomalous magnetic
moment. 
However, the departure of the quark-mass dependent curve from the lattice data displayed in 
Fig.~\ref{fig:SSE_r2v_cons} indicates that the leading one-loop SSE formula for
$\kappa_v\cdot(r_2^v)^2$ of Eqs.~(\ref{eqn:r2umd-SSE},~\ref{eqn:r2umd-SSE-core}) should only be
trusted for pion masses much less than the currently available $300\text{ MeV}$. Judging from
the steep slopes displayed by both the curves for the Dirac and Pauli radii as opposed to the
almost mass-independent nature of the lattice data, it is conceivable that the leading one-loop
SSE formulae may only be applicable at pion masses well below $300\text{ MeV}$, as already suggested in Ref.~\cite{Gockeler:2003ay}.

The anomalous magnetic moment still has two free parameters, $E_1^r$ and $\kappa_v^0$. 
In addition to the physical point, we need another data point to determine both parameters. 
We choose to use our $m_\pi = 355\text{ MeV}$ result in the determination, since this point is the most accurately calculated and its relatively large pion mass makes it less susceptible to finite volume effects. 
The resulting curve (the solid line) is given in Fig.~\ref{fig:SSE_kappa_cons}. 
For comparison, we also show the curve using the leading-order SSE formula in Eq.~(\ref{eqn:kappa}) (the dashed line). 
In this case, only the experimental point is included to determine $\kappa_v^0$.  
We can see that the dashed line deviates greatly from the lattice data. 
This is not surprising, as the dominating contribution to $\kappa_v$ is the term linear in $m_\pi$,
the coefficient of which is determined by $\frac{g_A^2 M_N}{4 \pi F_\pi^2}$. 
This is clearly not the case in our data. 
Regarding the limit of applicability of Eq.~(\ref{eqn:kappa-SSE}) (which includes the dominant
next-to-leading one-loop corrections to the strict ${\mathcal O}(\epsilon^3)$ SSE result of 
Eq.~(\ref{eqn:kappa})), the plot in Fig.~\ref{fig:SSE_kappa_cons} does not give us a clear
indication up to which pion mass the formula can be quantitatively employed.
Furthermore, we observe that the ``normalized'' anomalous magnetic moments display a flat
pion-mass dependence around $2.5$ nuclear magnetons. 
The new dynamical DWF data extend this ``plateau'' of the \emph{normalized} magnetic 
moments---which was already observed at much larger pion masses in the quenched simulation of 
Ref.~\cite{Gockeler:2003ay}---now into the region of pion masses as low as $300\text{ MeV}$.
Surprisingly, we can find no indication of a rise in the magnetic moment at these low pion
masses, although the onset of such a rise had been anticipated for pion masses around 
$300\text{ MeV}$ in the fit results of Ref.~\cite{Gockeler:2003ay} (see Fig.~11).

Overall, these curves show much stronger curvatures than our lattice results. 
Even with pion masses as light as $300\text{ MeV}$, the ${\mathcal O}(\epsilon^3)$ SSE formulae
do not seem to be consistent with our data.  
There are several possible explanations for the inconsistencies. 
One is that the pion masses in our simulations are still too heavy for the SSE formula 
at this order to be applicable, and the higher-order contributions may not be negligible 
in this range. 
The other possibility is that our results still suffer from uncontrolled systematic errors, 
such as finite volume effects, especially at the light pion masses. 
This will be discussed later in Sect.~\ref{sect:sys_errs}.
We want to point out that our limited number of data points is not sufficient to constrain 
the chiral fits, which clearly demonstrates the need for calculations at lighter pion masses. 
Thus we do not regard our results in Tab.~\ref{tab:fit-SSE} as conclusive. 
Rather, we take it as an indication of the difficulty of chirally extrapolating 
currently available lattice data. 

Also plotted in Fig.~\ref{fig:SSE_cons} are our domain wall results at $m_\pi = 330\text{ MeV}$ at a
coarser lattice spacing~\cite{Allton:2008pn} ($a\approx 0.114$ fm), as well as our updated
mixed-action calculations~\cite{LHPC_mixedaction_nucleonstr_2008} at a lattice spacing of about 0.124 fm. These
results are roughly consistent with the fine domain wall results, indicating that the discretization errors may be small. 
 
\subsubsection{Chiral extrapolations using covariant baryon chiral perturbation theory}

In this section we apply a different formulation of $SU(2)$ chiral
effective field theory in the baryon sector, without explicit $\Delta\,(1232)$
degrees of freedom: covariant Baryon ChPT as introduced in Ref.~\cite{Gasser:1987rb}
with a modified version of infrared regularization ($\overline{IR}$-scheme). 
For details about the formalism and differences from the standard infrared regularization 
introduced by Becher and Leutwyler~\cite{Becher:1999he}, we refer the reader to 
Refs.~\cite{Dorati:2007bk,Gail:2007phd,Hemmert:CBChPT}. 
The expressions for the $m_\pi$-dependence of the mean squared isovector Dirac and Pauli
radii and the isovector anomalous magnetic moment have been derived in~\cite{Gail:2007phd} 
up to order $p^4$, {\it i.e.} at the next-to-leading one-loop accuracy 
and are collected below\footnote{
  In Ref.~\cite{Gail:2007phd} the form factor slopes $\rho_1^v$ and $\rho_2^v$ are used, 
  which are related to our notation for $r_1^v$ and $r_2^v$ by 
  $\rho_1^v = \frac{1}{6}(r_1^v)^2$ and $\rho_2^v =\frac{1}{6}\kappa_v\cdot (r_2^v)^2$. 
}. 

For the isovector mean squared Dirac radius, the expression is given as 
\begin{eqnarray}
\label{eqn:r1v-cbchipt}
(r_1^v)^2 &=& B_{c1} + \left[(r_1^v)^2\right]^{(3)} + \left[(r_1^v)^2\right]^{(4)}
  + {\mathcal O}(m_\pi^2),
\end{eqnarray}
where
\begin{align}
&
B_{c1} = -12 d_6^r(\lambda), 
\\
&
\begin{aligned}
\left[(r_1^v)^2\right]^{(3)} 
  = & -\frac{1}{16\pi^2F_\pi^2M^4} 
    \begin{aligned}[t]
      \Bigg[
        & 7g_A^2 M^4 + 2(5g_A^2+1)M^4 \log\frac{m_\pi}{\lambda}+M^4-15g_A^2m_\pi^2M^2 \\
        & + g_A^2 m_\pi^2(15m_\pi^2-44M^2)\log\frac{m_\pi}{M}  
      \Bigg] 
    \end{aligned}\\
    & + \frac{g_A^2m_\pi}{16\pi^2F_\pi^2M^4\sqrt{4M^2-m_\pi^2}} \left[
      15 m_\pi^4 - 74 m_\pi^2 M^2 + 70 M^4\right] 
      \mathrm{arccos}\left(\frac{m_\pi}{2M} \right), 
\end{aligned}      
\\
&
\begin{aligned}
\left[(r_1^v)^2\right]^{(4)} =  
  -\frac{3c_6g_A^2m_\pi^2}{16\pi^2F_\pi^2M_0^4\sqrt{4M_0^2-m_\pi^2}}\Bigg[ 
    & m_\pi(m_\pi^2-3M_0^2)\mathrm{arccos}\left( \frac{m_\pi}{2M_0} \right) \\
    & + \sqrt{4M_0^2-m_\pi^2}\left[M_0^2+(M_0^2-m_\pi^2)\log\frac{m_\pi}{M_0}\right] 
    \Bigg].
\end{aligned}
\end{align}
The terms contributing up to and including ${\mathcal O}(p^i)$ are denoted by the
superscript $(i)$.
Without any loss of generality, the regularization scale
$\lambda$ is set equal to $M_0$, the nucleon mass in the chiral limit.
The low-energy constants $d_6$ and $c_6$ appear, respectively, in the third- and
second-order $\pi N$ effective Lagrangian. 
The mass function $M$ must be identified with $M_0$  if one truncates the previous expression 
at ${\mathcal O}(p^3)$, whereas at order $p^4$, according to Ref.~\cite{Gail:2007phd}, 
$M$ should be replaced by~\cite{Dorati:2007bk}
\begin{eqnarray}
\label{eqn:MN}
M_N(m_\pi) &=& M_0 - 4c_1 m_\pi^2 + \frac{3 g_A^2 m_\pi^3}{32\pi^2
  F_\pi^2 \sqrt{4-\frac{m_\pi^2}{M_0^2}}} \left ( -4 +
  \frac{m_\pi^2}{M_0^2} + 4c_1 \frac{m_\pi^4}{M_0^3} \right )
\mathrm{arccos}\left (\frac{m_\pi}{2M_0}\right )
\nonumber \\
& &  - \frac{3 m_\pi^4}{128\pi^2
  F_\pi^2}\left [ \left (\frac{6 g_A^2}{M_0}-c_2\right ) + 4\left
    (\frac{g_A^2}{M_0}-8c_1+c_2+4c_3\right)\log\left
    (\frac{m_\pi}{\lambda} \right)
\right ] \nonumber \\
& & + 4 e_1^r(\lambda)
m_\pi^4-\frac{3c_1g_A^2m_\pi^6}{8\pi^2F_\pi^2M_0^2}\log\left
  (\frac{m_\pi}{M_0}\right ).
\end{eqnarray}
where $c_1$, $c_2$ and $c_3$ are second-order low-energy constants 
and $e_1^r(\lambda)$ denotes an effective coupling consisting of a combination of 
fourth order low energy constants. 
In our current analysis, we always include terms up to ${\mathcal O}(p^4)$, 
hence $M$ in all the CBChPT expressions presented here should be identified with $M_N(m_\pi)$.

The pion mass dependence of the isovector Pauli radius is given by
\begin{eqnarray}
\label{eqn:kappav-r2v-cbchipt}
\kappa_v(m_\pi)\cdot (r_2^v)^2 = \frac{M_N}{M_0}\left ( B_{c2}+[\kappa_v\cdot(r_2^v)^2]^{(3)} + 
  [\kappa_v\cdot (r_2^v)^{2}]^{(4)} \right)
  + {\mathcal O}(m_\pi)
,
\end{eqnarray}
where\footnote{
We note that $\mathcal{C}$ in Eq.(\ref{eqn:r2umd-SSE-core}) is equivalent to $e_{74}^r(\lambda)$.
}
\begin{align}
&
B_{c2} = 24M_0 e_{74}^r(\lambda), 
\\
&
\begin{aligned}
\left[\kappa_v\cdot(r_2^v)^{2}\right]^{(3)} = 
  & \frac{g_A^2M_0}{16\pi^2F_\pi^2M^5(m_\pi^2-4M^2)}\Bigg[ 
    -124M^6+105m_\pi^2M^4-18m_\pi^4M^2 \\
  & + 6(3m_\pi^6-22M^2m_\pi^4+44M^4m_\pi^2-16M^6)\log\frac{m_\pi}{M} \bigg] \\
  & +\frac{g_A^2M_0}{8\pi^2F_\pi^2M^5m_\pi(4M^2-m_\pi^2)^{3/2}}\Bigg[
    9m_\pi^8-84M^2m_\pi^6+246M^4m_\pi^4 \\
  &- 216M^6m_\pi^2+16M^8\Bigg] \mathrm{arccos}\left(\frac{m_\pi}{2M}\right),
\end{aligned}
\\
&
\begin{aligned}
\left[\kappa_v\cdot(r_2^v)^2\right]^{(4)} = 
  & -\frac{g_A^2c_6m_\pi^3}{16\pi^2F_\pi^2M_0^4(4M_0^2-m_\pi^2)^{3/2}}
    \left[4m_\pi^4-27m_\pi^2M_0^2+42M_0^4\right] \mathrm{arccos}\left(\frac{m_\pi}{2M_0}\right) \\
  & +\frac{1}{16\pi^2F_\pi^2M_0^4(m_\pi^2-4M_0^2)}\Bigg[
    16c_4M_0^7+52g_A^2M_0^6-4c_4m_\pi^2M_0^5-14c_6g_A^2m_\pi^2M_0^4 \\
  & - 13g_A^2m_\pi^2M_0^4+8(3g_A^2-c_4M_0)(m_\pi^2-4M_0^2)M_0^4\log\frac{m_\pi}{M_0}
    +4c_6g_A^2m_\pi^4M_0^2 \\
  & - g_A^2(m_\pi^2-4M_0^2)(4c_6m_\pi^4-3c_6m_\pi^2M_0^2+24M_0^4)\log\frac{m_\pi}{M_0}
  \Bigg]. 
\end{aligned}
\end{align}

For the isovector anomalous magnetic moment, the ${\mathcal O}(p^4)$ CBChPT expression is 
\begin{eqnarray}
\label{eqn:kappav-cbchipt}
\kappa_v = \frac{M_N}{M_0}\left [ c_6 -16M_0 m_\pi^2 e_{106}^r(\lambda) + 
  \delta\kappa_v^{(3)} + \delta\kappa_v^{(4)} \right ]
  + {\mathcal O}(m_\pi^3)
  , 
\end{eqnarray}
where
\begin{align}
&
\begin{aligned}
\delta\kappa_v^{(3)} = 
  & \frac{g_A^2m_\pi^2M_0}{8\pi^2F_\pi^2M^3}\left[ 
    (3m_\pi^2 - 7M^2) \log\frac{m_\pi}{M}  - 3M^2 \right] \\
  & - \frac{g_A^2 m_\pi M_0}{8\pi^2 F_\pi^2 M^3\sqrt{4M^2-m_\pi^2}} \left[
      3m_\pi^4-13M^2m_\pi^2+8M^4 \right] 
    \mathrm{arccos}\left (\frac{m_\pi}{2M} \right ),
\end{aligned}
\\
&
\begin{aligned}
\delta\kappa_v^{(4)} = 
  & -\frac{m_\pi^2}{32\pi^2F_\pi^2M_0^2}
  \begin{aligned}[t]
  \Bigg[
    & 4g_A^2 (c_6+1)M_0^2-g_A^2(5c_6m_\pi^2+28M_0^2)\log\frac{m_\pi}{M_0} \\
    & + 4M_0^2(2c_6g_A^2 + 7g_A^2 + c_6 -4c_4M_0)\log\frac{m_\pi}{\lambda}
  \Bigg]
  \end{aligned} \\
  & -\frac{g_A^2 c_6m_\pi^3}{32\pi^2F_\pi^2M_0^2\sqrt{4M_0^2-m_\pi^2}}(5m_\pi^2-16M_0^2)
      \mathrm{arccos}\left(\frac{m_\pi}{2M_0}\right).
\end{aligned}
\end{align}
Note that $\frac{M_N}{M_0}c_6$ is equivalent to $\kappa_v^0$ in Eq.(\ref{eqn:kappa-SSE}).

In our chiral extrapolations, we treat $g_A$, $F_\pi$, $c_2$, $c_3$ and $c_4$ as
input parameters. The available information about the chiral limit values of
$g_A$ and $F_\pi$ have been discussed in the previous section. 
We set the second-order couplings consistently 
with Refs.~\cite{Bernard:1996gq,Fettes:1998ud,Entem:2002sf}\footnote{
  For a discussion about the value of $c_3$ see \cite{Procura:2006bj, Khan:2003cu}.
}. 
We summarize these values in Tab.~\ref{tab:BChPT_input}.

\begin{table}[htbp]
\centering
\caption{Input values for the covariant baryon chiral fits.}\label{tab:BChPT_input}

\begin{tabular}{ccccc}
\hline
\hline
$g_A$ & $F_\pi$ [GeV] &  $c_2$ [GeV$^{-1}$] & $c_3$ [GeV$^{-1}$]  &$c_4$ [GeV$^{-1}$] \\
\hline
1.2 & 0.0862 & 3.2 & -3.4 & 3.5 \\
\hline
\hline 
\end{tabular}
\end{table}

\begin{figure}[htbp]
\centering
\includegraphics[width=0.5\textwidth,clip]{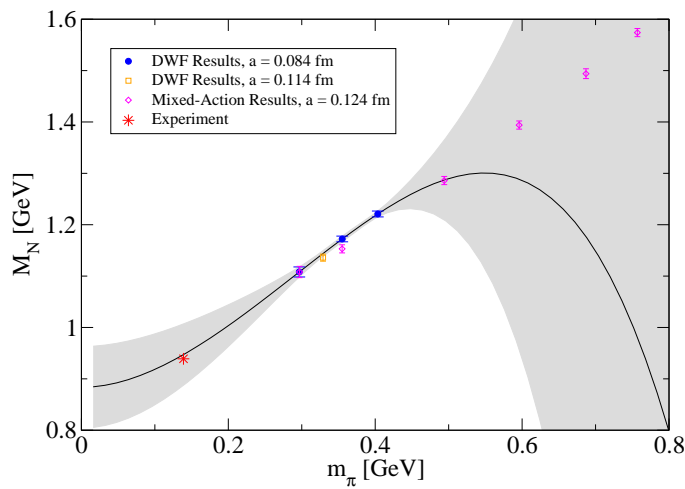}
\caption{
  \label{fig:MN_cov_extrap} 
  Chiral extrapolation for the nucleon mass using the ${\mathcal O}(p^4)$  
  CBChPT formula in Eq.~(\ref{eqn:MN}). 
  The solid line is the fit to only the fine domain wall data (solid circles). 
  The square is the coarse domain wall result, and the diamonds are the mixed-action results 
  from Ref.~\cite{WalkerLoud:2008bp}. 
}
\end{figure}

We determine $M_0$, $c_1$ and $e_1^r(\lambda)$ appearing in $M_N(m_\pi)$ by fitting 
the nucleon masses from the three fine DWF ensembles to Eq.~(\ref{eqn:MN}).  
The fit values are tabulated in Tab.~\ref{tab:MN-fit}  and the resulting fit curve 
is shown in Fig.~\ref{fig:MN_cov_extrap}.  
The fit (denoted as ``Lattice only'' in the table) is in excellent agreement with the
physical nucleon mass, 
but the small number of data points included in the fit gives substantial statistical errors. 
To better constrain the value of $M_0$, which is needed in the subsequent fits, we also fit 
the data with the experimental point as a constraint (denoted as ``Lattice+Exp.'').  
The results are again shown in Tab.~\ref{tab:MN-fit}. 
The two fits give consistent results, and we will use central values of $M_0$, $c_1$ 
and $e_1^r(\lambda)$ determined from the ``Lattice+Exp.'' fit subsequently. 

For comparison, we also plot the coarse ($a=0.114$ fm) domain wall result 
at $m_\pi \approx 330\text{ MeV}$, as well as the mixed-action results~\cite{WalkerLoud:2008bp} 
at $a=0.124\text{ fm}$ in Fig.~\ref{fig:MN_cov_extrap}. 
We see that these results are qualitatively very consistent, indicating the discretization errors 
are small.

\begin{table}[htbp]
\centering
\caption{
  \label{tab:MN-fit}
  Low-energy constants determined from the fit to the pion mass dependence of the nucleon mass 
  using the ${\mathcal O}(p^4)$ CBChPT expression. 
  Only the domain wall results on the fine lattices are included in the ``Lattice only" fit,
  and in the ``Lattice+Exp'' fit we impose that the curve goes through the physical point.
}
\begin{tabular}{c|r@{.}lr@{.}lr@{.}l}
\hline\hline
Fit  & \multicolumn{2}{c}{$M_0$ [GeV]} & \multicolumn{2}{c}{$c_1$ [GeV$^{-1}$]} & 
  \multicolumn{2}{c}{$e_1^r$($\lambda=1\text{ GeV}) [\text{GeV}^{-3}]$} \\
\hline
Lattice only    & 0&883(79) &  $-$1&01(26) & 1&1(1.3)  \\
Lattice + Exp.  & 0&8726(29) & $-$1&049(40) & 0&90(32)\\
\hline
\hline
\end{tabular}
\end{table}

\begin{figure}[htbp]
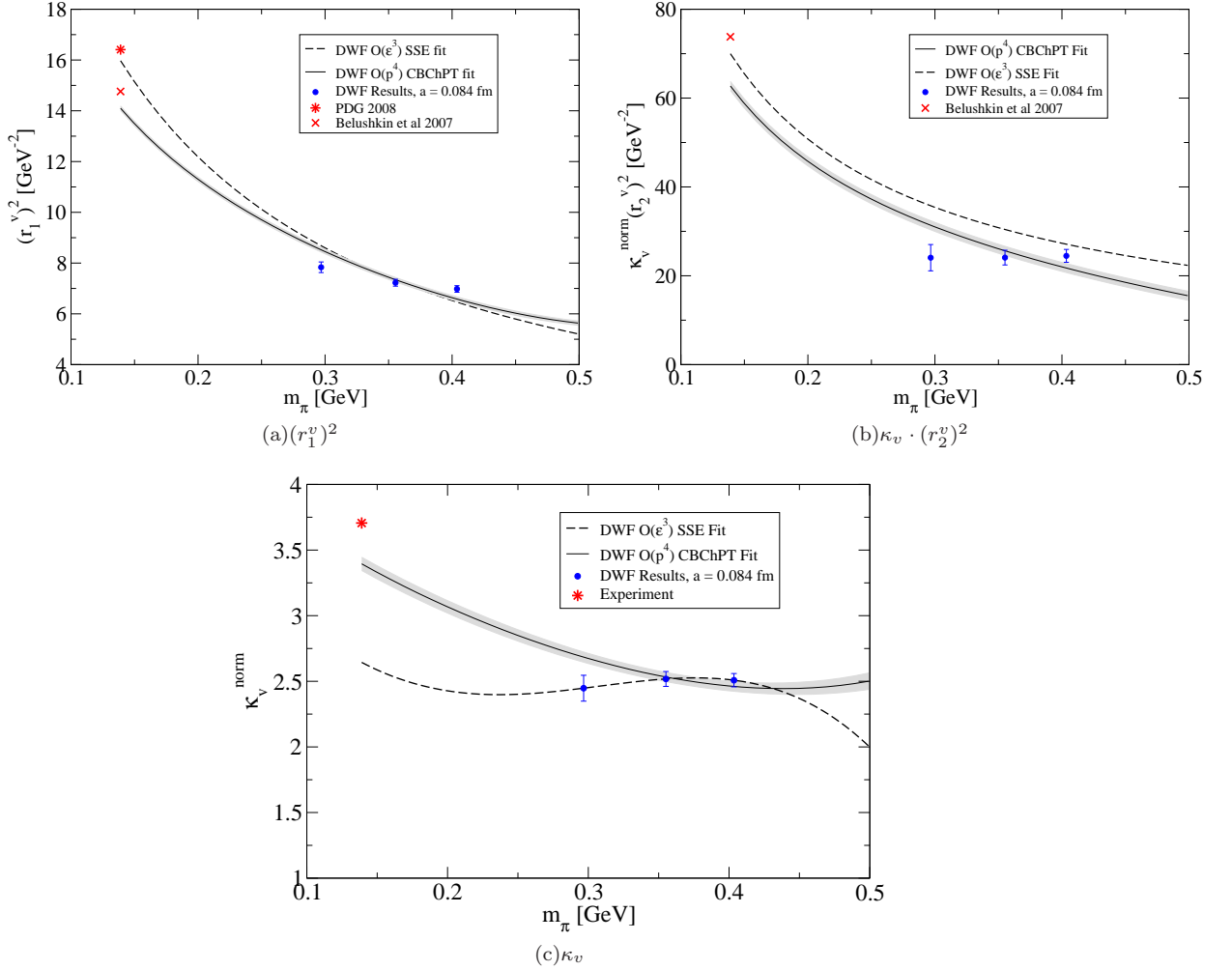

\hfill
\centering
\subfigure[$(r_1^v)^2$]{
\includegraphics[width=0.45\textwidth,clip]{figs/chi_extrap/r1v_cov}
\label{fig:cov_r1v}
}
\subfigure[$\kappa_v\cdot(r_2^v)^2$]{
\includegraphics[width=0.45\textwidth,clip]{figs/chi_extrap/r2v_cov}
\label{fig:cov_r2v}
}\\
\subfigure[$\kappa_v$]{
\includegraphics[width=0.5\textwidth,clip]{figs/chi_extrap/kappa_cov}
\label{fig:cov_kappa}
}
\caption{\label{fig:cov_sim_fit}Simultaneous fit to $(r_1^v)^2$, $\kappa_v\cdot(r_2^v)^2$ and
  $\kappa_v$ using the covariant baryon formula (solid lines). The dashed lines show the SSE fits without the constant term for $\kappa_v\cdot (r_2^v)^2$.}
\end{figure}

We determine the remaining four low-energy constants, $c_6$, $d_6^r(\lambda)$, $e_{74}^r(\lambda)$ and
$e_{106}^r(\lambda)$, from a simultaneous fit to $(r_1^v)^2$, $\kappa_v\cdot (r_2^v)^2$ and $\kappa_v$ using
${\mathcal O}(p^4)$ CBChPT expressions presented previously, with the results shown in
Tab.~\ref{tab:BChPT}. 
The large $\chi^2$/dof value indicates that the ${\mathcal O}(p^4)$ CBChPT does not
describe our data either. 
We compare the chiral extrapolations using both the CBChPT formula and 
the ${\mathcal O}(\epsilon^3)$ SSE formula in Fig.~\ref{fig:cov_sim_fit}. 
The solid curves with error bands are the results of the CBChPT simultaneous fit, 
and the dashed curves are the SSE fits using Eqs.~(\ref{eqn:r1umd-SSE}),~(\ref{eqn:r2umd-SSE}) 
and~(\ref{eqn:kappa-SSE}) as described in Sect.~\ref{sec:SSE}. 
It appears that both the SSE and CBChPT expressions are not compatible with our data, 
but since many of the low-energy constants in CBChPT are fixed from phenomenology or 
the nucleon mass, the fit is better constrained than that using the $\mathcal{O}(\epsilon^3)$ SSE expressions. This is especially important for $\kappa_v$, for which the SSE expression involves more parameters than currently available lattice data. Nevertheless, both formulations fail to describe our data at this mass range.

\begin{table}[htbp]
\centering
\caption{
\label{tab:BChPT}
  Fit parameters for the simultaneous fit to $(r_1^v)^2$, $\kappa_v\cdot (r_2^v)^2$ 
  and $\kappa_v$ using the ${\mathcal O}(p^4)$ covariant baryon formula. 
  We have set $\lambda = M_0$.
}

\begin{tabular}{ccccc}
\hline
\hline
$\chi^2$/dof &$c_6$ & $d_6^r(\lambda)$  [GeV$^{-2}$]  &  $e_{74}^r$  $(\lambda)$ [GeV$^{-2}$] & $e_{106}^r$ $( \lambda)$ [GeV$^{-3}$] \\
\hline
7.3(2.4) & 4.290(46) &0.839(7) &  1.350(45) & $-$0.132(37) \\
\hline
\hline
\end{tabular}
\end{table}

%% file: text/lattice_res/isoscalar_ff.tex
Since we have not calculated the disconnected contributions to the three-point functions 
for the form factors, in this section we give results for the isoscalar form
factors as defined in Eq.~(\ref{eqn:def_ff_isoscalar}) from the connected diagrams only. 
The renormalized results (using the renormalization factors discussed
in Sect.~\ref{sect:renorm}) in terms of the quark flavor content 
$F_{1,2}^{u+d}(Q^2)= F_{1,2}^u(Q^2)+F_{1,2}^d(Q^2) \equiv 3 F_{1,2}^s(Q^2)$ are presented 
in Tab.~\ref{tab:ff_ml004}--\ref{tab:ff_ml008}. 
First, we study the $Q^2$ dependence of both the isoscalar Dirac and Pauli form factors 
using phenomenological models, and then discuss briefly the chiral extrapolations of the results.

\subsection{$Q^2$ dependence}

Unlike the isovector Dirac form factor, $F_1^{u+d}(0)$ is not set to the known value of 3. Thus we perform dipole fits to $F_1^{u+d}(Q^2)$ separately to each ensemble using the formula 
in Eq.~(\ref{eq:two-par-dipole}).  Similar to the isovector case (see Sect.~\ref{sect:q2_dependence}), the dipole Ansatz describes 
the data reasonably well at small $Q^2$ values, typically below $0.6\text{ GeV}^2$. 
As large $Q^2$ values are included in the fit, the fit quality becomes worse, 
but the fit parameters do not change significantly. Furthermore, the fitted values of $F_1^{u+d}(0)$ are very consistent with the expected value of 3. 

\begin{figure}[htbp]
\centering
  \includegraphics[width=0.5\textwidth,clip]{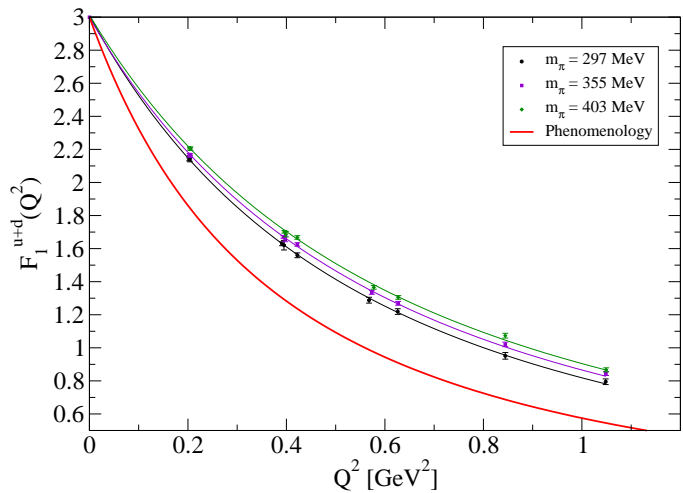}
  \caption{\label{fig:F1upd-dipole}The isoscalar Dirac form factor,
    $F_1^{u+d}(Q^2)$, with dipole fits. 
    The thick solid (red) curve is
    a phenomenological fit to experimental data~\cite{Kelly:2004hm}.  
}
\end{figure}

To demonstrate the quality of the fits, in Fig.~\ref{fig:F1upd-dipole} we show the dipole
fits to all the $Q^2$ values.
One can see that the data are reasonably well described by the fit curves.
Also plotted is the phenomenological fit to experimental data using the parameterization in Ref.~\cite{Kelly:2004hm},
although we note that no error estimate is provided and the empirical analysis involves many
potential systematic errors discussed in the introduction. 
To determine the isoscalar mean squared Dirac radii, we follow the same reasoning as in
Sect.~\ref{sect:q2_dependence} and obtain them from the dipole fits with a cut at 
$Q^2 \leq 0.5$ GeV$^2$. 
The results are shown in Tab.~\ref{tab:isoscalar-radii}.

\begin{figure}[htbp]
\centering
  \includegraphics[width=0.5\textwidth,clip]{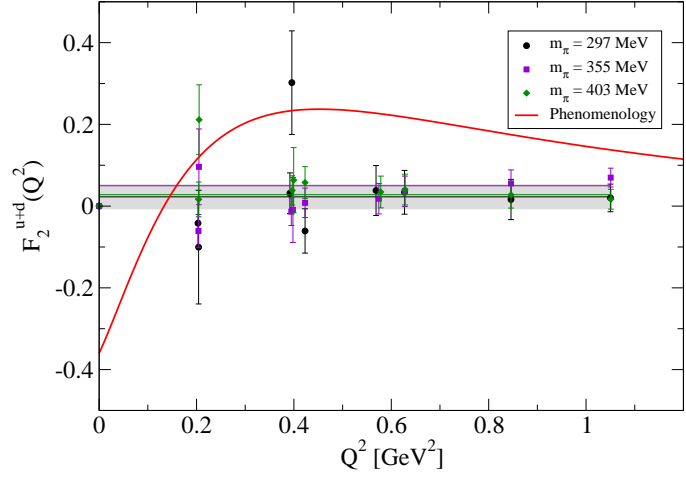}
  \caption{\label{fig:F2upd-const}The isoscalar Pauli form factor,
    $F_2^{u+d}(Q^2)$, with constant fits. Only the error band for
    the fit to the $m_\pi = 297$ MeV ensemble is shown.
    The thick solid (red) curve is a phenomenological fit to experimental data~\cite{Kelly:2004hm}.     
}
\end{figure}

In experiments, the isoscalar Pauli form factor shows a notable bump at
$Q^2 \approx 0.4$ GeV$^2$ (solid curve in Fig.~\ref{fig:F2upd-const}), 
although again there are no error estimates.
Our data are too noisy to distinguish this feature at this moment. 
In fact, the results, shown  in Fig.~\ref{fig:F2upd-const}, are rather flat. 
We show the constant fits to each ensemble separately, and find that the constants
are consistent with zero within two standard deviations. 
The error band corresponds to the constant fit to the $m_\pi = 297$ MeV data. 

If we restrict the fits to only the small $Q^2$ region ($\leq 0.5$ GeV$^2$), we are able to
perform linear fits to the data and obtain both $\kappa_s\cdot (r_2^s)^2$ (from the slope)
and $\kappa_s$ (from the intercept), the results of which are
also shown in Tab.~\ref{tab:isoscalar-radii}\footnote{
  Like in the isovector case, the anomalous magnetic moment quoted here is normalized 
  to the physical nuclear magneton according to Eq.(\ref{eq:kappa-norm}).
}. 

\begin{table}[htbp]
\centering
\caption{\label{tab:isoscalar-radii}
  Results for the isoscalar Dirac and Pauli mean squared radii and the anomalous magnetic moment. 
  A dipole fit with a $Q^2$ cutoff at 0.5 GeV$^2$ is used to determine $(r_1^s)^2$.  Linear fits to
  $F_2^s$ with $Q^2 \leq 0.5$ GeV$^2$ are used to determine $\kappa_s\cdot(r_2^s)^2$ and $\kappa_s$. 
  The results shown below have been normalized to the physical nuclear magneton.
}
\begin{tabular}{c|r@{.}lr@{.}l|r@{.}lr@{.}lr@{.}l}
\hline\hline
$m_\pi\text{ [MeV]}$ & \multicolumn{2}{c}{$\chi^2$/dof} & 
  \multicolumn{2}{c|}{$(r_1^s)^2$ [GeV$^{-2}$] } & 
  \multicolumn{2}{c}{$\chi^2$/dof} & 
  \multicolumn{2}{c}{$\kappa_s^\text{norm}\cdot(r_2^s)^2$ [GeV$^{-2}$]} & 
  \multicolumn{2}{c}{$\kappa_s^\text{norm}$} \\
\hline
297 & 0&12(35) & 11&00(13) & 3&3(2.1) & $-$0&55(55)& $-$0&038(37) \\
355 & 0&97(98) & 10&34(8) & 1&4(1.4) & $-$0&39(29)& $-$0&030(22) \\
403 & 1&7(1.3) &  9&90(8) & 2&2(1.7) & $-$0&07(28) & 0&011(21) \\
\hline
\hline
\end{tabular}
\end{table}

\subsection{Chiral extrapolations}
\subsubsection{Chiral extrapolations using ${\mathcal O}(\epsilon^3)$ small scale expansion}

\begin{figure}[htbp]
  \centering
  \includegraphics[width=0.5\textwidth,clip]{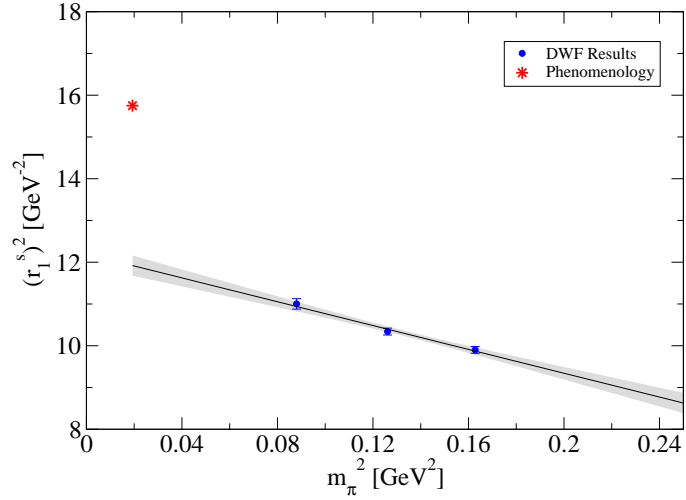}
  \caption{\label{fig:r1s-linear}
    The isoscalar Dirac radius and linear extrapolation. 
    The star indicates the phenomenological value obtained in Ref.~\cite{Mergell:1995bf}.
  }
\end{figure}

\begin{figure}[htbp]
  \centering
  \includegraphics[width=0.5\textwidth,clip]{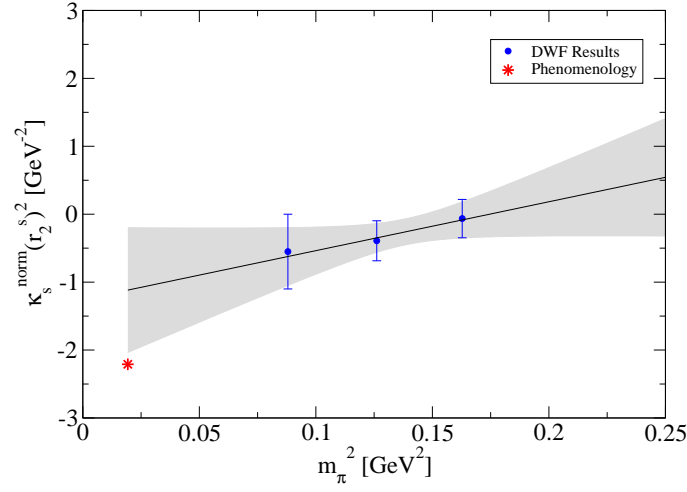}
  \caption{\label{fig:r2s-linear}
    The isoscalar Pauli radius and linear extrapolation. 
    The star indicates the phenomenological value $18.4\text{ GeV}^{-2}$ 
    obtained in Ref.~\cite{Mergell:1995bf}. 
  }
\end{figure}

As is well known in ChPT (e.g. see the discussion in~\cite{Bernard:1998gv}), chiral 
dynamics in the isoscalar form factors of the nucleon starts at the 3-pion cut, i.e. at 2-loop level, 
corresponding to ${\cal O}(\epsilon^5)$ in the power-counting of SSE. 
Hence, although the  ${\mathcal O}(\epsilon^3)$ SSE expressions for the pion mass and momentum
transfer dependence of the isoscalar Dirac and Pauli form
factors have also been derived in~\cite{Bernard:1998gv} and given as 
\begin{eqnarray}
\nonumber
F_1^{s}(Q^2) & = & 1 + \tilde{B}_1 \frac{Q^2}{(4\pi F_\pi)^2}, \\
\nonumber
F_2^s(Q^2) & = & \kappa_s ,
\end{eqnarray}
they cannot be utilized for chiral 
extrapolations. 
Therefore, in this section, we simply extrapolate linearly in $m_\pi^2$ the mean squared Dirac radius
to the physical point.
This is shown in Fig.~\ref{fig:r1s-linear}, where we can see that the linear extrapolation 
gives a result at the physical pion mass which is much lower than the empirical value. Similarly, we perform a linear extrapolation for $\kappa_s \cdot (r_2^s)^2$, which is shown in Fig.~\ref{fig:r2s-linear}.

\begin{figure}[htbp]
  \centering
  \includegraphics[width=0.5\textwidth,clip]{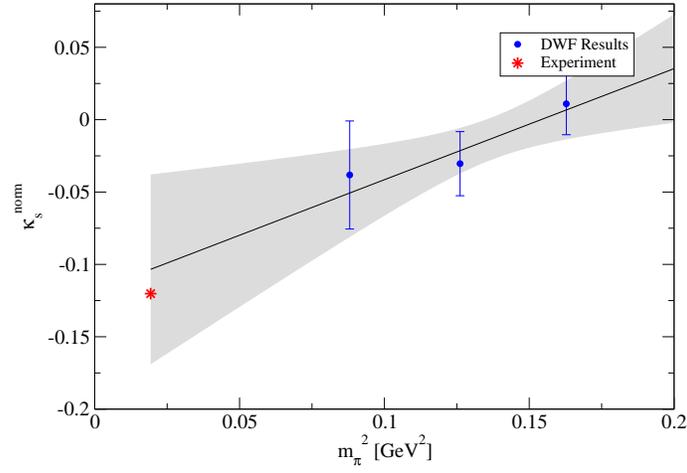}
  \caption{\label{fig:kappas}The isoscalar anomalous magnetic moment, $\kappa_s$ and linear extrapolation. The star  indicates the experimental value~\cite{Amsler:2008uo}.}
\end{figure}

For $\kappa_s$ beyond order $\epsilon^3$, additional terms arise including a term linear in the quark mass. 
Following Ref.~\cite{Hemmert:2002uh}, we write
\begin{eqnarray}
\kappa_s & = & \kappa_s^0 - 8 E_2 M_N m_\pi^2, 
\end{eqnarray}
where $\kappa_s^0$ and $E_2$ are two unknown LECs. This linear dependence describes our data well, as is shown in Fig.~\ref{fig:kappas}. 

\subsubsection{Chiral extrapolations in covariant baryon chiral
  perturbation theory}
 
 The CBChPT formulae up to ${\mathcal O}(p^4)$ for $(r_1^s)^2$, $(r_2^s)^2$ and $\kappa_s$ have
also been derived in~\cite{Gail:2007phd}. 
We collect them here for completeness.
We note, however, that the next-to-leading one-loop CBChPT results for the isoscalar form factors 
of the nucleon as presented in this section---just as in the case of the leading one-loop
SSE-analysis discussed in the previous section---do not contain their dominant chiral dynamics 
arising from the 3-pion cut. 
Such effects would only become visible at the two-loop level, i.e. starting 
at ${\mathcal O}(p^5)$ in CBChPT. 
The results presented here are therefore to be interpreted with care, as several important 
contributions with potentially large impact on the chiral extrapolation functions 
are not included at this order. 
For the isoscalar mean squared Dirac radius, the CBChPT expression is given by
\begin{equation}
(r_1^s)^2  =  B_{c1}^s + \left[(r_1^s)^2\right]^{(3)}+\left[(r_1^s)^2\right]^{(4)},
\label{eq:cov_r1s}
\end{equation}
where
\begin{align}
& 
B_{c1}^s = -24 d_7, 
\\
&
\begin{aligned} 
\left[(r_1^s)^2\right]^{(3)} =
    \frac{3g_A^2m_\pi^2}{16\pi^2 F_\pi^2 M^4(m_\pi^2-4M^2)} \Bigg[ 
    & 5 m_\pi^2 M^2 - 18 M^4 \nonumber 
      +\frac{m_\pi(5m_\pi^4-34M^2m_\pi^2+54M^4)}{\sqrt{4M^2-m_\pi^2}}
  \mathrm{arccos}\left ( \frac{m_\pi}{2M} \right ) \nonumber \\
    & -(m_\pi^2-4M^2)(5m_\pi^2-4M^2) \log\frac{m_\pi}{M}\Bigg], 
\end{aligned}
\\
&
\left[(r_1^s)^2\right]^{(4)} = \frac{9g_A^2\kappa_s^0 m_\pi^2}{16\pi^2F_\pi^2
   M_0^4}
\Bigg[ M_0^2 +
  (M_0^2-m_\pi^2)\log\frac{m_\pi}{M_0}
+ \frac{m_\pi(m_\pi^2-3M_0^2)}{\sqrt{4M_0^2-m_\pi^2}}\mathrm{arccos}\left( 
    \frac{m_\pi}{2M_0} \right ) \Bigg].
\end{align}
Here, again, when the expression is truncated at ${\mathcal O}(p^3)$, $M$ should
be identified with $M_0$, while at ${\mathcal O}(p^4)$, it should be replaced by
$M_N(m_\pi)$ in Eq.~(\ref{eqn:MN}). 
Similarly, for $\kappa_s\cdot (r_2^s)^2$, we have
\begin{eqnarray}
\kappa_s\cdot(r_2^s)^2 &=& \frac{M_N}{M_0}\left ( B_{c2}^s +
  \left[\kappa_s\cdot(r_2^s)^2\right]^{(3)} + \left[\kappa_s\cdot(r_2^s)^2\right]^{(4)} \right ), 
\label{eq:cov_r2s}
\end{eqnarray}
with 
\begin{align}
&
B_{c2}^s = 48 M_0 e_{54},
\\
&
\begin{aligned}
\left[\kappa_s\cdot(r_2^s)^2\right]^{(3)} = 
  \frac{3g_A^2 m_\pi^2 M_0}{16 \pi^2 F_\pi^2 M^5 (4 M^2 - m_\pi^2)} \Bigg[ 
    & \frac{m_\pi (6m_\pi^4 - 40 M^2 m_\pi^2 + 60 M^4)}{\sqrt{4M^2-m_\pi^2}} 
      \mathrm{arccos}\left( \frac{m_\pi}{2M} \right) \\
    & - 2 \left(10M^4-3m_\pi^2M^2+(4M^2-m_\pi^2)(2M^2-3m_\pi^2)\log\frac{m_\pi}{M}\right) 
    \Bigg],
\end{aligned}
\\
&
\begin{aligned}
\left[\kappa_s\cdot(r_2^s)^2\right]^{(4)} = 
  \frac{3\kappa_s^0 g_A^2 m_\pi^2}{16\pi^2F_\pi^2M_0^4(m_\pi^2-4M_0^2)} \Bigg[ 
    & - \frac{m_\pi (4m_\pi^4-27 M_0^2 m_\pi^2 + 42 M_0^4)}{\sqrt{4M_0^2-m_\pi^2}} 
        \mathrm{arccos}\left( \frac{m_\pi}{2M_0} \right ) \\
    & + 14 M_0^4 - 4m_\pi^2 M_0^2 + (m_\pi^2 - 4M_0^2)(4m_\pi^2-3M_0^2) 
        \log\frac{m_\pi}{M_0} 
    \Bigg].
\end{aligned}
\end{align}

The CBChPT expression for the isoscalar anomalous magnetic moment is written as 
\begin{eqnarray}
\kappa_s &=& \frac{M_N}{M_0} \left [ \kappa_s^0 - 16M_0 m_\pi^2
  e_{105}^r(\lambda) + \delta \kappa_s^{(3)} + \delta \kappa_s^{(4)}
\right ],
\label{eq:cov_kappas}
\end{eqnarray}
where 
\begin{align}
&
\delta\kappa_s^{(3)} = 
  -\frac{3g_A^2 m_\pi^2 M_0}{8 \pi^2 F_\pi^2 M^3}
    \left[ \frac{m_\pi ( m_\pi^2 - 3 M^2)}{\sqrt{4M^2 - m_\pi^2}}
          \mathrm{arccos}\left ( \frac{m_\pi}{2M} \right ) + M^2 + (M^2 -
      m_\pi^2) \log \frac{m_\pi}{M} \right]\, ,
\\
&
\begin{aligned}
\delta\kappa_s^{(4)} = 
  \frac{3 g_A^2 m_\pi^2}{32\pi^2 F_\pi^2 M_0^2} \Bigg[ 
    & 4 M_0^2 + \kappa_s^0(3m_\pi^2-4M_0^2)\log\frac{m_\pi}{M_0}
     - \kappa_s^0 \frac{m_\pi(3m_\pi^2 - 8 M_0^2)}{\sqrt{4M_0^2 -
    m_\pi^2} }\mathrm{arccos}\left( \frac{m_\pi}{2M_0} \right) 
    \Bigg].
\end{aligned}
\end{align}

\begin{figure}[htbp]
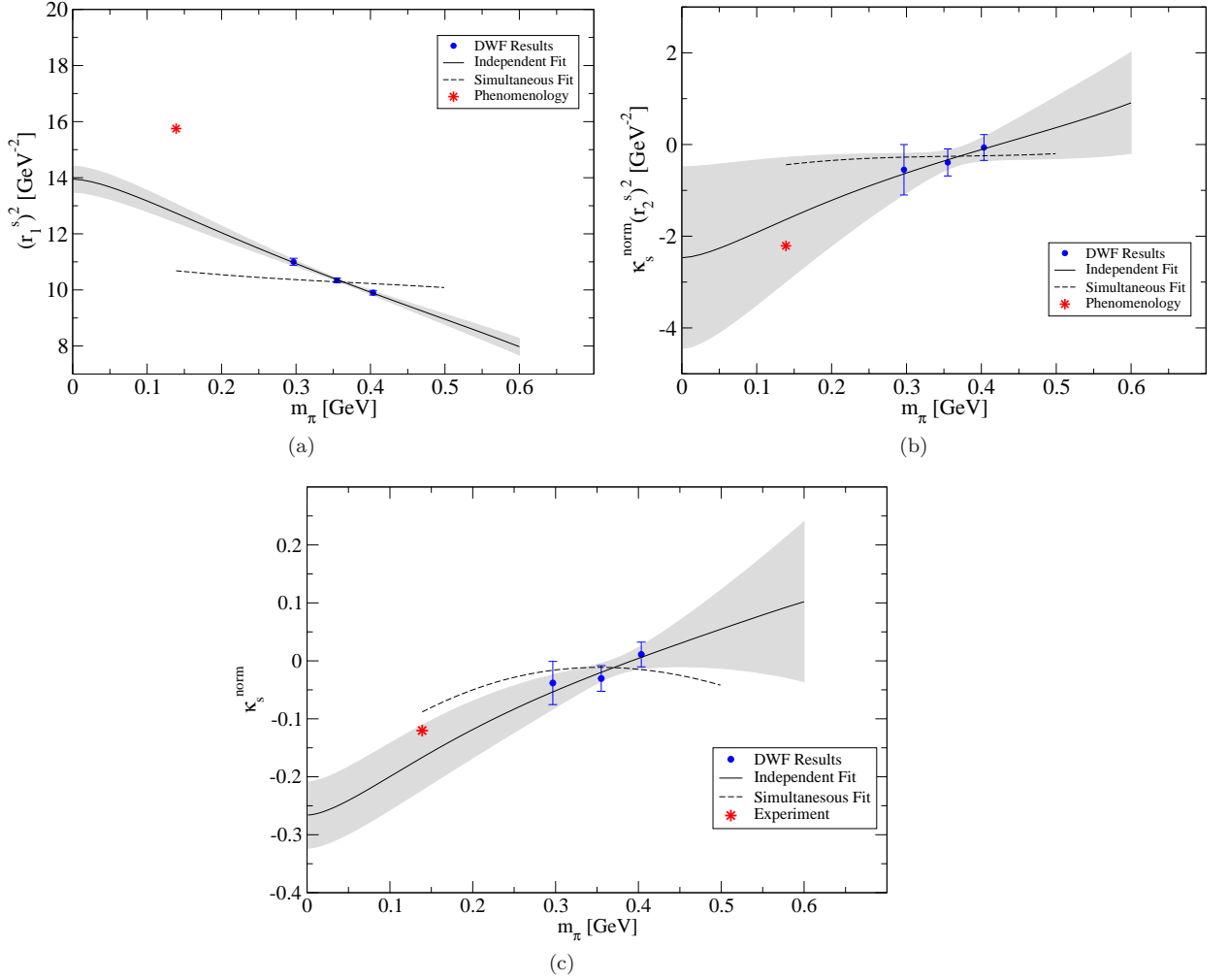

\hfill
\centering
\subfigure[]{
\includegraphics[width=0.45\textwidth,clip]{figs/chi_extrap/r1s_cov}
\label{fig:cov_r1s}
}
\subfigure[]{
\includegraphics[width=0.45\textwidth,clip]{figs/chi_extrap/r2s_cov}
\label{fig:cov_r2s}
}\\
\subfigure[]{
\includegraphics[width=0.5\textwidth,clip]{figs/chi_extrap/kappas_cov}
\label{fig:cov_kappas}
}
\caption{\label{fig:isoscalar_cov_fit} 
  ${\mathcal O}(p^4)$ CBChPT fits to $(r_1^s)^2$, $\kappa_s\cdot(r_2^s)^2$ and $\kappa_s$. 
  The solid lines are fits to the three quantities   separately with the resulting fit 
  parameters summarized in Tab.~\ref{tab:isoscalar_cov_ind}. 
  The dashed lines are simultaneous fits with the parameters summarized 
  in Tab.~\ref{tab:isoscalar_cov_sim}.
}
\end{figure}

As in the isovector case, we use the values in Tab.~\ref{tab:BChPT_input} as input in the 
extrapolations, leaving $\kappa_s^0$, $d_7$, $e_{54}$ and $e_{105}^r(\lambda)$ 
as free parameters. 
Since $(r_1^s)^2$, $\kappa_s\cdot (r_2^s)^2$ and $\kappa_s$ all contain the low-energy constant 
$\kappa_s^0$, naively we should perform a simultaneous fit to all three quantities, 
as we have done for the isovector case. However, as stated earlier, the dominant chiral dynamics for the isoscalar quantities appears at $\mathcal{O}(p^5)$. We do not expect these $\mathcal{O}(p^3)$ expressions to describe our data. In fact, the simultaneous fit to these three quantities gives a $\chi^2$/dof of about 9 (see Tab.~\ref{tab:isoscalar_cov_sim}), showing the difficulty in fitting these quantities consistently. Looking closely at each quantity separately, we find that independent fits to $(r_1^s)^2$, 
$\kappa_s\cdot(r_2^s)^2$ and $\kappa_s$ lead to an inconsistency in the estimation of the 
common parameter $\kappa_s^0$, as shown in Tab.~\ref{tab:isoscalar_cov_ind}. For demonstrative purposes, we compare the resulting fit curves from the simultaneous fit and the independent fits in Fig.~\ref{fig:isoscalar_cov_fit}, from which we see that the independent fits provide reasonable extrapolations for the data, while the simultaneous fit misses the data points badly, indicating inconsistencies of the CBChPT expressions at this order. We also note that the extrapolated value for $(r_1^s)^2$ at the physical pion mass is about 20\% lower than the phenomenological value. These observations lead us to conclude that the CBChPT expressions at $\mathcal{O}(p^3)$ are not applicable in the pion mass range of our calculation.  
Of course, since we have not included the disconnected diagrams in our calculations,
there are uncontrolled systematic errors which may affect the pion mass dependence. 
Further investigations are required to draw definitive conclusions for these isoscalar quantities. 

\begin{table}[htbp]
\centering
\caption{\label{tab:isoscalar_cov_sim}
  Fit parameters from the \emph{simultaneous} fit to $(r_1^s)^2$, $\kappa_s\cdot(r_2^s)^2$ 
  and $\kappa_s$ using Eqs.~(\ref{eq:cov_r1s}),~(\ref{eq:cov_r2s}) and~(\ref{eq:cov_kappas}).} 
\begin{tabular}{ccccc}
\hline
\hline
$\chi^2$/dof & $\kappa_s^0$ & $d_7$ & $e_{54}$ & $e_{105}^r(\lambda = M_0)$\\
8.5(2.6) & $-$0.172(23) & $-$0.458(24) & $-$0.0159(41) & 0.598(26) \\
\hline
\hline
\end{tabular}
\end{table}

\begin{table}[htbp]
\centering
\caption{\label{tab:isoscalar_cov_ind}
  Fit parameters from \emph{independent} fits to $(r_1^s)^2$, $\kappa_s\cdot(r_2^s)^2$ 
  and $\kappa_s$ using Eqs.~(\ref{eq:cov_r1s}),~(\ref{eq:cov_r2s}) and~(\ref{eq:cov_kappas}). }
\begin{tabular}{cccc}
\hline\hline
 & $\chi^2$/dof & $\kappa_s^0$ & $d_7$ \\
$(r_1^s)^2$ & 0.2(9) & 2.67(44) & $-$0.581(19) \\
\hline
& $\chi^2$/dof & $\kappa_s^0$ & $e_{54}$ \\
$\kappa_s\cdot (r_2^s)^2$ & 0.08(55) & 1.6(2.0) & $-$0.055(44) \\
\hline
& $\chi^2$/dof & $\kappa_s^0$ & $e_{105}^r(\lambda = M_0)$ \\
$\kappa_s$ & 0.4(1.3) & $-$0.247(53) & 0.506(63) \\
\hline\hline
\end{tabular}
\end{table}

%% file: text/syst_errors/me_plateau_study.tex
\begin{figure}[ht]
\centering
\includegraphics[width=\textwidth]{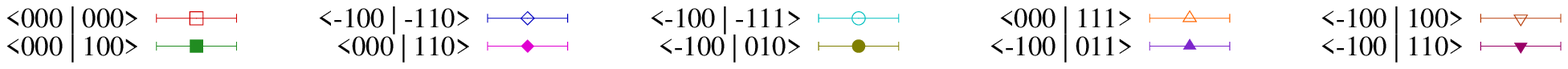}\\
\includegraphics[width=.49\textwidth]{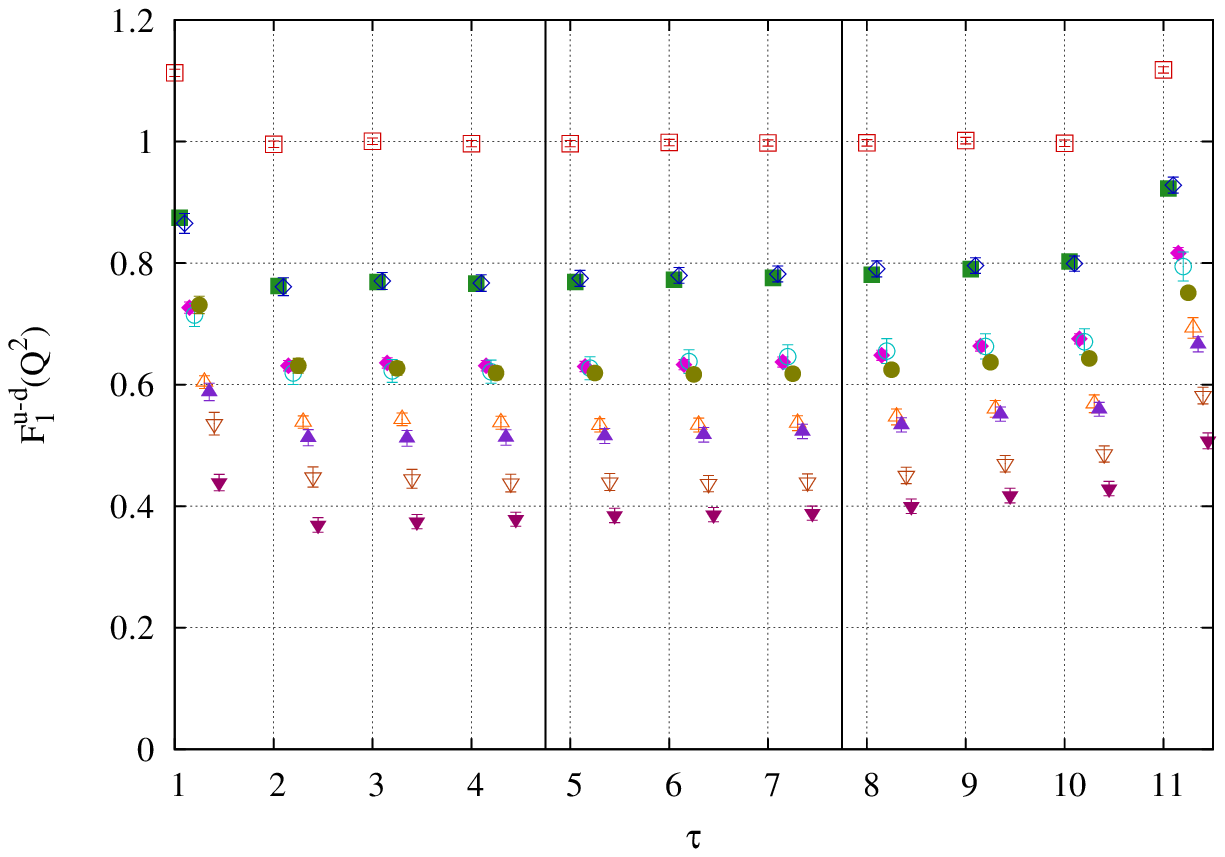}
\includegraphics[width=.49\textwidth]{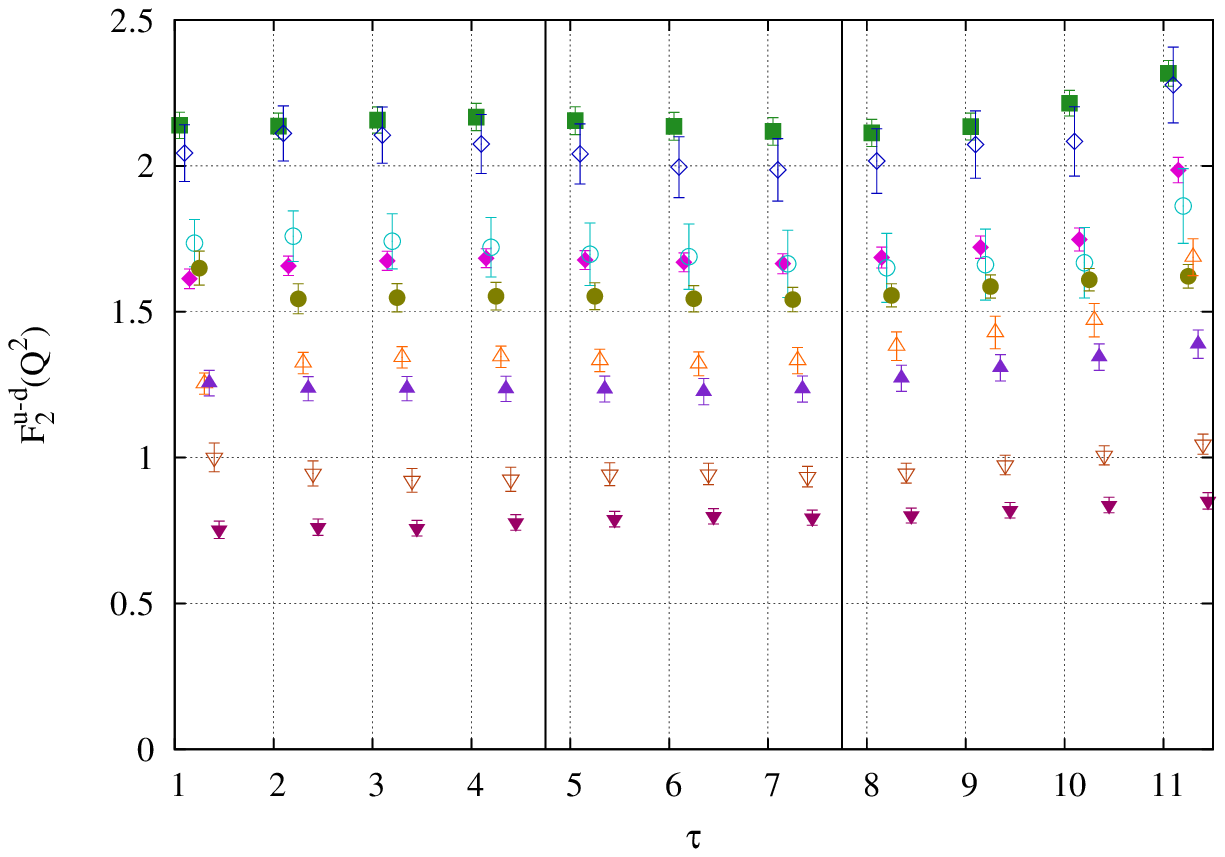}
\caption{\label{fig:f1_plateau_q2nonzero}
  Nucleon form factor plateaus for the lightest $m_\pi=297\,\mathrm{MeV}$ ensemble.}
\end{figure}

The correlation functions may have systematic bias due to the excited and/or unphysical
oscillating states \cite{Syritsyn:2007mp, Lin:2008uz, Ohta:2008kd}. 
To control it, we solve the overdetermined system separately for each location of the
operator and examine the plateau for the form factors.  Examples are shown 
in Fig.~\ref{fig:f1_plateau_q2nonzero}. 
Due to the tuning of the quark sources, the contaminations
from states other than ground are suppressed and do not contribute to the matrix element
plateaus close to their centers. 

To put quantitative bounds on possible bias, we study the excited states in the nucleon correlators.
The nucleon two-point correlation functions have very precise information on the
presence of the non-ground state contamination. 
For example, with our current statistics the parameters of a fit with three states
are well constrained: 
\begin{equation}
\label{eqn:twopt_ansatz_2exp_osc}
C_{\text{2pt}}(t ; P) = 
  Z_0(P) e^{-E_0 t} + 
  Z_1(P) e^{-E_1 t} + 
  (-1)^t Z_{\text{osc}}(P) e^{-E_{\text{osc}} t},\quad Z_{0,1} > 0.
\end{equation}
Having estimated the energy gap $\Delta E_{10}(P) = E_1(P) - E_0(P)$ and the magnitude 
of the contamination $Z_1(P)/Z_0(P)$, 
one can put bounds on the excited state contribution to the matrix elements computed from the
two- and three-point lattice nucleon correlators.

\begin{figure}[ht]
\centering
  \begin{minipage}{0.49\textwidth}
    \includegraphics[width=\textwidth]{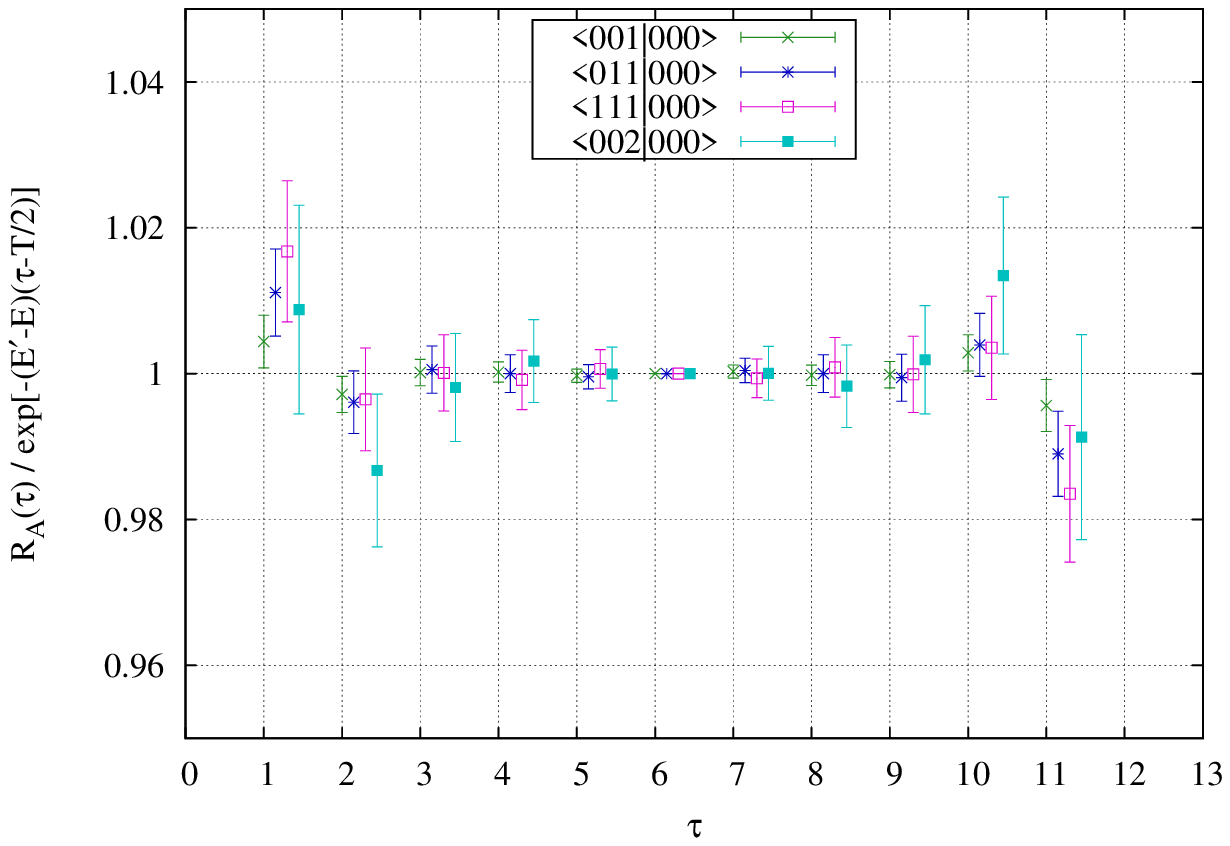}
  \end{minipage}
  \begin{minipage}{0.49\textwidth}
    \includegraphics[width=\textwidth]{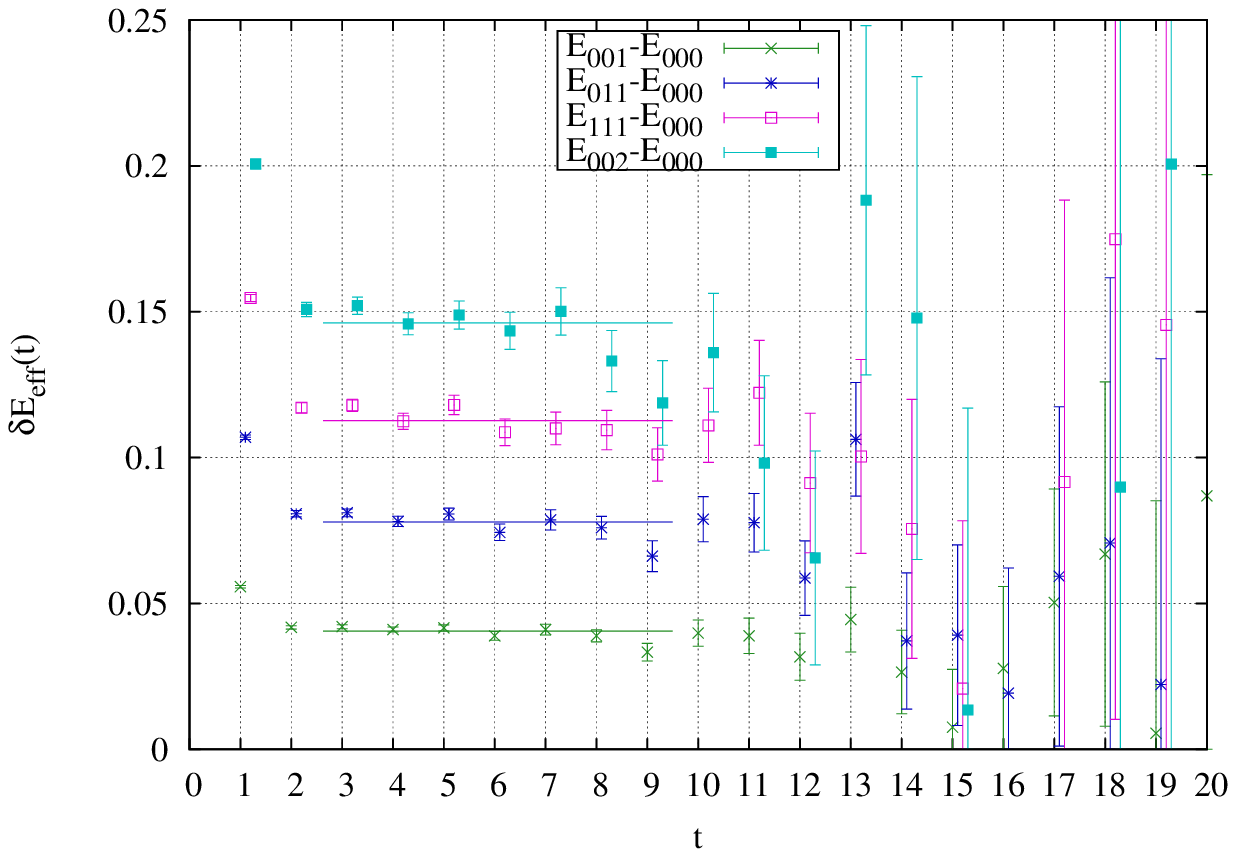}
  \end{minipage}
\caption{\label{fig:eff_energy_plots}
 The left panel shows the ratio $A(\tau)$ in  Eq.~(\ref{eqn:newratio}).  The right panel shows the effective energy difference (\ref{eqn:eff_energy_diff}) in lattice units
  and the fit values of $ E^\prime - E $ used in the left figure.
  The degree to which  the contaminations to all $\vec P^\prime\ne\vec0$ two-point correlators
  are  canceled by the contamination to the $\vec P=\vec0$ correlator is remarkable  }
\end{figure}

The ratio formula (\ref{eqn:me_ratio}) for physical matrix elements
has two factors: $R^{V^\mu} \equiv  R_N R_A $. 
Excited states can potentially contribute to either one. 
First, we study the asymmetry ratio, $R_A$, defined in Eq.~(\ref{eqn:RA}). 
As was pointed out above, this factor compensates the asymmetric $\tau$-dependence in $R_N$, and in the absence of excited states it would be equal to 
$\exp\left[-(E^\prime - E)(\tau-T/2)\right]$.
Although this factor involves different two-point functions, their excited state contributions
appear to cancel each other 
 to a large extent, as shown in Fig.~\ref{fig:eff_energy_plots}.
The left panel of Fig.~\ref{fig:eff_energy_plots} shows the ratio of $R_A$  to the exponential result in the absence of excited states
\begin{equation}
\label{eqn:newratio}
   \frac{  R_A(\tau) } {e^{-(E^\prime -E) (\tau - T/2) }  } =
 \frac{
  \sqrt{\frac{C_\text{2pt}(T-\tau, P)C_\text{2pt}(\tau, P^\prime)}{C_\text{2pt}(T-\tau, P^\prime)C_\text{2pt}(\tau, P)}}   } {e^{-(E^\prime -E) (\tau - T/2)   }} ,
\end{equation}
where  $(E^\prime - E)$ in the denominator is determined by the best fit to $R_A$ in the range $ 3 \le \tau \le 6$ .  The fact that this ratio is unity to within 1\% over a plateau from $ 3 \le \tau \le 9$  indicates that excited state contributions are negligible.  Furthermore, the right panel of Fig.~\ref{fig:eff_energy_plots}
shows the effective ground state energy difference

\begin{equation}
\label{eqn:eff_energy_diff}
\delta E^\text{eff}(t) =
  \log\left[\frac{C_{\text{2pt}}(t,P^\prime)}{C_{\text{2pt}}(t+1,P^\prime)} / 
                     \frac{C_{\text{2pt}}(t,P)}{C_{\text{2pt}}(t+1,P)}\right] ,
\end{equation}
  which in the absence of any excited state contaminants, would simply be 
$\delta E^\text{eff}(t)  =   ( E^\prime - E)$.  For comparison, the values of $ E^\prime - E$
determined above are plotted on the same graph, and
agree nicely in the fiducial range $2\le \tau\le10$.
Thus, we neglect small contaminations from this factor.

\begin{figure}[ht]
\centering
\includegraphics[width=.5\textwidth]{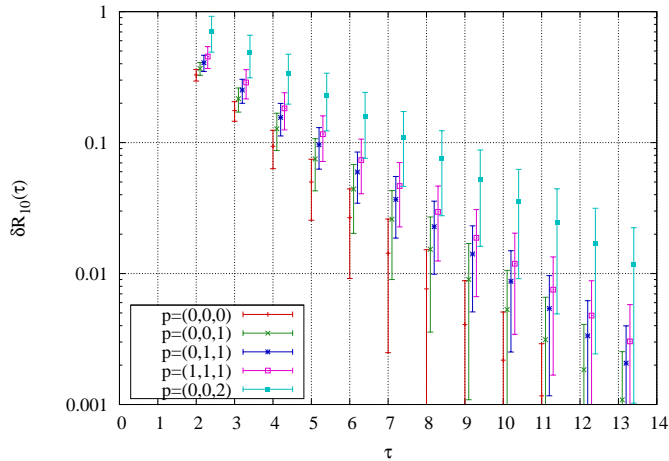}
\caption{\label{fig:deltaR10}
  Suppression factor for the excited state contributions $\delta R_{10}(\tau)$ (\ref{eqn:deltaR10}),
  as estimated from  fittingthe two-point function. 
  Note that the actual matrix element ${\mathcal O}_{10}$ is not included in the plotted value,
  which therefore shows only the relative fall-off of the exponential tail contamination.
  Note also that the factor for the $p=(0,0,2)$ state is substantially larger than for 
  the other states shown.
}
\end{figure}

Second, we estimate the contribution to $R_N$ defined in Eq.~(\ref{eqn:RN}) assuming only
one excited state and no oscillating term\footnote{
  We neglect the contribution of oscillating states because they decay even faster than  
  excited states.
}:
\begin{gather}
\nonumber
C_{\text{3pt}}(\tau, T) \approx 
  \left. C_{\text{3pt}}(\tau, T)\right|_{0}
  \left[1 + 
    \sqrt{\frac{Z_1}{Z_0}}
      \frac{{\mathcal O}_{0^\prime1}}{{\mathcal O}_{0^\prime0}} e^{-\Delta E\tau} +
    \sqrt{\frac{Z_1^\prime}{Z_0^\prime}}
      \frac{{\mathcal O}_{1^\prime0}}{{\mathcal O}_{0^\prime0}} e^{-\Delta E^\prime(T-\tau)} +
    \sqrt{\frac{Z_1^\prime Z_1}{Z_0^\prime Z_0}}
      \frac{{\mathcal O}_{1^\prime1}}{{\mathcal O}_{0^\prime0}} 
      e^{-\Delta E^\prime(T-\tau) - \Delta E\tau}\right],
\\
\label{eqn:plateau_contam}
\begin{split}
\frac{C_{\text{3pt}}(\tau,T)}{\sqrt{C_{\text{2pt}}(T) C_{\text{2pt}}^\prime(T)}} \approx
  \left(\frac{C_{\text{3pt}}(\tau,T)}{\sqrt{C_{\text{2pt}}(T) C_{\text{2pt}}^\prime(T)}}\right)_{0}
  \times \bigg[ 
  1 & +
      \frac{{\mathcal O}_{0^\prime1}}{{\mathcal O}_{0^\prime0}}\delta R_{10}(\tau) + 
      \frac{{\mathcal O}_{1^\prime0}}{{\mathcal O}_{0^\prime0}}\delta R_{10}^\prime(T-\tau)
  \\& +
      \frac{{\mathcal O}_{1^\prime1}}{{\mathcal O}_{0^\prime0}} 
          \delta R_{10}(\tau) \delta R_{10}^\prime(T-\tau)
      - \frac12\left(\delta R_{11} + \delta R_{11}^\prime\right) 
  \bigg],
\end{split}
\end{gather}
where
\begin{gather}
\label{eqn:deltaR10}
\delta R_{10}^{(\prime)}(\tau) = 
  \sqrt{\frac{Z_1^{(\prime)}}{Z_0^{(\prime)}}} e^{-\Delta E^{(\prime)}\tau},
\quad\quad
\delta R_{11}^{(\prime)} = 
  \frac{Z_1^{(\prime)}}{Z_0^{(\prime)}} e^{-\Delta E^{(\prime)}T} =
  \left[\delta R_{10}^{(\prime)}(T/2)\right]^2,
\end{gather}
and we have expanded to leading order assuming 
 that $\delta R_{11}^{(\prime)} \ll 1$.
The value of the suppression factor $\delta R_{10}^{(\prime)}(\tau)$ is shown 
in Fig.~\ref{fig:deltaR10}.
Its values are estimated from the fit parameters $Z_{0,1}$, $E_{1,0}$ in 
Eq.~(\ref{eqn:twopt_ansatz_2exp_osc}), and the errors are computed using the Jackknife procedure.
Note that $\delta R_{10}^{(\prime)}(\tau)$ falls off steeply with $\tau$. 
As a result, its contribution can be easily detected and removed by fitting the plateau with
\begin{equation}
\label{eqn:threept_fit}
R^{\mathcal O}(\tau) \approx C_0 + C_1 e^{-\Delta E} + C_1^\prime e^{-\Delta E^\prime(T-\tau)}
\end{equation}
From Fig.~\ref{fig:deltaR10} one may estimate the last two terms in the contamination 
formula~(\ref{eqn:plateau_contam}),
suppressed by $\delta R_{11}^{(\prime)}$ and $\delta R_{10}(\tau)\delta R_{10}^\prime(T-\tau)$.
If one further assumes that the excited state matrix elements are at most of the same 
order as the ground state elements,
$\frac{{\mathcal O}_{1^\prime1}}{{\mathcal O}_{0^\prime0}} \lesssim 1$, 
the effect of the last two terms in Eq.~(\ref{eqn:plateau_contam}) is well below 1\%.
It is also worth noting that higher momentum matrix elements with $\vec p=(0,0,2)$ would
contain substantially larger contamination, as compared to lower momenta.
Such matrix elements are excluded from our analysis.

\begin{figure}[h!]
\centering
\includegraphics[width=.49\textwidth]{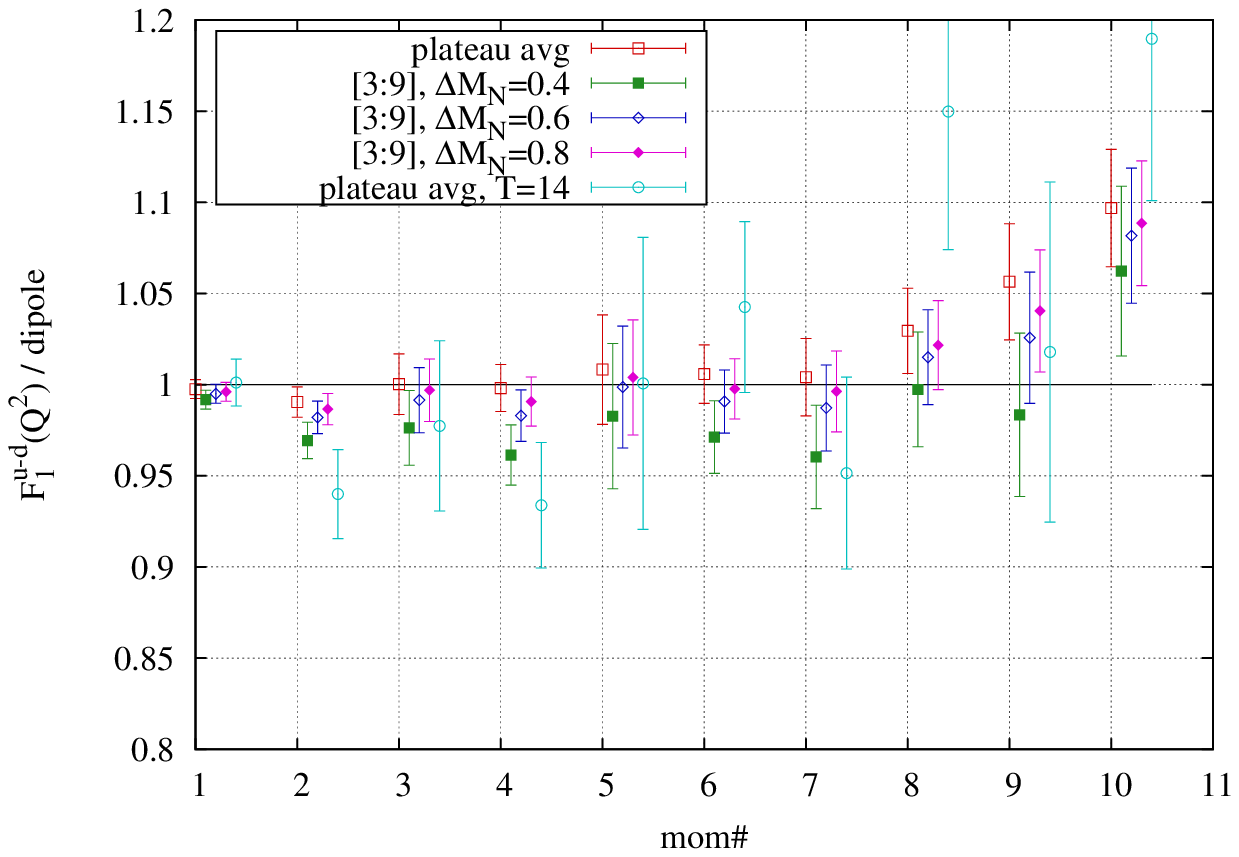}
\includegraphics[width=.49\textwidth]{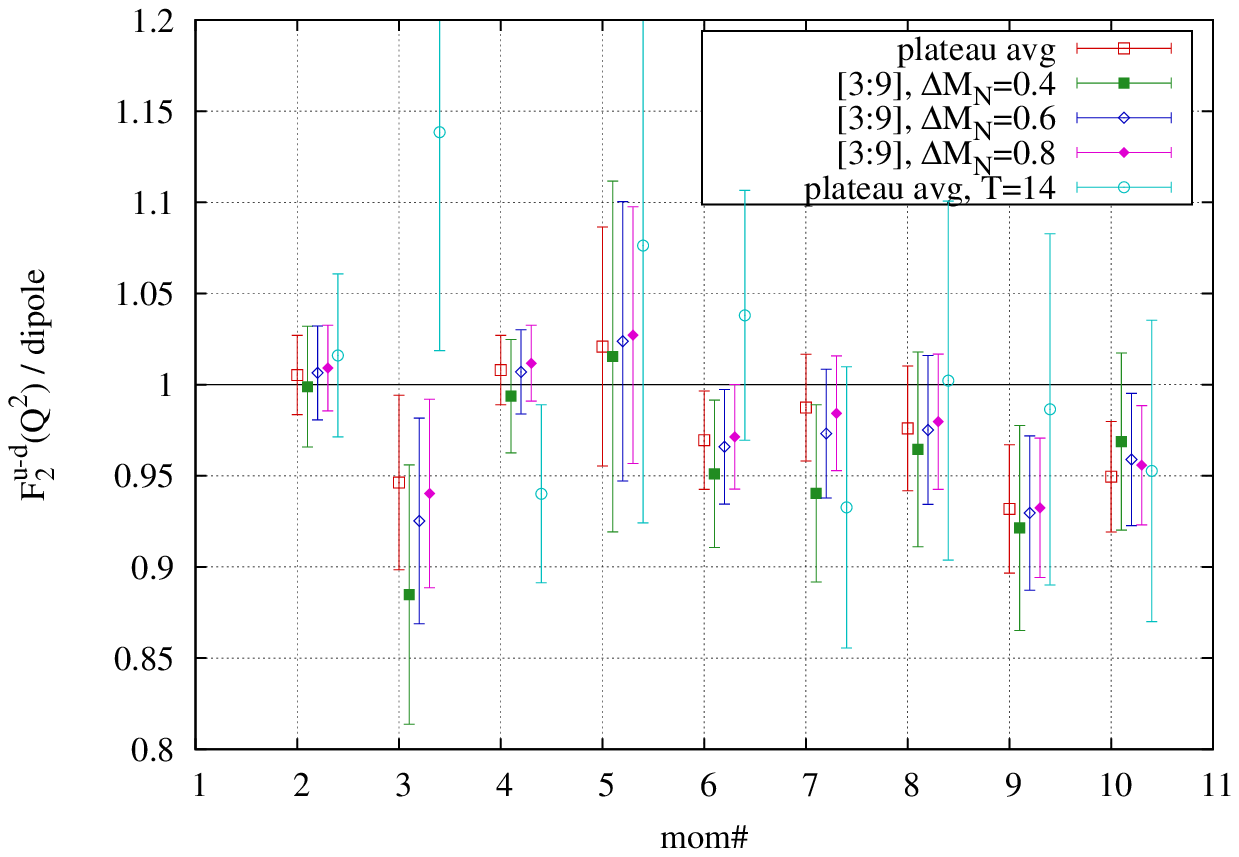}
\caption{\label{fig:comp-pltx-av-fits-indep-dt14}
  Comparison of form factors extracted from plateau averages and from fitting plateaus with
  formula (\ref{eqn:threept_fit}) for the ensemble with the lightest pion mass 
  $m_\pi=297\text{ MeV}$.
  The result is stable with variation of the mass gap $\Delta M_N$, which means that the 
  contamination is small. 
  All but the last group of points use source-sink separation $T=12$.
  The last group, calculated for 330 gauge configurations, uses the larger 
  separation $T=14$ and independent backward propagators.
  Each form factor value is divided by the central value of the dipole fit.
  Tab.~\ref{tab:mom_list} lists the momentum combinations corresponding to each index on the
  horizontal axis.
}
\end{figure}

Finally, we compare the form factors extracted using the plateau average and fitting the 
$\tau$-dependence to Eq.~(\ref{eqn:threept_fit}).
Due to the uncertainty in the two-point correlator fitting parameters, 
we perform fits for a range of mass gaps $\Delta M_N=0.4$, $0.6$ and $0.8$,
which bracket the fitted values from different fitting ranges 
and fitting with or without the oscillating term in Eq.~(\ref{eqn:twopt_ansatz_2exp_osc}).
The energy gaps $\Delta E$ for the $\vec P\ne0$ states are computed using the continuum
dispersion formula. 
The result is statistically independent of the mass gap value used 
(see Fig.~\ref{fig:comp-pltx-av-fits-indep-dt14}) and is stable when fitting inside the region $2\le\tau\le 10$. 
The complete consistency between conventional plateau averages and results for which excited state contaminants are explicitly included in the analysis and separated from the physical ground state contribution clearly indicates the absence systematic errors from excited state contaminants in our present results.

\begin{figure}[h!]
\centering
\includegraphics[width=.49\textwidth]{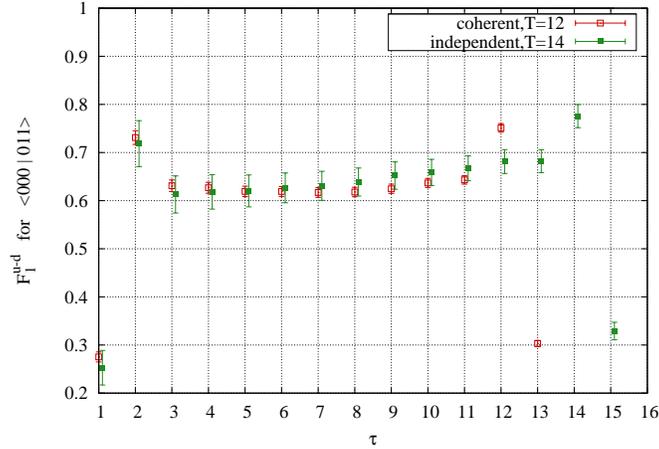}
\caption{\label{fig:comp-dt12-dt14-F1}
  Comparison of $F_1^{u-d}$ plateau using coherent backward propagators with $T=12$ and
  independent backward propagators with $T=14$. 
  The momentum transfer $Q^2$ corresponds to $\langle000 | 011\rangle$.
}
\end{figure}

In addition,   we have also compared results with two different source-sink separations, $T$ = 12 and $T$ = 14.  
If the coherent sink technique were ever to introduce additional noise into the calculation, 
one would expect it to be worst for the larger $T$, for which the first adjacent unwanted sink is closer. 
Hence, in the case of  $T$ = 14, we have used independent sinks to check that this is not a problem. 
One of the typical plateaus comparing $T=12$ and $T=14$ separations is plotted
in Fig.~\ref{fig:comp-dt12-dt14-F1} and shows agreement within statistics.
Separations 12 and 14 are also compared  in Fig.~\ref{fig:comp-pltx-av-fits-indep-dt14}, where we show each of the form factors computed 
on a subset of the $am_q=0.004$ ensemble using independent backward propagators 
and the larger source-sink separation $T=14$.
The agreement of results that use two different separations and techniques
directly indicates that our method does not suffer from the systematic effects due to 
excited states or the coherent propagator technique.


%% file: text/syst_errors/finite_vol.tex
Ideally, we would like to control systematic errors arising from volume dependence using an
 effective field theory that describes the dependence of the observables of interest 
as a function of spatial volume and pion mass, 
and a set of calculations of the lattice observables with a specified action for a range of volumes 
at pion masses for which the effective field theory is applicable.
Verifying  that the effective theory fits the measured volume dependence with 
low energy constants that are consistent with other lattice and phenomenological constraints would then
assure solid theoretical and computational control of finite volume effects. 
To date, this program has not been carried out completely for form factors\footnote{
  We note that the Regensburg group has recently started the extension of the ${\mathcal O}(p^4)$ 
  CBChPT calculation for the isovector form factors of the nucleon to a finite volume in 
  the p-regime~\cite{Greil:2009-in-prep}.
} 
with any lattice action,
and finite volume effects have been a convenient excuse for any disagreements with experiment.  
Hence, it is useful to examine the available data for domain wall fermions and to assess 
what quantitative evidence there is for or against significant finite volume corrections.
  
To examine volume dependence carefully, it is important to only compare lattice calculations at different volumes that use precisely the same action and computational methodology.  In this context, we believe it can be seriously misleading to argue on the basis of plots containing a variety of calculations with different actions, analysis techniques, renormalization schemes, etc. In addition, we will find it useful to  
 distinguish between forward matrix elements, for which  we have applied detailed effective field theory formulae to finite volume corrections\cite{Edwards:2005ym}, and off-diagonal matrix elements, for which we have not yet done so.
Due to the high computational cost of domain wall fermions, the only high precision calculations of nucleon observables we are aware of with two large volumes for light quarks are the mixed action calculations for $355\text{ MeV}$ pion mass in volumes of spatial extent $2.5\text{ fm}$ and $3.5\text{ fm}$,   corresponding to $m_\pi L = 4.4$ and~$6.2$ respectively~\cite{Bratt:2008uf,LHPC_mixedaction_nucleonstr_2008}. Thus, we will base our computational arguments on these results. We begin with the forward matrix element corresponding to the axial charge, which was first calculated for these two lattice volumes in Ref.~\cite{Edwards:2005ym}. In that work, the values of the axial charge calculated at the two volumes agreed within their statistical errors of $5\%$, which when combined quadratically indicated that the volume dependence was less than $7\%$.  
A new high statistics calculation~\cite{LHPC_mixedaction_nucleonstr_2008} yields the calculated fractional difference between $g_A$ at $3.5\text{ fm}$ and $2.5\text{ fm}$ of $(-0.4\pm1.9) \%$ at $m_\pi = 355\text{ MeV}$.
 In effective field theory, the loop integrals over momenta arising in an infinite volume are replaced by sums over discrete momenta in a periodic finite volume, and explicit expressions are available for the axial charge in a finite volume~\cite{Beane:2004rf,Detmold:2005pt,Khan:2006de}. Using the low energy parameters determined from fitting the data in Ref.~\cite{Edwards:2005ym}, chiral perturbation theory specifies that the fractional difference between $g_A$ at $3.5\text{ fm}$ and $2.5\text{ fm}$   is $+0.4\%$.  This is consistent with the observation in Ref.~\cite{Lin:2008uz} that for a reasonable value of $c_A \sim 1.5$, the finite volume corrections are quite small. 
 Hence, for the forward matrix element $g_A$, all the evidence is consistent in indicating that the systematic error arising from a spatial box of extent $2.5\text{ fm}$ is of the order of one percent.

Since we have not performed a similar analysis of chiral perturbation in a finite volume for form factors at finite momentum transfer,  our only recourse at present is comparison of  the mixed action numerical results.  Because we are most interested in chiral extrapolation of rms radii, we have focussed on the most accurate means of calculating the slope at zero momentum transfer.  Thus, we take the momentum combination $(1,0,0)2 \pi /L$, $(0,0,0)$, on each lattice and fit the form factor at the resulting momentum transfer with a one-parameter dipole formula, from which we determine the slope at the origin. Comparing the results for $r_1^v$ and quadratically combining the errors for the two independent calculations, we find that the fractional difference between $r_1^v$ at $3.5\text{ fm}$ and $2.5\text{ fm}$ is $(3.7\pm2.6) \%$  at $m_\pi=355\text{ MeV}$\footnote{
  We note that as discussed in Ref.~\cite{LHPC_mixedaction_nucleonstr_2008}, dipole fits of all 
  the form factor data out to some fixed cutoff yield discrepancies in the dipole fits 
  between the two volumes that increase as the cutoff increases, but this comparison focusses 
  on other features of the form factor besides the radius that we seek to chirally extrapolate.
}.  Further evidence suggesting finite volume corrections to radii are small is the fact that in Ref.\cite{Yamazaki:2009zq}, even decreasing the lattice size to $1.8\text{ fm}$ yields small changes in  $r_1^v$ and  $r_2^v$. Both by virtue of the fact that Monte Carlo calculations of the slope are intrinsically noisier than for forward matrix elements and fact that we have not performed an effective field theory analysis of the finite volume corrections, our control of the volume dependence of form factor radii is worse than for $g_A$.  Whereas the error may also be on the order of one percent as in $g_A$, we cannot completely exclude a result at the upper limit of the error bars of the order of $6\%$. 
Thus, presently,  it cannot be  excluded that a rapidly
growing finite-size effect below $m_\pi=355\text{ MeV}$ is affecting the pion mass
dependence of $r_1^v$ in our data. This will have to be resolved in the future.

%% file: text/comparison.tex
We briefly compare our results and conclusions with  those of previous calculations.
We start with the isovector Dirac and Pauli radii, $r^v_1$ and $r^v_2$.
Previous calculations using Wilson fermions had reached pion masses of about $400\text{ MeV}$.
Both the quenched~\cite{Gockeler:2003ay,Alexandrou:2006ru} and $N_\text{f}=2$ unquenched
results~\cite{Alexandrou:2006ru} showed a mild pion mass dependence for  $r^v_1$ and $r^v_2$.
A recent calculation on $N_\text{f}=2+1$ domain-wall fermion configurations 
at $a=0.114$fm extended the range of pion masses down to $330\text{ MeV}$~\cite{Yamazaki:2009zq}. 
These results show that the very mild upward trend of  $r^v_1$ and  $r^v_2$ extends down to
that pion mass. 
Summary plots comparing results for Wilson fermions with zero and two flavors 
and domain-wall fermions for zero, two, and 2+1 flavors  are given in Figs.~14 and~19 
of Ref.~\cite{Yamazaki:2009zq}. 
For the case of $r^v_1$,  which has smaller statistical errors, for each action, 
the data tend to lie on straight lines with comparable small slope and some scatter 
in normalization, with perhaps a hint that the $N_\text{f}=2$ calculations, performed on
box sizes $1.9\text{ fm}$, lie somewhat low.
The $r^v_2$ data also appear to lie on straight lines with similar small slope, 
albeit with larger scatter.
Our results for   $r^v_1$ and $r^v_2$, which extend down to $m_\pi=300\text{ MeV}$,
also show a small pion mass dependence, and are consistent within statistical errors 
with the 2+1 flavor domain wall results on $a=0.114\text{ fm}$ lattices.
We conclude that this flat behavior, surprising as it is from the chiral effective theory
point of view, is genuine. 
The one-loop SSE formulae of Sect.~\ref{sec:SSE} cannot accommodate this ``flat'' pion mass 
dependence in the radii down to such low values of the pion mass $\sim 300\text{ MeV}$, 
with or without the inclusion of  a higher order ``core'' term. 
Indeed, the curves shown in Figs.~\ref{fig:SSE_r1v_cons},~\ref{fig:SSE_r2v_cons} indicate 
that the SSE calculation would have favored an upward trend in the extracted 
isovector radii which should have become visible in the pion-mass range 
studied in this work, consistent with the expectations drawn in Ref.~\cite{Burkardt:2000za}. 
The only explanation for this behavior available at the moment is that the leading one-loop 
SSE calculation is only valid for pion masses $< 300\text{ MeV}$.

As for the anomalous magnetic moment $\kappa_v$, our results are in very good agreement 
with those obtained with $N_\text{f}=2$ dynamical Wilson fermions in~\cite{Gockeler:2007hj}, 
with recent $N_{\rm f}=2$ twisted-mass results \cite{Alexandrou:2008rp} and 
with the recent $N_{\rm f}=2+1$ domain-wall calculation~\cite{Yamazaki:2009zq}.
We remark that our Fig.~\ref{fig:SSE_cons} displays the ``normalized'' anomalous magnetic moment 
$\kappa^\text{norm}$, while Fig. 17 of Ref.~\cite{Yamazaki:2009zq} shows the magnetic moment
normalized by the quark-mass dependent nucleon mass, $\kappa^\text{lat}$. 
The difference between the two figures\footnote{
  The numerical difference between the lattice data and the experimental value is smaller 
  in the case of $\kappa_{\rm lat}$.
}
reflects the $m_\pi$ dependence of the nucleon mass, which is quite strong 
(see Fig.~\ref{fig:MN_cov_extrap}). 
All in all, for $\kappa_v$ too, the calculated pion mass dependence is rather mild, 
and results at lower pion masses will have to bend upwards rather sharply if they are 
to agree with the experimental value.

%% file: text/conclusions.tex
We have presented lattice calculations of nucleon form factors with $N_f $ = 2+1 flavors of dynamical domain wall fermions on fine $32^3 \times 64$ lattices with $a$ = 0.084 fm at pion masses of 297, 355, and 403 MeV that achieve a new level of precision in both statistical and systematic errors.  Statistical errors have been reduced by using from 3600 to 7064 measurements of operators at a given mass by performing 8 measurements per lattice and verifying their statistical independence.  Statistical errors and error correlations have been carefully analyzed in our overdetermined analysis, which combines as many stochastically distinct measurements of the same physical form factors as practical.

Because of the high level of statistical precision, we have carefully investigated and controlled potential sources of systematic error.  We have ruled out systematic errors arising from the source-sink separation in two different ways.  First, we have derived analytic expressions for the contamination by excited states, and, using lattice data from two-point correlation functions, have shown quantitatively that the coefficients of excited state admixtures in these expressions yield negligible contributions to the observables of interest.   In addition, we compared explicit calculations with source-sink separations $T = 12$ and $T = 14$ and have shown that results from the two source-sink separations are indeed statistically consistent, as expected from the excited state analysis.
We have verified that even in the worst case---the lightest pion mass and maximum source-sink 
separation---results calculated using the time-saving coherent sink technique are consistent 
with results calculated with conventional independent sinks.  
By comparison with companion calculations on a coarse lattice with $a=0.114\text{ fm}$,  
we have verified that lattice spacing errors are small.    
Finally, based on the overall consistency between the recent high-statistics mixed action results 
and the current work, we have presented two arguments that finite volume corrections 
to the present calculations in a volume of spatial extent $2.5\text{ fm}$ are small.   
First, the forward matrix element $g_A$ changes by a very small amount when the spatial extent 
is changed from $2.5$ to $3.5\text{ fm}$: $+1.0\%$ in chiral perturbation theory and 
$(-0.4 \pm 1.9)\%$ in explicit lattice calculations. 
For the more complicated case of off-diagonal matrix elements, the measured fractional change 
in $r_1^v$ is $(3.7 \pm 2.6)\%$.

The high precision of the calculated form factors is shown in 
Figs.~\ref{fig:data_fit_ratio},\ref{fig:Ge_Gm_data_fit_ratio}, where, in order to see 
the discrepancies with dipole fits,  we plotted the ratio of the lattice calculations 
to the best dipole fits on an expanded scale.  
This precise data enabled us to  extract the Dirac radius, $r_1^v$,  Pauli radius $r_2^v$,  
and anomalous magnetic moment $\kappa_v$ with much smaller errors than in earlier calculations 
and to study chiral extrapolations to correspondingly higher precision.  
In contrast to earlier studies in which the lattice error bars were sufficiently large that 
the data appeared to be consistent with chiral perturbation theory, in this work 
we have shown that the $m_\pi$ dependence of the lattice results for 
$\left(r_1^v\right)^2$, $\left(r_2^v\right)^2$, and $\kappa_v$ at the three masses 
$297$, $355$, and $403\text{ MeV}$ cannot be simultaneously fit by either 
${\cal O}(\epsilon^3)$  SSE or NNLO CBChPT. 
The data points for $\left(r_1^v\right)^2$ rise too slowly with decreasing $m_\pi$ and the data 
for $\left(r_2^v\right)^2$ are too flat to be fit by either the SSE or CBChPT curves 
that rise smoothly with decreasing  $m_\pi$ to approach the experimental results. 
Since there happen to be three free parameters in SSE to fit the 3 measured values of $\kappa_v$, 
the SSE can actually fit the anomalous magnetic moment, but CBChPT, which is 
physically constrained to rise with decreasing $m_\pi$, is also seriously in conflict 
with the lattice measurements of $\kappa_v$ at the accessible masses.  
Similarly, we were unable to simultaneously fit the isoscalar quantities 
$\left(r_{1,2}^s\right)^2$ and $\kappa_s$, which, to this order of ChPT, have fewer parameters.
 
 With the present data at these three pion masses, we see three possible explanations for the discrepancy with chiral perturbation theory.  One possibility is an outright error somewhere in the lattice calculations. However, by virtue of meticulous checks,  key calculations with independent codes, and the qualitative similarity of our results to those of other groups~\cite{Alexandrou:2006ru,Gockeler:2007hj,Alexandrou:2008rp,Ohta:2008kd,Yamazaki:2009zq}, we believe this is unlikely.  A second possibility is that finite volume effects are significantly larger than the estimates we obtained from our  355 MeV mixed action studies for spatial sizes $2.5\text{ fm}$ and $3.5\text{ fm}$.  This possibility clearly warrants further study of chiral perturbation theory for off-forward matrix elements in a finite volume and careful high statistics studies in a series of volumes. The third possibility is that chiral perturbation theory at the present order is not applicable for this range of $m_\pi$.   Indeed, significant problems have previously been encountered in describing the $m_\pi$ dependence  of baryon masses, and one observes, for example, that the highly linear dependence of the nucleon mass on $m_\pi$ seen in a variety of lattice calculations with different actions can only arise from an apparently unnatural cancellation of analytic and non-analytic terms in chiral perturbation theory~\cite{WalkerLoud:2008bp}.  This possibility clearly warrants lattice calculations at a series of lower values of $m_\pi$ all the way down to the physical pion mass.  
 
 The last two possibilities each raise very interesting and important questions in hadron structure.
 Given the high computational cost of chiral fermions relative to improved Wilson fermions and the fact that there are no crucial operator mixing problems in form factors necessitating exact chiral symmetry on the lattice, it appears that the most expeditious means of understanding the volume dependence and behavior down to the physical pion mass 
will be with an appropriate form of an improved isotropic Wilson action. Such calculations are clearly essential for further progress in understanding the fundamental structure of the nucleon.

%% file: text/app_tables/ff_tables.tex

\begin{table}[ht]
\centering
\caption{Renormalized results for the Dirac and Pauli form factors from the $am_l=0.004$ ensemble
  with $m_\pi \approx 297$ MeV.}\label{tab:ff_ml004}
\begin{tabular}{r@{.}lr@{.}lr@{.}lr@{.}lr@{.}lr@{.}lr@{.}lr@{.}lr@{.}lr@{.}l}
\hline \hline 
\multicolumn{2}{c}{$(aQ)^2$} & \multicolumn{2}{c}{$Q^2$ [GeV$^2$]} & 
  \multicolumn{2}{c}{$F_1^u$} & \multicolumn{2}{c}{$F_1^d$} & 
  \multicolumn{2}{c}{$F_1^{u+d}$} & \multicolumn{2}{c}{$F_1^{u-d}$} &
  \multicolumn{2}{c}{$F_2^u$} & \multicolumn{2}{c}{$F_2^d$} & 
  \multicolumn{2}{c}{$F_2^{u+d}$} & \multicolumn{2}{c}{$F_2^{u-d}$} \\ 
\hline 
0&000000 & 0&000 & 2&004(5) &  1&004(3) & 3&008(7) & 1&000(5) & 
  \multicolumn{2}{c}{---}&  \multicolumn{2}{c}{---} &  
  \multicolumn{2}{c}{---} & \multicolumn{2}{c}{---}\\ 
0&037025 & 0&203 & 1&457(7) &  0&683(4) & 2&140(9) & 0&774(7) & 1&050(46)&  $-$1&092(27) &  $-$0&042(59) & 2&142(46)\\ 
0&037235 & 0&204 & 1&462(13) &  0&681(8) & 2&142(17) & 0&781(13) & 0&956(106)&  $-$1&056(60) &  $-$0&100(139) & 2&013(102)\\ 
0&071421 & 0&392 & 1&132(9) &  0&497(5) & 1&629(12) & 0&635(8) & 0&853(35)&  $-$0&822(23) &  0&031(50) & 1&675(32)\\ 
0&072155 & 0&396 & 1&130(21) &  0&491(12) & 1&621(29) & 0&639(19) & 0&995(104)&  $-$0&693(55) &  0&302(127) & 1&688(108)\\ 
0&077106 & 0&423 & 1&089(11) &  0&470(6) & 1&560(15) & 0&619(10) & 0&745(42)&  $-$0&806(26) &  $-$0&061(54) & 1&550(43)\\ 
0&103678 & 0&569 & 0&912(13) &  0&376(7) & 1&288(18) & 0&536(11) & 0&685(44)&  $-$0&647(26) &  0&038(61) & 1&332(39)\\ 
0&114341 & 0&627 & 0&870(13) &  0&350(7) & 1&219(17) & 0&520(12) & 0&634(41)&  $-$0&601(26) &  0&034(54) & 1&235(43)\\ 
0&154213 & 0&846 & 0&696(15) &  0&256(8) & 0&951(20) & 0&440(13) & 0&479(36)&  $-$0&463(23) &  0&016(49) & 0&943(36)\\ 
0&191447 & 1&050 & 0&591(13) &  0&204(7) & 0&795(17) & 0&387(11) & 0&408(25)&  $-$0&388(16) &  0&020(34) & 0&796(25)\\ 
\hline \hline 
\end{tabular}
\end{table}


\begin{table}[ht]
\centering
\caption{Renormalized results for Dirac and Pauli form factors from the $am_l=0.006$ ensemble with $m_\pi
  \approx 355$ MeV.} \label{tab:ff_ml006}
\begin{tabular}{r@{.}lr@{.}lr@{.}lr@{.}lr@{.}lr@{.}lr@{.}lr@{.}lr@{.}lr@{.}l}
\hline \hline 
\multicolumn{2}{c}{$(aQ)^2$}  & \multicolumn{2}{c}{$Q^2$ [GeV$^2$]} & 
  \multicolumn{2}{c}{$F_1^u$}  & \multicolumn{2}{c}{$F_1^d$} & 
  \multicolumn{2}{c}{$F_1^{u+d}$} & \multicolumn{2}{c}{$F_1^{u-d}$} & 
  \multicolumn{2}{c}{$F_2^u$}  & \multicolumn{2}{c}{$F_2^d$} & 
  \multicolumn{2}{c}{$F_2^{u+d}$} & \multicolumn{2}{c}{$F_2^{u-d}$} \\ 
\hline 
0&000000 & 0&000 & 2&000(3) &  1&000(2) & 2&999(4) & 1&000(3) & 
  \multicolumn{2}{c}{---} &  \multicolumn{2}{c}{---} &  
  \multicolumn{2}{c}{---} &  \multicolumn{2}{c}{---}\\ 
0&037176 & 0&204 & 1&478(4) &  0&691(2) & 2&169(6) & 0&788(4) & 1&132(31)&  $-$1&192(17) &  $-$0&061(40) & 2&324(30)\\ 
0&037348 & 0&205 & 1&471(8) &  0&691(4) & 2&162(10) & 0&779(8) & 1&226(71)&  $-$1&130(38) &  0&096(93) & 2&356(66)\\ 
0&071948 & 0&395 & 1&162(6) &  0&509(3) & 1&671(8) & 0&653(5) & 0&909(25)&  $-$0&923(15) &  $-$0&014(34) & 1&832(24)\\ 
0&072559 & 0&398 & 1&155(12) &  0&503(7) & 1&659(17) & 0&652(11) & 0&866(62)&  $-$0&875(33) &  $-$0&009(79) & 1&741(59)\\ 
0&077106 & 0&423 & 1&134(8) &  0&491(4) & 1&625(10) & 0&643(7) & 0&878(28)&  $-$0&870(16) &  0&008(36) & 1&748(28)\\ 
0&104730 & 0&574 & 0&946(9) &  0&389(5) & 1&336(12) & 0&557(7) & 0&745(28)&  $-$0&727(17) &  0&018(37) & 1&472(27)\\ 
0&114455 & 0&628 & 0&904(9) &  0&365(4) & 1&269(11) & 0&540(8) & 0&693(29)&  $-$0&656(16) &  0&036(37) & 1&349(28)\\ 
0&154213 & 0&846 & 0&736(10) &  0&285(5) & 1&021(14) & 0&452(9) & 0&560(25)&  $-$0&505(15) &  0&056(33) & 1&065(25)\\ 
0&191561 & 1&051 & 0&619(9) &  0&225(4) & 0&844(11) & 0&393(7) & 0&476(17)&  $-$0&407(11) &  0&070(23) & 0&883(17)\\ 
\hline \hline 
\end{tabular}
\end{table}


\begin{table}[ht]
\centering
\caption{ Renormalized results for Dirac and Pauli form factors from the $am_l=0.008$ ensemble with $m_\pi
  \approx 403$ MeV.} \label{tab:ff_ml008}
\begin{tabular}
{r@{.}lr@{.}lr@{.}lr@{.}lr@{.}lr@{.}lr@{.}lr@{.}lr@{.}lr@{.}l}
\hline \hline 
\multicolumn{2}{c}{$(aQ)^2$} & \multicolumn{2}{c}{$Q^2$ [GeV$^2$]} & 
  \multicolumn{2}{c}{$F_1^u$}  & \multicolumn{2}{c}{$F_1^d$} & 
  \multicolumn{2}{c}{$F_1^{u+d}$} & \multicolumn{2}{c}{$F_1^{u-d}$} & 
  \multicolumn{2}{c}{$F_2^u$}  & \multicolumn{2}{c}{$F_2^d$} & 
  \multicolumn{2}{c}{$F_2^{u+d}$} & \multicolumn{2}{c}{$F_2^{u-d}$} \\ 
\hline 
0&000000 & 0&000 & 2&006(3) &  1&006(1) & 3&012(3) & 1&000(2) &  
  \multicolumn{2}{c}{---} & \multicolumn{2}{c}{---} & 
  \multicolumn{2}{c}{---} & \multicolumn{2}{c}{---} \\ 
0&037277 & 0&204 & 1&502(4) &  0&706(2) & 2&208(5) & 0&796(4) & 1&210(32)&  $-$1&193(19) &  0&016(43) & 2&403(31)\\ 
0&037427 & 0&205 & 1&499(8) &  0&706(4) & 2&204(11) & 0&793(7) & 1&342(65)&  $-$1&131(39) &  0&211(85) & 2&473(65)\\ 
0&072306 & 0&397 & 1&180(6) &  0&521(4) & 1&702(8) & 0&659(5) & 0&965(26)&  $-$0&926(17) &  0&038(36) & 1&891(26)\\ 
0&072839 & 0&400 & 1&168(12) &  0&519(6) & 1&687(16) & 0&649(10) & 0&985(60)&  $-$0&922(36) &  0&063(80) & 1&908(59)\\ 
0&077106 & 0&423 & 1&160(8) &  0&508(4) & 1&667(11) & 0&652(7) & 0&918(29)&  $-$0&861(19) &  0&058(39) & 1&779(29)\\ 
0&105450 & 0&578 & 0&965(8) &  0&401(5) & 1&366(11) & 0&564(7) & 0&783(29)&  $-$0&749(18) &  0&035(39) & 1&532(28)\\ 
0&114533 & 0&628 & 0&924(8) &  0&381(5) & 1&305(11) & 0&543(8) & 0&710(27)&  $-$0&669(18) &  0&041(38) & 1&379(26)\\ 
0&154213 & 0&846 & 0&771(11) &  0&302(6) & 1&073(14) & 0&469(9) & 0&567(24)&  $-$0&539(16) &  0&028(33) & 1&106(24)\\ 
0&191639 & 1&051 & 0&633(9) &  0&234(5) & 0&867(12) & 0&399(8) & 0&459(17)&  $-$0&442(12) &  0&017(24) & 0&901(17)\\ 
\hline \hline 
\end{tabular}
\end{table}


\begin{table}[ht]
\centering
\caption{Comparison of fit Ans\"atze to the isovector Dirac form factors $F_1^{u-d}$ for all three ensembles with different $Q^2$ cutoffs.}
\label{tab:F1_umd_fit_comp}

\begin{tabular}{c||c|c||c|c}
\hline
\hline
\multicolumn{5}{c}{$am_l = 0.004$} \\
\hline
& \multicolumn{2}{c||}{Dipole} & \multicolumn{2}{c}{Tripole} \\
\hline
$Q^2$ cutoff [GeV$^2$] & $\chi^2$/dof & $M_D^{-2}\, [\text{GeV}^{-2}]$ & $\chi^2$/dof &
$M_T^{-2}\, [\text{GeV}^{-2}]$ \\
\hline
0.3 & 0.2(6) & 0.670(22) & 0.2(6) & 0.436(14)\\
\hline
0.4 & 0.3(6) & 0.659(19) & 0.8(9) & 0.424(12) \\
\hline
0.5 & 0.5(6) & 0.653(17) & 1.0(9) & 0.418(11) \\
\hline
0.6 & 0.4(5) & 0.652(17) & 1.0(8) & 0.417(11) \\
\hline
0.7 & 0.5(5) & 0.649(17) & 1.2(8) & 0.414(11) \\
\hline
0.9 & 0.9(7) & 0.638(16) & 1.9(1.0) & 0.404(10) \\
\hline
1.1 & 1.4(8) & 0.632(16) & 3.0(1.1) & 0.398(10) \\
\hline
\hline 
\multicolumn{5}{c}{$am_l = 0.006$} \\
\hline
&\multicolumn{2}{c||}{Dipole} & \multicolumn{2}{c}{Tripole} \\
\hline
$Q^2$ cutoff [GeV$^2$] & $\chi^2$/dof & $M_D^{-2}\, [\text{GeV}^{-2}]$ & $\chi^2$/dof &
$M_T^{-2}\, [\text{GeV}^{-2}]$ \\
\hline
0.3 & 0.5(1.0) & 0.625(13) & 0.5(1.0) & 0.407(8) \\
\hline
0.4 & 1.8(1.3) & 0.610(12) & 3.3(1.8) & 0.393(7) \\
\hline
0.5 & 2.8(1.5) & 0.602(11) & 4.8(1.9) & 0.386(7) \\
\hline
0.6 & 2.3(1.2) & 0.602(11) & 4.2(1.7) & 0.386(7) \\
\hline
0.7 & 2.1(1.1) & 0.601(11) & 3.8(1.5) & 0.385(7) \\
\hline
0.9 & 2.0(1.0) & 0.597(11) & 4.1(1.4) & 0.379(7) \\
\hline
1.1 & 2.0(9) & 0.595(11) & 4.8(1.5) & 0.375(7) \\
\hline
\hline
\multicolumn{5}{c}{$am_l = 0.008$} \\
\hline
 &\multicolumn{2}{c||}{Dipole} & \multicolumn{2}{c}{Tripole} \\
\hline
$Q^2$ cutoff [GeV$^2$] & $\chi^2$/dof & $M_D^{-2}\, [\text{GeV}^{-2}]$ & $\chi^2$/dof &
$M_T^{-2}\, [\text{GeV}^{-2}]$ \\
\hline
0.3 & 0.09(42) & 0.592(13) & 0.09(42) & 0.386(8) \\
\hline
0.4 & 0.3(5) & 0.588(12) & 1.0(1.0) & 0.380(7)  \\
\hline
0.5 & 0.9(9) & 0.582(11) & 1.9(1.2) & 0.374(7)  \\
\hline
0.6 & 1.0(8) & 0.579(11) & 2.2(1.2) & 0.371(7)  \\
\hline
0.7 & 0.9(7) & 0.579(11) & 2.0(1.1) & 0.370(7) \\
\hline
0.9 & 1.1(7) & 0.575(10) & 2.7(1.2) & 0.366(6) \\
\hline
1.1 & 1.0(7) & 0.575(10) & 2.7(1.1) & 0.365(6)\\
\hline
\hline
\end{tabular}
\end{table}


\begin{table}[ht]
\centering
\caption{Comparison of fit Ans\"atze to the isovector Pauli form factors $F_2^{u-d}$ for all three ensembles with different $Q^2$ cutoffs.}
\label{tab:F2_umd_fit_comp}

\begin{tabular}{c||c|c|c||c|c|c}
\hline
\hline
\multicolumn{7}{c}{$am_l = 0.004$} \\
\hline
& \multicolumn{3}{c||}{Dipole} & \multicolumn{3}{c}{Tripole} \\
\hline
$Q^2$ cutoff [GeV$^2$] & $\chi^2$/dof & $F_2(0)$ & $M_D^{-2}\, [\text{GeV}^{-2}]$ & $\chi^2$/dof
& $F_2(0)$ & $M_T^{-2}\, [\text{GeV}^{-2}]$ \\
\hline
0.5 & 1.2(1.3) & 2.89(12) & 0.820(70) & 1.2(1.3) & 2.85(11) & 0.505(40) \\
\hline
0.6 & 1.1(1.1) & 2.92(11) & 0.846(63) &1.0(1.0) & 2.87(10) & 0.516(36) \\
\hline
0.7 & 0.9(8) & 2.93(11) & 0.847(60) & 0.8(8) & 2.87(10) & 0.513(33)  \\
\hline
0.9 & 0.9(8) & 2.98(9) & 0.888(46) & 0.7(7) & 2.89(8) & 0.526(15)\\
\hline
1.1 & 0.8(7) & 2.97(9) & 0.881(41) & 0.9(7) & 2.85(8) & 0.509(21)  \\
\hline
\hline
\multicolumn{7}{c}{$am_l = 0.006$} \\
\hline
 & \multicolumn{3}{c||}{Dipole} & \multicolumn{3}{c}{Tripole} \\
\hline
$Q^2$ cut [GeV$^2$] & $\chi^2$/dof & $F_2(0)$ & $M_D^{-2}\, [\text{GeV}^{-2}]$ & $\chi^2$/dof &
$F_2(0)$ & $M_T^{-2}\, [\text{GeV}^{-2}]$ \\
\hline
0.5 & 1.7(1.5) & 3.14(7) & 0.797(39) & 1.6(1.5) & 3.10(7) & 0.492(22) \\
\hline
0.6 & 1.4(1.2) & 3.16(7) & 0.810(35) & 1.2(1.1) & 3.10(6) & 0.495(20) \\
\hline
0.7 & 1.5(1.1) & 3.18(7) & 0.825(33) & 1.1(1.0) & 3.12(6) & 0.501(19) \\
\hline
0.9 & 1.4(1.0) & 3.22(6) & 0.851(26) & 1.0(8)  & 3.13(5) & 0.505(14) \\
\hline
1.1 & 1.3(9) & 3.24(5) & 0.861(22) & 1.0(7) & 3.11(5) & 0.499(12)  \\
\hline
\hline
\multicolumn{7}{c}{$am_l=0.008$}\\
\hline
 & \multicolumn{3}{c||}{Dipole} & \multicolumn{3}{c}{Tripole} \\
\hline
$Q^2$ cut [GeV$^2$] & $\chi^2$/dof & $F_2(0)$ & $M_D^{-2}\, [\text{GeV}^{-2}]$ & $\chi^2$/dof &
$F_2(0)$ & $M_T^{-2}\, [\text{GeV}^{-2}]$ \\
\hline
0.5 & 2.2(1.7) & 3.26(7) & 0.813(34) & 2.1(1.7) & 3.21(6) & 0.501(19) \\
\hline
0.6 & 1.6(1.3) & 3.26(6) & 0.813(33) & 1.7(1.3) & 3.20(6) & 0.497(19) \\
\hline
0.7 & 2.5(1.4) & 3.29(6) & 0.841(31) & 2.1(1.3) & 3.22(6) & 0.511(17)  \\
\hline
0.9 & 2.1(1.2) & 3.31(5) & 0.851(22) & 1.8(1.1) & 3.22(5) & 0.506(12) \\
\hline
1.1 & 2.0(1.1) & 3.32(5) & 0.862(20) & 1.6(1.0) & 3.21(5) & 0.502(10) \\
\hline
\hline
\end{tabular}
\end{table}

%% file: text/app_smearing/smearing.tex
Since careful optimization of the interpolating field for the nucleon source is crucial for the
high precision calculations described in this work, in this appendix we describe in detail our
optimization procedure and record the optimal parameters in two commonly used conventions.

We have two objectives in constructing sources for propagators that will be optimal 
for calculating hadronic matrix elements.  
The first is to maximize the overlap between the interpolating field acting on the QCD vacuum  
and the hadronic ground state.
The second is to  minimize fluctuations arising from the source itself.   
Let $\bar{N}$ denote an interpolating field with the quantum numbers of the hadron, 
$|\Psi \rangle = {\mathcal C}^{-1/2}\bar{N}|\Omega \rangle$ denote the normalized state obtained 
by its action on the vacuum, and $|n\rangle$ denote the $n^{th}$ eigenstate of the hadron 
(projected to zero momentum in the present discussion). 
Then, maximizing $ | \langle 0| \Psi\rangle |^2$ minimizes the contributions of excited states 
to the measurement of the hadronic matrix element of an operator $\cal O$ 
\begin{equation}
\langle N(t_3){\cal O}(t_2) N(t_1) \rangle = 
  {\mathcal C} \sum_{n,m} \langle \Psi |n\rangle \langle n | {\cal O}|m\rangle 
               \langle m | \Psi \rangle e^{-E_n(t_3-t_2) -E_m (t_2-t_1)},
\end{equation}
and hence enables one to reduce the source-sink separation while controlling contamination from
excited states as discussed in Sect.~\ref{sect:excited_states_systematics}.

The first objective is met by using smeared propagators and treating the rms radius of the smearing as a variational parameter.  Although similar effects can be accomplished with gauge fixed sources, we use gauge invariant sources of the  Wuppertal, or equivalently, Gaussian  form by smearing a delta function source over the three spatial dimensions of the source time slice.

Wuppertal smearing of a point source at the origin of time slice $t$ is defined in the MIT USQCD software as

\begin{equation}
\label{eq:wuppertal}     
\psi(x,t) = \left( 1 + \alpha \sum_{i=1}^3 
      \left[U(x, i) \delta_{x+\hat i,y} + U^\dag(x-\hat i, i)\delta_{x-\hat i,y}\right]   
    \right)^N \delta_{y,0}  \, \, ,
\end{equation}

and Gaussian smearing is defined in  Chroma software as 

\begin{equation}
\label{wupsmear} 
\begin{aligned}
\psi(x,t) &= \left( 1 - \frac{\sigma^2 \nabla^2}{4N}  \right)^N \psi(x,t)  \\
  &= \left(1 - \frac{3 \sigma^2}{2N}\right)^N   
     \left(1 + \frac{\sigma^2 /4N}{1 - 3 \sigma^2/2N} 
     \sum_{i=1}^3 \left[U(x, i) \delta_{x+\hat i,y} + U^\dag(x-\hat i, i) \delta_{x-\hat i,y}\right]
    \right)^N \delta_{y,0}  \, \, . 
\end{aligned}
\end{equation}

The Chroma  and MIT parameters are related by
\begin{eqnarray}
\alpha &=& \frac{\sigma^2/4N}{1-3 \sigma^2/2N} \,, \\
\sigma^2 & = & \frac{2N\alpha}{3 \alpha + 1/2} \, .
\end{eqnarray}

Note that there is an instability for $\alpha < 0$, since the sign of the source generated in Eq.~(\ref{eq:wuppertal}) is  then $(-1)^{x+y+z}$, and the resulting spatially oscillating source has an extremely poor overlap with the physical ground state.  In terms of the Gaussian parameters, the instability arises for $ N < 3 \sigma^2 /2 $.  

Because the smeared sources contain link variables $U$, the statistical fluctuations in correlation functions using these sources are larger than those arising from point sources.  To attain our second objective of minimizing the fluctuations arising from the source itself, it is highly advantageous to perform APE smearing of the gauge links used in generating the source on the time slice of the source.  In each iteration of APE smearing, each link is replaced by a linear combination of itself and the sum of staples within that time slice, and projected back onto SU(3) as follows
\begin{equation}
U^{(N+1)}_{x,i} = \mathrm{Proj}_{SU(3)} \left[ U^N_{x,i } + \beta \sum_{j \ne i}^3 U^N_{x,j} U^N_{x+j,
i} U^{N \dag}_{x+i,j}    \right] \label{apesmear},
\end{equation}
and the APE smearing is repeated N times.  An alternative notation is
\begin{equation}
U^{(N+1)}_{x,i} = \mathrm{Proj}_{SU(3)} \left[ A U^N_{x,i } +  \sum_{j \ne i}^3 U^N_{x,j} U^N_{x+j, i}
U^{N \dag}_{x+i,j}    \right] ,
\end{equation}
so that 
\begin{equation}
A = 1/ \beta \label{apesmear2}.
\end{equation}

A convenient measure of the  smearing of the source $\psi(x,t)$  in Eq.~(\ref{eq:wuppertal}) is the rms radius
\begin{equation}
r_\text{rms} = \langle r^2 \rangle^{\frac{1}{2}} = \left[  \frac{\int d^3x |\vec x|^2 \psi^*(\vec x,t) \psi(\vec x,t)}{\int d^3x  \psi^*(\vec x,t) \psi(\vec x,t)}   \right]^{1/2} , \label{rmsdef}
\end{equation}
and Fig.~2 of Ref.\,\cite{Dolgov:2002zm}
 shows how $r_\text{rms}$ depends on the parameters $N$ and $\alpha$. 
As one expects from  the fact that
smearing is a random walk governed by the gauge fields, the rms radius is
approximately proportional to $\sqrt{N}$. 
Since the size of the source is
nearly independent of $\alpha$ for $\alpha > 3$, at which point the constant
term in Eq.~(\ref{eq:wuppertal}) becomes negligible relative to the hopping
term, in all our calculations, we use $\alpha = 3$, which provides the maximum $ r_\text{rms}$  for a given number of smearing steps $N$.

It is simplest to think about optimization criteria for Wilson fermions, for which one can 
construct a transfer matrix and correct propagators such that the two-point correlation function 
has quarks and antiquarks properly normal ordered at zero time separation\cite{Luscher:1976ms}. 
In this case, the source may be  optimized straightforwardly by maximizing the overlap 
between the normalized state created by the action of the source  
$|\Psi^{(r)} \rangle = {\mathcal C}^{-1/2} \bar{N}^{(r)} |\Omega \rangle$, 
where the source $\bar{N}^{(r)}$ has rms radius $r$, 
and the normalized ground state of the nucleon $| 0 \rangle$.
Denoting the momentum projected normalized eigenstates of the nucleon by $| n \rangle $
and their energies by $E_n$, the momentum projected two-point correlation function may be expanded:
\begin{equation}
C^{(r)}(t) = \int d^3x \langle N^{(r)}(x,t) \bar{N}^{(r)}(0,0) \rangle = 
  {\mathcal C} \sum_n \left|\langle \Psi^{(r)}|n\rangle\right|^2 e^{-E_nt}\label{twoptfn},
\end{equation}
where ${\mathcal C}$ is an unknown normalization constant. 
Since one can directly measure the correlation function at zero time separation
\begin{equation}
\label{Acoef}
A^{(r)} = C^{(r)}(0) =  
  {\mathcal C} \sum_n \left|\langle \Psi^{(r)} |n\rangle\right|^2
,
\end{equation}
and reliably fit the large $t$ behavior of the correlation function to extract the
ground state contribution
\begin{equation}
\label{Bcoef}
B^{(r)} =  {\mathcal C}\left|\langle  \Psi^{(r)} |0\rangle\right|^2, 
\end{equation}
the probability that the source contains the nucleon
ground state is given by 
\begin{equation}
\label{prob}
{\mathcal P}^{(r)} = \frac{B^{(r)}}{A^{(r)}} =
  \left|\langle \Psi^{(r)} |0\rangle\right|^2.
\end{equation}

Using this criterion, Bratt\cite{Bratt:2008ur} has recently shown that optimizing the source size with 4-component nucleon sources yields a maximum overlap of 35\%, projecting onto the upper two components (in the Bjorken-Drell convention for which these components yield the non-relativistic limit) increases the overlap to 50\%, and  APE smearing of the gauge links in the source further increases the overlap to 80\%.

For domain wall fermions,  which do not have a local transfer matrix, we consider the following generalization of Eqs.~(\ref{Acoef}-\ref{prob}), which compares the ratio of the correlation function and extrapolated ground state contribution at time $t$ instead of time 0:

\begin{eqnarray}
\label{Acoef2} 
A^{(r)}(t) &  = & C^{(r)}(t) , \\
\label{Bcoef2}
B^{(r)}(t) & = &  {\mathcal C}\left|\langle  \Psi^{(r)} |0\rangle\right|^2 e^{-E_0 t} , \\
\label{prob2}
{\mathcal P}^{(r)}(t) & =  &\frac{B^{(r)}(t)}{A^{(r)}(t)} .
\end{eqnarray}
This ratio, ${\mathcal P}^{(r)}(t)$, ranges from the overlap ${\mathcal P}^{(r)} $ at $ t=0$ 
to 1 in the limit $t \to \infty$. 
We expect that for small $t$,  it is still a good measure of the presence of excited state 
components in the source and should have a maximum close to the maximum in ${\cal P}^{(r)}$.  
This expectation is borne out in the case of Wilson fermions, and we note  that this criterion 
gets even better as the lattice spacing decreases. 
Since  we are only interested in the dependence of ${\mathcal P}^{(r)}(t)  $ on the rms radius 
$r$ and the absolute normalization for $ t \ne  0$  has no physical significance, 
it suffices to calculate the following ratio for large $t_0$
\begin{equation}
\frac{C^{(r)}(t_0)}{C^{(r)}(t)} \xrightarrow{t_0 \to \infty}   
  \frac{{\mathcal C} |\langle \Psi^{(r)}|0\rangle|^2 e^{-E_0 t_0}}{C^{(r)}(t)} =
  {\mathcal P}^{(r)}(t) e^{E_0 (t_0 - t)}.
\end{equation}
For each value of $t$, it is convenient to normalize the curve such that its maximum value is unity. Hence, defining the  rms radius at the maximun as  $r^*$, our  final criterion for optimizing the smearing is the ratio
\begin{equation}
\label{finrat}
R^{(r)}(t) = \frac{C^{(r)}(t_0) / C^{(r)}(t)}{C^{(r^*)}(t_0) / C^{(r^*)}(t)}
.
\end{equation}
Equation~(\ref{finrat}) has the computational advantages that all oscillating terms in the time dependence of the correlation functions cancel out of the ratios and that jacknife or bootstrap analysis enables accurate measurements on small ensembles.

\begin{figure}[t]
\centering
\vspace*{-1.1cm}
\includegraphics[width=17pc,angle=-90,scale=.85]{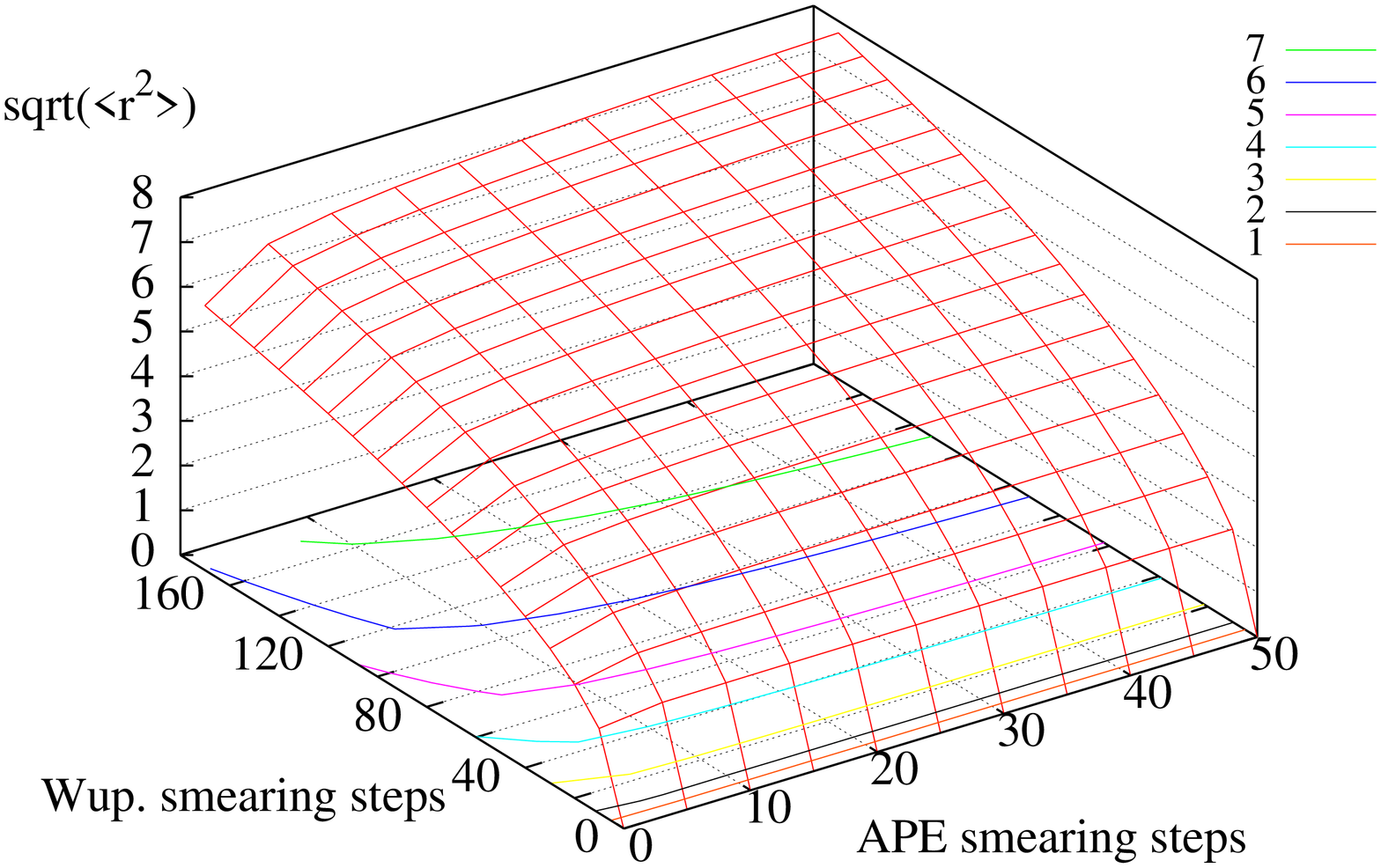}
\includegraphics[width=17pc,angle=-90,scale=.85]{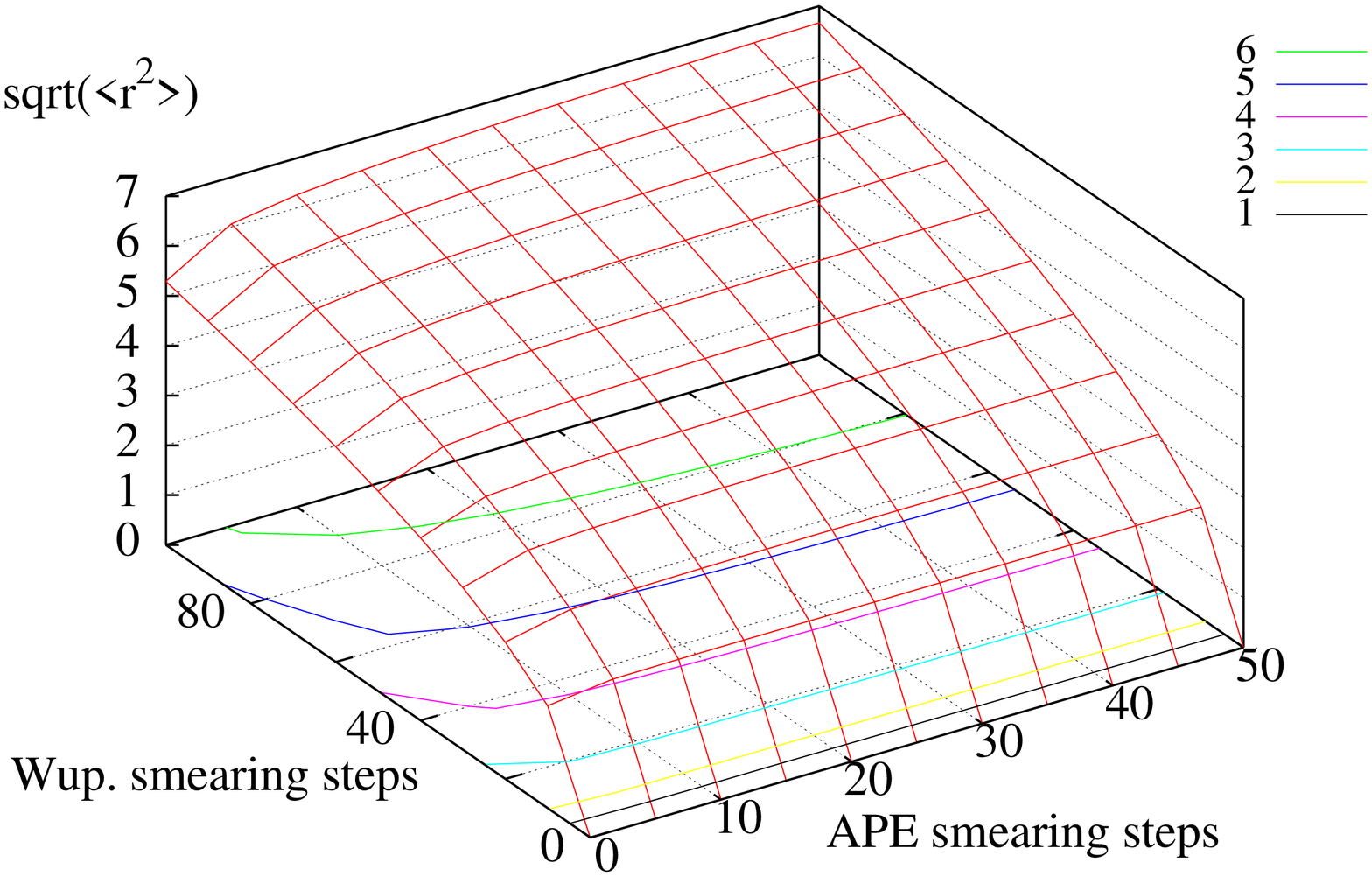}
\vspace*{-.5cm}
\caption{\label{rms32}The rms radius of a gauge invariant smeared source as a function of the coefficient $\alpha$ and number of smearing steps $N$ defined in Eq.~(\ref{eq:wuppertal}) for the coarse (left panel) and fine (right panel) lattices. The  curves projected in the horizontal plane show the numbers of smearing steps required for the specific values of rms radii shown in the key.}
\end{figure}

We now show the results of optimizing the ratio  $R^{(r)}(t)$  on a coarse $24^3 \times 64$ 
domain wall lattice with $a = 0.114\text{ fm}$, $m_\pi = 420\text{ MeV}$, $m_s = 0.04$, 
and $m_u = 0.01$, using 32 configurations and on a fine $32^3 \times 64$ domain wall lattice 
with $a = 0.081\text{ fm}$, $m_\pi = 310\text{ MeV}$, $m_s = 0.03$, and $m_u = 0.004$ 
using 33 configurations.  
We included both APE smearing, with $\beta = 0.3509 $, and Wuppertal smearing with $\alpha = 3$. Because APE smearing smoothes the links, the rms radius obtained from a given number of Wuppertal steps changes with the number of APE steps, becoming slightly larger as the number of APE smears increases.  Figure~\ref{rms32} shows the rms radius calculated as a function of both the number of APE and Wuppertal steps for both lattice spacings.

 \begin{figure}[t]
\centering
\vspace*{-.5cm}
\raisebox{ 0cm}{\includegraphics[width=17pc,angle=-90,scale=.85]{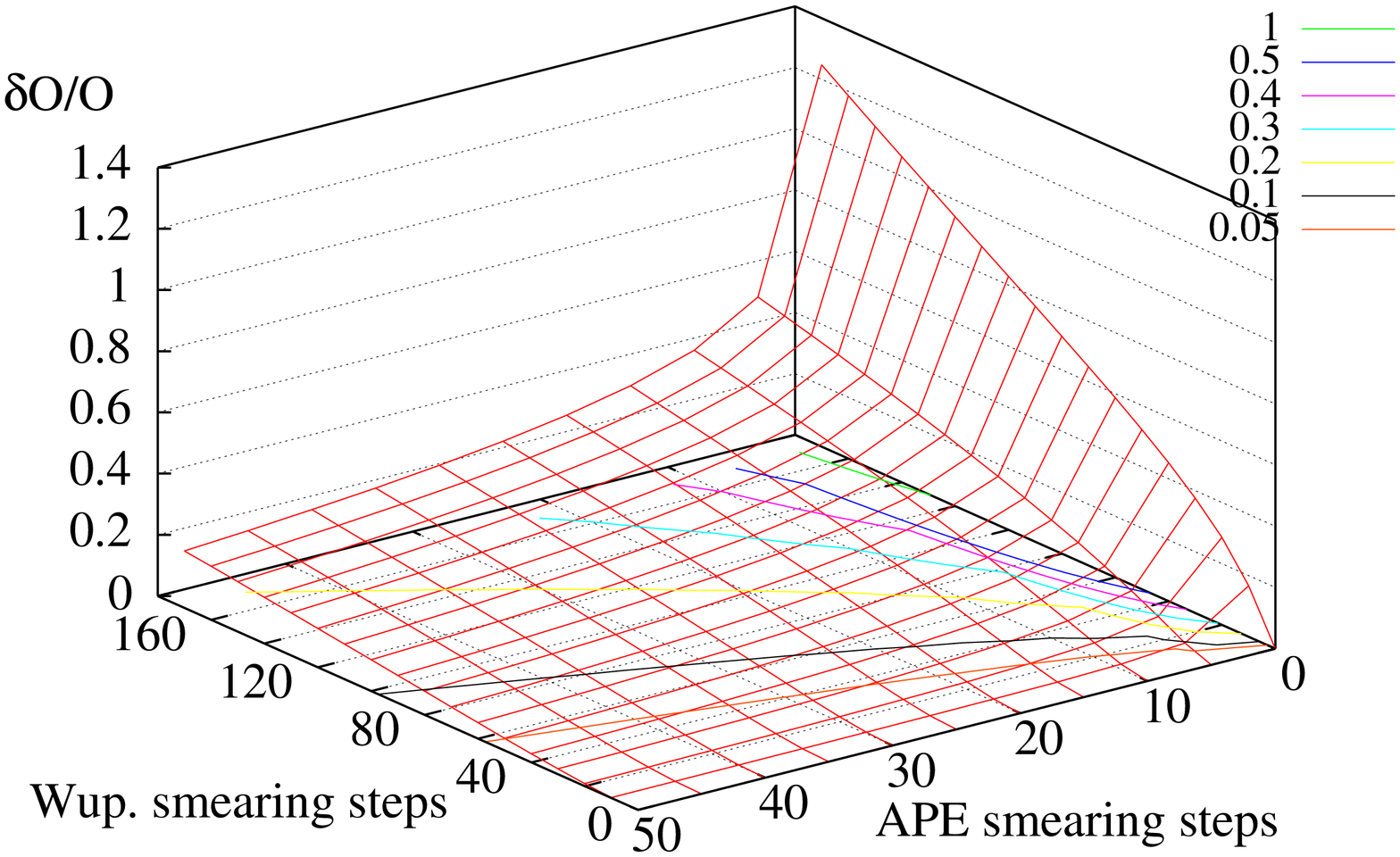}}
\raisebox{ 0cm}{\includegraphics[width=17pc,angle=-90,scale=.85]{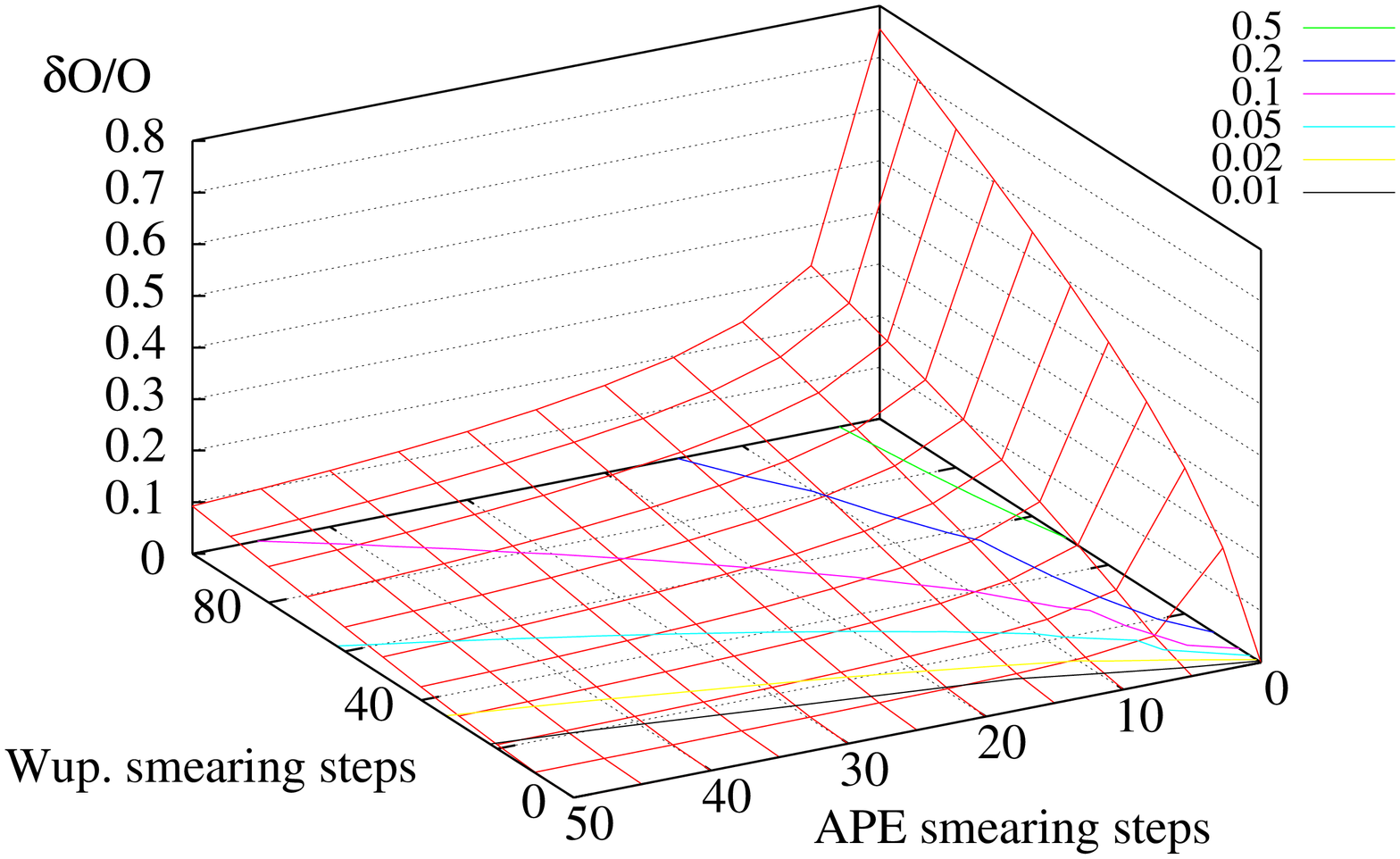}}
\vspace*{-.7cm}
\caption{\label{fluct32}The source variance ${\delta {\cal O}} / { {\cal O}} $ as a function of the rms radius and number of APE smears $N$ for the  coarse (left panel) and fine (right panel) lattices. The curves projected in the horizontal plane show the numbers of smearing steps required for the specific values of source variance shown in the key.} 
\end{figure}
 \begin{figure}[t]
\centering
\vspace*{-.1cm}
\raisebox{ 0cm}{\includegraphics[width=17pc,angle=-90,scale=.85]{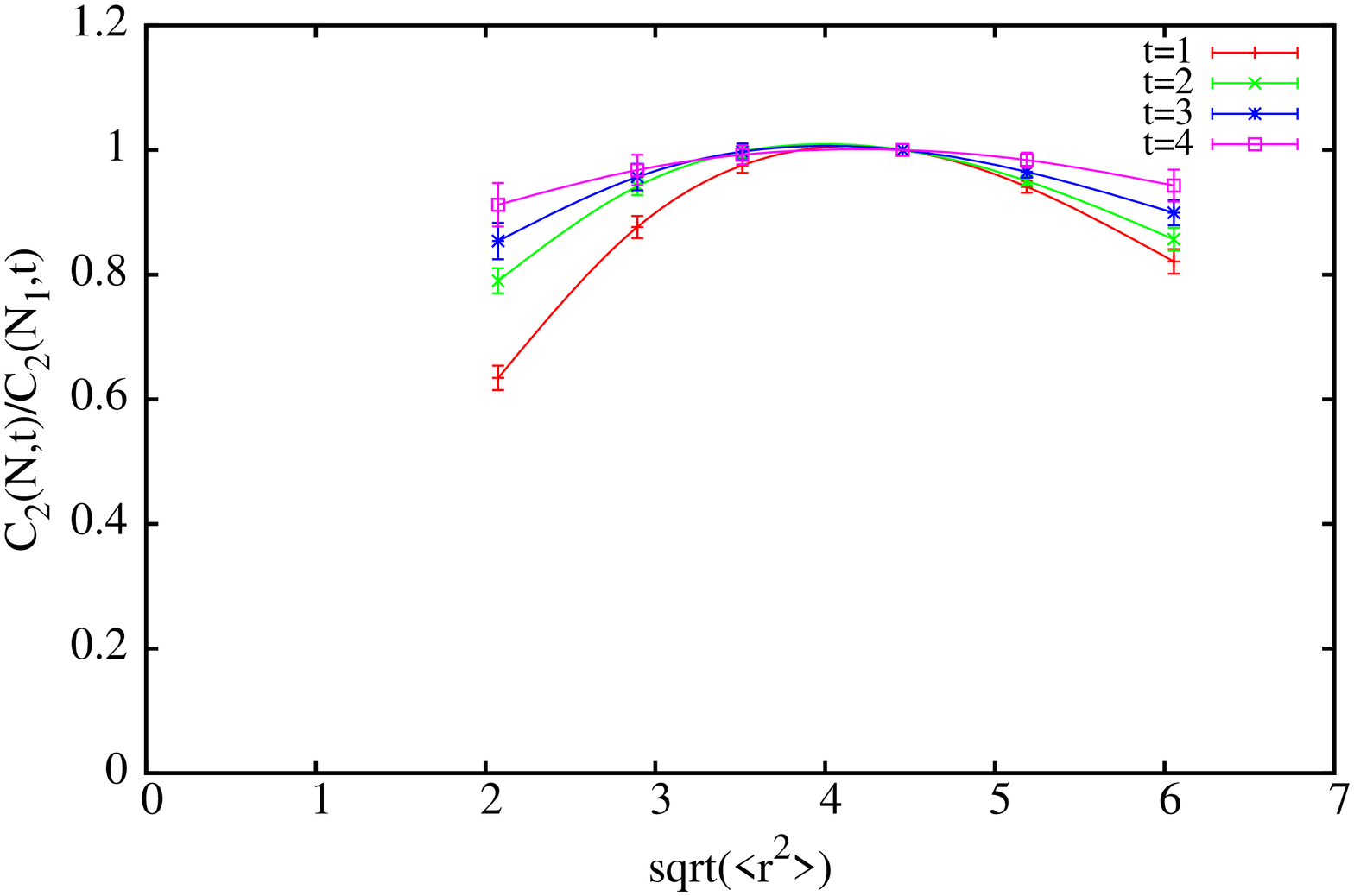}}
\raisebox{ 0cm}{\includegraphics[width=17pc,angle=-90,scale=.85]{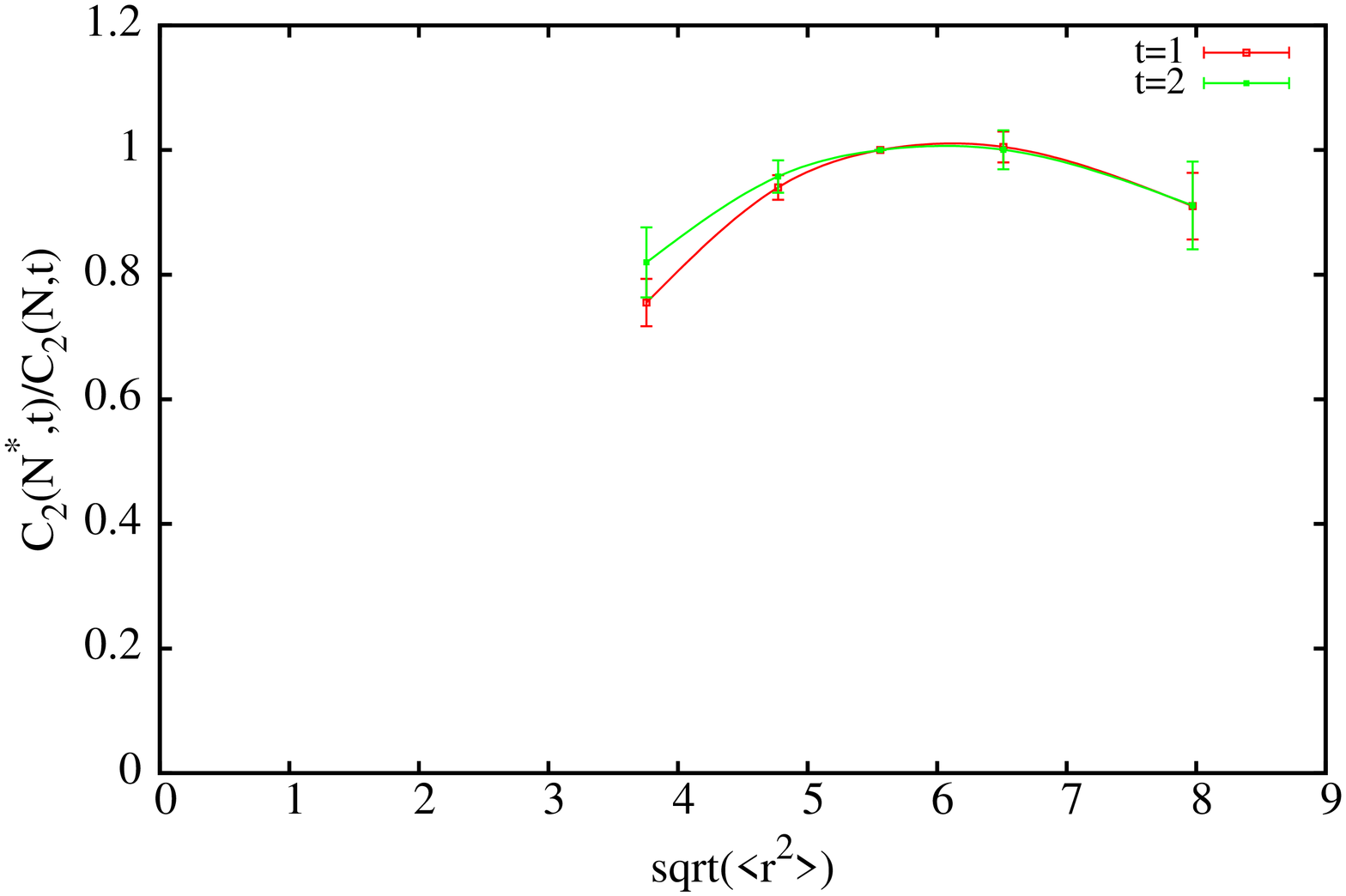}}
\vspace*{-.2cm}
\caption{\label{overlap32}
The left panel shows the ratio  $R^{(r)} (t)$  for $t$ = 1,2,3, and 4  as a function of $r$   on the coarse  lattice.  The solid curves are splines passing through the mean values to guide the eye.  This graph provides a robust determination of the optimal rms radius $r$ = 4.0 lattice units, corresponding to $N$ = 40 Wuppertal smearing steps.
The right panel shows the analogous ratio  $R^{(r)} (t)$  for $t$ = 1 and 2  as a function of $r$  on the fine lattice. } 
\end{figure}

Since fluctuations in the normalization of the source directly contribute to the overall fluctuations in correlation functions, it is desirable to use APE smearing to smooth the spatial links used in generating the source and thereby diminish the fluctuations.  A simple measure of these fluctuations is the relative fluctuation  
$\frac{\delta {\cal O}}{ {\cal O}} =  \frac{\langle ( {\cal O} - \langle  {\cal O}\rangle)^2\rangle ^{\frac{1}{2}}}{\langle  {\cal O} \rangle }$, where $ {\cal O}$ is the rms radius 
defined in Eq.~(\ref{rmsdef}).   Figure~\ref{fluct32} shows the dramatic effect that APE smearing has in reducing these fluctuations for both lattice spacings.  Since the incremental benefit of successive smearing becomes small beyond 25 smearing steps, we have chosen to use 25 steps throughout.  Note that for the largest number of Wuppertal steps, this reduces the noise by a factor of more than 5 in each case.  

 Figure~\ref{overlap32} shows the primary result of the calculation for both lattice spacings.  For the coarse lattice,
 the ratio $R^{(r)}(t)$ is calculated at six values of the number of Wuppertal steps, $N = 10, 20, 30, 50, 70, 100$, corresponding to rms radii, $r =  2.07,\, 2.89,\, 3.51,\, 4.46,\, 5.19,\text{ and }6.06$ lattice units respectively. We chose $r^* = 4.46\text{ fm}$ and calculated bootstrap error bars using 32 configurations. Instead of normalizing at a single value of $t_0$ as in Eq.~(\ref{finrat}), the errors in the ratios in Fig.~\ref{overlap32} were further reduced by normalizing to an exponential fit to each correlation function in the region $ t = [ 6 - 12]$. These results are completely consistent with those of a single $t_0$, but display the shape of the maxima more precisely.
 Note that for all four values $t = 1, 2, 3$, and $4$, the curves are accurately determined and
 the ratio $R^{(r)}(t)$ has a maximum at approximately the same point, $r = 4.0$, corresponding to $N = 40$. Thus, we believe our optimization criterion is  robust and statistically accurate for domain wall fermions.

For the fine lattices, the ratio $R^{(r)}(t)$ is calculated at 5 values of the number of 
Wuppertal steps, $N =  30, 50, 70, 100$, and $150$,  corresponding to  
rms radii $r =  3.76, 4.77, 5.56, 6.51$, and $7.77$  lattice units respectively. 
We chose $r^* = 5.56$, normalized by exponential fits to each correlation function in the region 
$t = [ 6 : 12]$,  calculated jackknife error bars, and only included $t = 1$ and 2 to avoid 
making the graph confusing due to the larger error bars.  
The maximum occurs at approximately  $r = 6.0$ lattice units,  corresponding to 
84 Wuppertal smearing steps. 
This result appears reasonable, since assuming a constant rms radius in physical units would imply 
that the the rms radius on the coarser lattice of $4.0$ lattice units would scale to 
$4 \times 0.123 / 0.093 = 5.3$ lattice units on the present lattice, 
and the pion mass on the finer lattice is somewhat lighter.

We summarize the final  parameters  for optimal sources used in this work in Tab.~\ref{tab1}, where the parameters are defined in Eqs.~(\ref{wupsmear} - \ref{apesmear2}).

\begin{table}[h]
\centering
\begin{tabular}{c|c|c|c|c|c|c|c}
\hline
\hline
lattice &\multicolumn{3}{c|}{APE smearing} &\multicolumn{3}{c|}{Wuppertal smearing} & size \\
\hline
$a$ (fm) & $\beta$ & A& $N_{APE}$ &$ \alpha$ & $\sigma$ & $ N_W $ & $\langle r^2\rangle^{1/2}$ \\  
\hline
0.114 & 0.3509 & 2.85 & 25 & 3 & 5.026 & 40 & 4.0  \\
0.084 & 0.3509 & 2.85 & 25 & 3 & 7.284 & 84 & 6.0  \\
\hline
\hline
\end{tabular}
\caption{Parameters for optimal sources.}
\label{tab1}
\end{table}

%% file: paper.bbl
\begin{thebibliography}{75}
\expandafter\ifx\csname natexlab\endcsname\relax\def\natexlab#1{#1}\fi
\expandafter\ifx\csname bibnamefont\endcsname\relax
  \def\bibnamefont#1{#1}\fi
\expandafter\ifx\csname bibfnamefont\endcsname\relax
  \def\bibfnamefont#1{#1}\fi
\expandafter\ifx\csname citenamefont\endcsname\relax
  \def\citenamefont#1{#1}\fi
\expandafter\ifx\csname url\endcsname\relax
  \def\url#1{\texttt{#1}}\fi
\expandafter\ifx\csname urlprefix\endcsname\relax\def\urlprefix{URL }\fi
\providecommand{\bibinfo}[2]{#2}
\providecommand{\eprint}[2][]{\url{#2}}

\bibitem[{\citenamefont{Burkardt}(2000)}]{Burkardt:2000za}
\bibinfo{author}{\bibfnamefont{M.}~\bibnamefont{Burkardt}},
  \bibinfo{journal}{Phys. Rev.} \textbf{\bibinfo{volume}{D62}},
  \bibinfo{pages}{071503} (\bibinfo{year}{2000}), \eprint{hep-ph/0005108}.

\bibitem[{\citenamefont{Burkardt}(2003)}]{Burkardt:2002hr}
\bibinfo{author}{\bibfnamefont{M.}~\bibnamefont{Burkardt}},
  \bibinfo{journal}{Int. J. Mod. Phys.} \textbf{\bibinfo{volume}{A18}},
  \bibinfo{pages}{173} (\bibinfo{year}{2003}), \eprint{hep-ph/0207047}.

\bibitem[{\citenamefont{Friedrich and Walcher}(2003)}]{Friedrich:2003iz}
\bibinfo{author}{\bibfnamefont{J.}~\bibnamefont{Friedrich}} \bibnamefont{and}
  \bibinfo{author}{\bibfnamefont{T.}~\bibnamefont{Walcher}},
  \bibinfo{journal}{Eur. Phys. J.} \textbf{\bibinfo{volume}{A17}},
  \bibinfo{pages}{607} (\bibinfo{year}{2003}), \eprint{hep-ph/0303054}.

\bibitem[{\citenamefont{Arrington et~al.}(2007)\citenamefont{Arrington,
  Melnitchouk, and Tjon}}]{Arrington:2007ux}
\bibinfo{author}{\bibfnamefont{J.}~\bibnamefont{Arrington}},
  \bibinfo{author}{\bibfnamefont{W.}~\bibnamefont{Melnitchouk}},
  \bibnamefont{and} \bibinfo{author}{\bibfnamefont{J.~A.} \bibnamefont{Tjon}},
  \bibinfo{journal}{Phys. Rev.} \textbf{\bibinfo{volume}{C76}},
  \bibinfo{pages}{035205} (\bibinfo{year}{2007}), \eprint{0707.1861}.

\bibitem[{\citenamefont{{H\"ohler} et~al.}(1976)}]{Hohler:1976ax}
\bibinfo{author}{\bibfnamefont{G.}~\bibnamefont{{H\"ohler}}}
  \bibnamefont{et~al.}, \bibinfo{journal}{Nucl. Phys.}
  \textbf{\bibinfo{volume}{B114}}, \bibinfo{pages}{505} (\bibinfo{year}{1976}).

\bibitem[{\citenamefont{Mergell et~al.}(1996)\citenamefont{Mergell,
  {Mei{\ss}ner}, and Drechsel}}]{Mergell:1995bf}
\bibinfo{author}{\bibfnamefont{P.}~\bibnamefont{Mergell}},
  \bibinfo{author}{\bibfnamefont{U.~G.} \bibnamefont{{Mei{\ss}ner}}},
  \bibnamefont{and} \bibinfo{author}{\bibfnamefont{D.}~\bibnamefont{Drechsel}},
  \bibinfo{journal}{Nucl. Phys.} \textbf{\bibinfo{volume}{A596}},
  \bibinfo{pages}{367} (\bibinfo{year}{1996}), \eprint{hep-ph/9506375}.

\bibitem[{\citenamefont{Belushkin
  et~al.}(2007{\natexlab{a}})\citenamefont{Belushkin, Hammer, and
  {Mei{\ss}ner}}}]{Belushkin:2006qa}
\bibinfo{author}{\bibfnamefont{M.~A.} \bibnamefont{Belushkin}},
  \bibinfo{author}{\bibfnamefont{H.~W.} \bibnamefont{Hammer}},
  \bibnamefont{and} \bibinfo{author}{\bibfnamefont{U.~G.}
  \bibnamefont{{Mei{\ss}ner}}}, \bibinfo{journal}{Phys. Rev.}
  \textbf{\bibinfo{volume}{C75}}, \bibinfo{pages}{035202}
  (\bibinfo{year}{2007}{\natexlab{a}}), \eprint{hep-ph/0608337}.

\bibitem[{\citenamefont{Bernauer}(2008)}]{Bernauer:2008zz}
\bibinfo{author}{\bibfnamefont{J.~C.} \bibnamefont{Bernauer}},
  \bibinfo{journal}{Lect. Notes Phys.} \textbf{\bibinfo{volume}{745}},
  \bibinfo{pages}{79} (\bibinfo{year}{2008}).

\bibitem[{\citenamefont{Milbrath et~al.}(1998)}]{Milbrath:1997de}
\bibinfo{author}{\bibfnamefont{B.~D.} \bibnamefont{Milbrath}}
  \bibnamefont{et~al.} (\bibinfo{collaboration}{Bates FPP}),
  \bibinfo{journal}{Phys. Rev. Lett.} \textbf{\bibinfo{volume}{80}},
  \bibinfo{pages}{452} (\bibinfo{year}{1998}), \eprint{nucl-ex/9712006}.

\bibitem[{\citenamefont{Pospischil et~al.}(2001)}]{Pospischil:2001pp}
\bibinfo{author}{\bibfnamefont{T.}~\bibnamefont{Pospischil}}
  \bibnamefont{et~al.} (\bibinfo{collaboration}{A1}), \bibinfo{journal}{Eur.
  Phys. J.} \textbf{\bibinfo{volume}{A12}}, \bibinfo{pages}{125}
  (\bibinfo{year}{2001}).

\bibitem[{\citenamefont{Gayou et~al.}(2002)}]{Gayou:2001qd}
\bibinfo{author}{\bibfnamefont{O.}~\bibnamefont{Gayou}} \bibnamefont{et~al.}
  (\bibinfo{collaboration}{Jefferson Lab Hall A}), \bibinfo{journal}{Phys. Rev.
  Lett.} \textbf{\bibinfo{volume}{88}}, \bibinfo{pages}{092301}
  (\bibinfo{year}{2002}), \eprint{nucl-ex/0111010}.

\bibitem[{\citenamefont{Gayou et~al.}(2001)}]{Gayou:2001qt}
\bibinfo{author}{\bibfnamefont{O.}~\bibnamefont{Gayou}} \bibnamefont{et~al.},
  \bibinfo{journal}{Phys. Rev.} \textbf{\bibinfo{volume}{C64}},
  \bibinfo{pages}{038202} (\bibinfo{year}{2001}).

\bibitem[{\citenamefont{Punjabi et~al.}(2005)}]{Punjabi:2005wq}
\bibinfo{author}{\bibfnamefont{V.}~\bibnamefont{Punjabi}} \bibnamefont{et~al.},
  \bibinfo{journal}{Phys. Rev.} \textbf{\bibinfo{volume}{C71}},
  \bibinfo{pages}{055202} (\bibinfo{year}{2005}), \eprint{nucl-ex/0501018}.

\bibitem[{\citenamefont{Arrington et~al.}(2004)}]{Arrington:2004hk}
\bibinfo{author}{\bibfnamefont{J.}~\bibnamefont{Arrington}}
  \bibnamefont{et~al.} (\bibinfo{year}{2004}), \eprint{nucl-ex/0408020}.

\bibitem[{\citenamefont{Collaboration}(2009)}]{olympus}
\bibinfo{author}{\bibfnamefont{T.~O.} \bibnamefont{Collaboration}},
  \bibinfo{journal}{http://web.mit.edu/olympus/}  (\bibinfo{year}{2009}).

\bibitem[{\citenamefont{{G\"ockeler} et~al.}(2005)}]{Gockeler:2003ay}
\bibinfo{author}{\bibfnamefont{M.}~\bibnamefont{{G\"ockeler}}}
  \bibnamefont{et~al.} (\bibinfo{collaboration}{QCDSF}),
  \bibinfo{journal}{Phys. Rev.} \textbf{\bibinfo{volume}{D71}},
  \bibinfo{pages}{034508} (\bibinfo{year}{2005}), \eprint{hep-lat/0303019}.

\bibitem[{\citenamefont{Alexandrou et~al.}(2006)\citenamefont{Alexandrou,
  Koutsou, Negele, and Tsapalis}}]{Alexandrou:2006ru}
\bibinfo{author}{\bibfnamefont{C.}~\bibnamefont{Alexandrou}},
  \bibinfo{author}{\bibfnamefont{G.}~\bibnamefont{Koutsou}},
  \bibinfo{author}{\bibfnamefont{J.~W.} \bibnamefont{Negele}},
  \bibnamefont{and} \bibinfo{author}{\bibfnamefont{A.}~\bibnamefont{Tsapalis}},
  \bibinfo{journal}{Phys. Rev.} \textbf{\bibinfo{volume}{D74}},
  \bibinfo{pages}{034508} (\bibinfo{year}{2006}), \eprint{hep-lat/0605017}.

\bibitem[{\citenamefont{Sasaki and Yamazaki}(2008)}]{Sasaki:2007gw}
\bibinfo{author}{\bibfnamefont{S.}~\bibnamefont{Sasaki}} \bibnamefont{and}
  \bibinfo{author}{\bibfnamefont{T.}~\bibnamefont{Yamazaki}},
  \bibinfo{journal}{Phys. Rev.} \textbf{\bibinfo{volume}{D78}},
  \bibinfo{pages}{014510} (\bibinfo{year}{2008}), \eprint{0709.3150}.

\bibitem[{\citenamefont{Tang et~al.}(2003)\citenamefont{Tang, Wilcox, and
  Lewis}}]{Tang:2003jh}
\bibinfo{author}{\bibfnamefont{A.}~\bibnamefont{Tang}},
  \bibinfo{author}{\bibfnamefont{W.}~\bibnamefont{Wilcox}}, \bibnamefont{and}
  \bibinfo{author}{\bibfnamefont{R.}~\bibnamefont{Lewis}},
  \bibinfo{journal}{Phys. Rev.} \textbf{\bibinfo{volume}{D68}},
  \bibinfo{pages}{094503} (\bibinfo{year}{2003}), \eprint{hep-lat/0307006}.

\bibitem[{\citenamefont{Boinepalli et~al.}(2006)\citenamefont{Boinepalli,
  Leinweber, Williams, Zanotti, and Zhang}}]{Boinepalli:2006xd}
\bibinfo{author}{\bibfnamefont{S.}~\bibnamefont{Boinepalli}},
  \bibinfo{author}{\bibfnamefont{D.~B.} \bibnamefont{Leinweber}},
  \bibinfo{author}{\bibfnamefont{A.~G.} \bibnamefont{Williams}},
  \bibinfo{author}{\bibfnamefont{J.~M.} \bibnamefont{Zanotti}},
  \bibnamefont{and} \bibinfo{author}{\bibfnamefont{J.~B.} \bibnamefont{Zhang}},
  \bibinfo{journal}{Phys. Rev.} \textbf{\bibinfo{volume}{D74}},
  \bibinfo{pages}{093005} (\bibinfo{year}{2006}), \eprint{hep-lat/0604022}.

\bibitem[{\citenamefont{{G\"ockeler} et~al.}(2007)}]{Gockeler:2007hj}
\bibinfo{author}{\bibfnamefont{M.}~\bibnamefont{{G\"ockeler}}}
  \bibnamefont{et~al.} (\bibinfo{collaboration}{QCDSF/UKQCD}),
  \bibinfo{journal}{PoS} \textbf{\bibinfo{volume}{LAT2007}},
  \bibinfo{pages}{161} (\bibinfo{year}{2007}), \eprint{0710.2159}.

\bibitem[{\citenamefont{Alexandrou}(2009)}]{Alexandrou:2009xk}
\bibinfo{author}{\bibfnamefont{C.}~\bibnamefont{Alexandrou}}
  (\bibinfo{year}{2009}), \eprint{0906.4137}.

\bibitem[{\citenamefont{Alexandrou et~al.}(2008)}]{Alexandrou:2008rp}
\bibinfo{author}{\bibfnamefont{C.}~\bibnamefont{Alexandrou}}
  \bibnamefont{et~al.} (\bibinfo{collaboration}{COL-NOTE = European Twisted
  Mass}) (\bibinfo{year}{2008}), \eprint{0811.0724}.

\bibitem[{\citenamefont{Lin et~al.}(2008)\citenamefont{Lin, Blum, Ohta, Sasaki,
  and Yamazaki}}]{Lin:2008uz}
\bibinfo{author}{\bibfnamefont{H.-W.} \bibnamefont{Lin}},
  \bibinfo{author}{\bibfnamefont{T.}~\bibnamefont{Blum}},
  \bibinfo{author}{\bibfnamefont{S.}~\bibnamefont{Ohta}},
  \bibinfo{author}{\bibfnamefont{S.}~\bibnamefont{Sasaki}}, \bibnamefont{and}
  \bibinfo{author}{\bibfnamefont{T.}~\bibnamefont{Yamazaki}},
  \bibinfo{journal}{Phys. Rev.} \textbf{\bibinfo{volume}{D78}},
  \bibinfo{pages}{014505} (\bibinfo{year}{2008}), \eprint{0802.0863}.

\bibitem[{\citenamefont{{H\"agler} et~al.}(2008)}]{Hagler:2007xi}
\bibinfo{author}{\bibfnamefont{P.}~\bibnamefont{{H\"agler}}}
  \bibnamefont{et~al.} (\bibinfo{collaboration}{LHPC}), \bibinfo{journal}{Phys.
  Rev.} \textbf{\bibinfo{volume}{D77}}, \bibinfo{pages}{094502}
  (\bibinfo{year}{2008}), \eprint{0705.4295}.

\bibitem[{\citenamefont{Bratt et~al.}(2008)}]{Bratt:2008uf}
\bibinfo{author}{\bibfnamefont{J.~D.} \bibnamefont{Bratt}} \bibnamefont{et~al.}
  (\bibinfo{collaboration}{LHPC}), \bibinfo{journal}{PoS}
  \textbf{\bibinfo{volume}{LATTICE2008}}, \bibinfo{pages}{141}
  (\bibinfo{year}{2008}), \eprint{0810.1933}.

\bibitem[{\citenamefont{Bratt et~al.}(2009)}]{LHPC_mixedaction_nucleonstr_2008}
\bibinfo{author}{\bibfnamefont{J.~D.} \bibnamefont{Bratt}} \bibnamefont{et~al.}
  (\bibinfo{collaboration}{LHPC}) (\bibinfo{year}{2009}), \bibinfo{note}{in
  preparation}.

\bibitem[{\citenamefont{Ohta and Yamazaki}(2008)}]{Ohta:2008kd}
\bibinfo{author}{\bibfnamefont{S.}~\bibnamefont{Ohta}} \bibnamefont{and}
  \bibinfo{author}{\bibfnamefont{T.}~\bibnamefont{Yamazaki}}
  (\bibinfo{collaboration}{for RBC and UKQCD}) (\bibinfo{year}{2008}),
  \eprint{0810.0045}.

\bibitem[{\citenamefont{Yamazaki et~al.}(2009)}]{Yamazaki:2009zq}
\bibinfo{author}{\bibfnamefont{T.}~\bibnamefont{Yamazaki}} \bibnamefont{et~al.}
  (\bibinfo{year}{2009}), \eprint{0904.2039}.

\bibitem[{\citenamefont{Syritsyn et~al.}(2008)}]{Syritsyn:2009np}
\bibinfo{author}{\bibfnamefont{S.~N.} \bibnamefont{Syritsyn}}
  \bibnamefont{et~al.}, \bibinfo{journal}{PoS}
  \textbf{\bibinfo{volume}{LATTICE2008}}, \bibinfo{pages}{169}
  (\bibinfo{year}{2008}), \eprint{0903.3063}.

\bibitem[{\citenamefont{Hemmert et~al.}(1998)\citenamefont{Hemmert, Holstein,
  and Kambor}}]{Hemmert:1997ye}
\bibinfo{author}{\bibfnamefont{T.~R.} \bibnamefont{Hemmert}},
  \bibinfo{author}{\bibfnamefont{B.~R.} \bibnamefont{Holstein}},
  \bibnamefont{and} \bibinfo{author}{\bibfnamefont{J.}~\bibnamefont{Kambor}},
  \bibinfo{journal}{J. Phys.} \textbf{\bibinfo{volume}{G24}},
  \bibinfo{pages}{1831} (\bibinfo{year}{1998}), \eprint{hep-ph/9712496}.

\bibitem[{\citenamefont{Dorati et~al.}(2008)\citenamefont{Dorati, Gail, and
  Hemmert}}]{Dorati:2007bk}
\bibinfo{author}{\bibfnamefont{M.}~\bibnamefont{Dorati}},
  \bibinfo{author}{\bibfnamefont{T.~A.} \bibnamefont{Gail}}, \bibnamefont{and}
  \bibinfo{author}{\bibfnamefont{T.~R.} \bibnamefont{Hemmert}},
  \bibinfo{journal}{Nucl. Phys.} \textbf{\bibinfo{volume}{A798}},
  \bibinfo{pages}{96} (\bibinfo{year}{2008}), \eprint{nucl-th/0703073}.

\bibitem[{\citenamefont{Gail}(2007)}]{Gail:2007phd}
\bibinfo{author}{\bibfnamefont{T.}~\bibnamefont{Gail}}, Ph.D. thesis,
  \bibinfo{school}{Technical University Munich} (\bibinfo{year}{2007}).

\bibitem[{\citenamefont{Becher and Leutwyler}(1999)}]{Becher:1999he}
\bibinfo{author}{\bibfnamefont{T.}~\bibnamefont{Becher}} \bibnamefont{and}
  \bibinfo{author}{\bibfnamefont{H.}~\bibnamefont{Leutwyler}},
  \bibinfo{journal}{Eur. Phys. J.} \textbf{\bibinfo{volume}{C9}},
  \bibinfo{pages}{643} (\bibinfo{year}{1999}), \eprint{hep-ph/9901384}.

\bibitem[{\citenamefont{Wang et~al.}(2007)\citenamefont{Wang, Leinweber,
  Thomas, and Young}}]{Wang:2007iw}
\bibinfo{author}{\bibfnamefont{P.}~\bibnamefont{Wang}},
  \bibinfo{author}{\bibfnamefont{D.~B.} \bibnamefont{Leinweber}},
  \bibinfo{author}{\bibfnamefont{A.~W.} \bibnamefont{Thomas}},
  \bibnamefont{and} \bibinfo{author}{\bibfnamefont{R.~D.} \bibnamefont{Young}},
  \bibinfo{journal}{Phys. Rev.} \textbf{\bibinfo{volume}{D75}},
  \bibinfo{pages}{073012} (\bibinfo{year}{2007}), \eprint{hep-ph/0701082}.

\bibitem[{\citenamefont{Wang et~al.}(2008)\citenamefont{Wang, Leinweber,
  Thomas, and Young}}]{Wang:2008vb}
\bibinfo{author}{\bibfnamefont{P.}~\bibnamefont{Wang}},
  \bibinfo{author}{\bibfnamefont{D.~B.} \bibnamefont{Leinweber}},
  \bibinfo{author}{\bibfnamefont{A.~W.} \bibnamefont{Thomas}},
  \bibnamefont{and} \bibinfo{author}{\bibfnamefont{R.~D.} \bibnamefont{Young}}
  (\bibinfo{year}{2008}), \eprint{0810.1021}.

\bibitem[{\citenamefont{Allton et~al.}(2008)}]{Allton:2008pn}
\bibinfo{author}{\bibfnamefont{C.}~\bibnamefont{Allton}} \bibnamefont{et~al.}
  (\bibinfo{collaboration}{RBC-UKQCD}), \bibinfo{journal}{Phys. Rev.}
  \textbf{\bibinfo{volume}{D78}}, \bibinfo{pages}{114509}
  (\bibinfo{year}{2008}), \eprint{0804.0473}.

\bibitem[{\citenamefont{Sharpe}(2007)}]{Sharpe:2007yd}
\bibinfo{author}{\bibfnamefont{S.~R.} \bibnamefont{Sharpe}}
  (\bibinfo{year}{2007}), \eprint{0706.0218}.

\bibitem[{\citenamefont{Blum et~al.}(2004)}]{Blum:2000kn}
\bibinfo{author}{\bibfnamefont{T.}~\bibnamefont{Blum}} \bibnamefont{et~al.},
  \bibinfo{journal}{Phys. Rev.} \textbf{\bibinfo{volume}{D69}},
  \bibinfo{pages}{074502} (\bibinfo{year}{2004}), \eprint{hep-lat/0007038}.

\bibitem[{\citenamefont{Gasser and Leutwyler}(1984)}]{Gasser:1983yg}
\bibinfo{author}{\bibfnamefont{J.}~\bibnamefont{Gasser}} \bibnamefont{and}
  \bibinfo{author}{\bibfnamefont{H.}~\bibnamefont{Leutwyler}},
  \bibinfo{journal}{Ann. Phys.} \textbf{\bibinfo{volume}{158}},
  \bibinfo{pages}{142} (\bibinfo{year}{1984}).

\bibitem[{\citenamefont{Bernard et~al.}(1998)\citenamefont{Bernard, Fearing,
  Hemmert, and {Mei{\ss}ner}}}]{Bernard:1998gv}
\bibinfo{author}{\bibfnamefont{V.}~\bibnamefont{Bernard}},
  \bibinfo{author}{\bibfnamefont{H.~W.} \bibnamefont{Fearing}},
  \bibinfo{author}{\bibfnamefont{T.~R.} \bibnamefont{Hemmert}},
  \bibnamefont{and} \bibinfo{author}{\bibfnamefont{U.~G.}
  \bibnamefont{{Mei{\ss}ner}}}, \bibinfo{journal}{Nucl. Phys.}
  \textbf{\bibinfo{volume}{A635}}, \bibinfo{pages}{121} (\bibinfo{year}{1998}),
  \eprint{hep-ph/9801297}.

\bibitem[{\citenamefont{Del~Debbio et~al.}(2007)\citenamefont{Del~Debbio,
  Giusti, {L\"uscher}, Petronzio, and Tantalo}}]{DelDebbio:2006cn}
\bibinfo{author}{\bibfnamefont{L.}~\bibnamefont{Del~Debbio}},
  \bibinfo{author}{\bibfnamefont{L.}~\bibnamefont{Giusti}},
  \bibinfo{author}{\bibfnamefont{M.}~\bibnamefont{{L\"uscher}}},
  \bibinfo{author}{\bibfnamefont{R.}~\bibnamefont{Petronzio}},
  \bibnamefont{and} \bibinfo{author}{\bibfnamefont{N.}~\bibnamefont{Tantalo}},
  \bibinfo{journal}{JHEP} \textbf{\bibinfo{volume}{02}}, \bibinfo{pages}{056}
  (\bibinfo{year}{2007}), \eprint{hep-lat/0610059}.

\bibitem[{\citenamefont{Dimopoulos et~al.}(2008)}]{Dimopoulos:2008sy}
\bibinfo{author}{\bibfnamefont{P.}~\bibnamefont{Dimopoulos}}
  \bibnamefont{et~al.} (\bibinfo{collaboration}{ETM}) (\bibinfo{year}{2008}),
  \eprint{0810.2873}.

\bibitem[{\citenamefont{Colangelo et~al.}(2001)\citenamefont{Colangelo, Gasser,
  and Leutwyler}}]{Colangelo:2001df}
\bibinfo{author}{\bibfnamefont{G.}~\bibnamefont{Colangelo}},
  \bibinfo{author}{\bibfnamefont{J.}~\bibnamefont{Gasser}}, \bibnamefont{and}
  \bibinfo{author}{\bibfnamefont{H.}~\bibnamefont{Leutwyler}},
  \bibinfo{journal}{Nucl. Phys.} \textbf{\bibinfo{volume}{B603}},
  \bibinfo{pages}{125} (\bibinfo{year}{2001}), \eprint{hep-ph/0103088}.

\bibitem[{\citenamefont{Bijnens et~al.}(1998)\citenamefont{Bijnens, Colangelo,
  and Talavera}}]{Bijnens:1998fm}
\bibinfo{author}{\bibfnamefont{J.}~\bibnamefont{Bijnens}},
  \bibinfo{author}{\bibfnamefont{G.}~\bibnamefont{Colangelo}},
  \bibnamefont{and} \bibinfo{author}{\bibfnamefont{P.}~\bibnamefont{Talavera}},
  \bibinfo{journal}{JHEP} \textbf{\bibinfo{volume}{05}}, \bibinfo{pages}{014}
  (\bibinfo{year}{1998}), \eprint{hep-ph/9805389}.

\bibitem[{\citenamefont{Kubis and {Mei{\ss}ner}}(2001)}]{Kubis:2000zd}
\bibinfo{author}{\bibfnamefont{B.}~\bibnamefont{Kubis}} \bibnamefont{and}
  \bibinfo{author}{\bibfnamefont{U.-G.} \bibnamefont{{Mei{\ss}ner}}},
  \bibinfo{journal}{Nucl. Phys.} \textbf{\bibinfo{volume}{A679}},
  \bibinfo{pages}{698} (\bibinfo{year}{2001}), \eprint{hep-ph/0007056}.

\bibitem[{\citenamefont{Bakeyev et~al.}(2004)}]{Bakeyev:2003ff}
\bibinfo{author}{\bibfnamefont{T.}~\bibnamefont{Bakeyev}} \bibnamefont{et~al.}
  (\bibinfo{collaboration}{QCDSF-UKQCD}), \bibinfo{journal}{Phys. Lett.}
  \textbf{\bibinfo{volume}{B580}}, \bibinfo{pages}{197} (\bibinfo{year}{2004}),
  \eprint{hep-lat/0305014}.

\bibitem[{\citenamefont{Kelly}(2004)}]{Kelly:2004hm}
\bibinfo{author}{\bibfnamefont{J.~J.} \bibnamefont{Kelly}},
  \bibinfo{journal}{Phys. Rev.} \textbf{\bibinfo{volume}{C70}},
  \bibinfo{pages}{068202} (\bibinfo{year}{2004}).

\bibitem[{\citenamefont{Hemmert and Weise}(2002)}]{Hemmert:2002uh}
\bibinfo{author}{\bibfnamefont{T.~R.} \bibnamefont{Hemmert}} \bibnamefont{and}
  \bibinfo{author}{\bibfnamefont{W.}~\bibnamefont{Weise}},
  \bibinfo{journal}{Eur. Phys. J.} \textbf{\bibinfo{volume}{A15}},
  \bibinfo{pages}{487} (\bibinfo{year}{2002}), \eprint{hep-lat/0204005}.

\bibitem[{\citenamefont{Colangelo and D\"urr}(2004)}]{Colangelo:2003hf}
\bibinfo{author}{\bibfnamefont{G.}~\bibnamefont{Colangelo}} \bibnamefont{and}
  \bibinfo{author}{\bibfnamefont{S.}~\bibnamefont{D\"urr}},
  \bibinfo{journal}{Eur. Phys. J.} \textbf{\bibinfo{volume}{C33}},
  \bibinfo{pages}{543} (\bibinfo{year}{2004}).

\bibitem[{\citenamefont{Walker-Loud et~al.}(2009)}]{WalkerLoud:2008bp}
\bibinfo{author}{\bibfnamefont{A.}~\bibnamefont{Walker-Loud}}
  \bibnamefont{et~al.}, \bibinfo{journal}{Phys. Rev.}
  \textbf{\bibinfo{volume}{D79}}, \bibinfo{pages}{054502}
  (\bibinfo{year}{2009}), \eprint{0806.4549}.

\bibitem[{\citenamefont{Bernard et~al.}(2005)\citenamefont{Bernard, Hemmert,
  and {Mei{\ss}ner}}}]{Bernard:2005fy}
\bibinfo{author}{\bibfnamefont{V.}~\bibnamefont{Bernard}},
  \bibinfo{author}{\bibfnamefont{T.~R.} \bibnamefont{Hemmert}},
  \bibnamefont{and} \bibinfo{author}{\bibfnamefont{U.-G.}
  \bibnamefont{{Mei{\ss}ner}}}, \bibinfo{journal}{Phys. Lett.}
  \textbf{\bibinfo{volume}{B622}}, \bibinfo{pages}{141} (\bibinfo{year}{2005}),
  \eprint{hep-lat/0503022}.

\bibitem[{\citenamefont{Amsler et~al.}(2008)}]{Amsler:2008uo}
\bibinfo{author}{\bibfnamefont{C.}~\bibnamefont{Amsler}} \bibnamefont{et~al.}
  (\bibinfo{collaboration}{Particle Data Group}), \bibinfo{journal}{Phys.
  Lett.} \textbf{\bibinfo{volume}{B667}}, \bibinfo{pages}{1}
  (\bibinfo{year}{2008}).

\bibitem[{\citenamefont{Hemmert et~al.}(2003)\citenamefont{Hemmert, Procura,
  and Weise}}]{Hemmert:2003cb}
\bibinfo{author}{\bibfnamefont{T.~R.} \bibnamefont{Hemmert}},
  \bibinfo{author}{\bibfnamefont{M.}~\bibnamefont{Procura}}, \bibnamefont{and}
  \bibinfo{author}{\bibfnamefont{W.}~\bibnamefont{Weise}},
  \bibinfo{journal}{Phys. Rev.} \textbf{\bibinfo{volume}{D68}},
  \bibinfo{pages}{075009} (\bibinfo{year}{2003}), \eprint{hep-lat/0303002}.

\bibitem[{\citenamefont{Edwards et~al.}(2006)}]{Edwards:2005ym}
\bibinfo{author}{\bibfnamefont{R.~G.} \bibnamefont{Edwards}}
  \bibnamefont{et~al.} (\bibinfo{collaboration}{LHPC}), \bibinfo{journal}{Phys.
  Rev. Lett.} \textbf{\bibinfo{volume}{96}}, \bibinfo{pages}{052001}
  (\bibinfo{year}{2006}), \eprint{hep-lat/0510062}.

\bibitem[{\citenamefont{Procura et~al.}(2007)\citenamefont{Procura, Musch,
  Hemmert, and Weise}}]{Procura:2006gq}
\bibinfo{author}{\bibfnamefont{M.}~\bibnamefont{Procura}},
  \bibinfo{author}{\bibfnamefont{B.~U.} \bibnamefont{Musch}},
  \bibinfo{author}{\bibfnamefont{T.~R.} \bibnamefont{Hemmert}},
  \bibnamefont{and} \bibinfo{author}{\bibfnamefont{W.}~\bibnamefont{Weise}},
  \bibinfo{journal}{Phys. Rev.} \textbf{\bibinfo{volume}{D75}},
  \bibinfo{pages}{014503} (\bibinfo{year}{2007}), \eprint{hep-lat/0610105}.

\bibitem[{\citenamefont{Khan et~al.}(2006)}]{Khan:2006de}
\bibinfo{author}{\bibfnamefont{A.~A.} \bibnamefont{Khan}} \bibnamefont{et~al.},
  \bibinfo{journal}{Phys. Rev.} \textbf{\bibinfo{volume}{D74}},
  \bibinfo{pages}{094508} (\bibinfo{year}{2006}), \eprint{hep-lat/0603028}.

\bibitem[{\citenamefont{Davidson et~al.}(1991)\citenamefont{Davidson,
  Mukhopadhyay, and Wittman}}]{Davidson:1991xz}
\bibinfo{author}{\bibfnamefont{R.~M.} \bibnamefont{Davidson}},
  \bibinfo{author}{\bibfnamefont{N.~C.} \bibnamefont{Mukhopadhyay}},
  \bibnamefont{and} \bibinfo{author}{\bibfnamefont{R.~S.}
  \bibnamefont{Wittman}}, \bibinfo{journal}{Phys. Rev.}
  \textbf{\bibinfo{volume}{D43}}, \bibinfo{pages}{71} (\bibinfo{year}{1991}).

\bibitem[{\citenamefont{Hemmert et~al.}(1997)\citenamefont{Hemmert, Holstein,
  and Kambor}}]{Hemmert:1996xg}
\bibinfo{author}{\bibfnamefont{T.~R.} \bibnamefont{Hemmert}},
  \bibinfo{author}{\bibfnamefont{B.~R.} \bibnamefont{Holstein}},
  \bibnamefont{and} \bibinfo{author}{\bibfnamefont{J.}~\bibnamefont{Kambor}},
  \bibinfo{journal}{Phys. Lett.} \textbf{\bibinfo{volume}{B395}},
  \bibinfo{pages}{89} (\bibinfo{year}{1997}).

\bibitem[{\citenamefont{Belushkin
  et~al.}(2007{\natexlab{b}})\citenamefont{Belushkin, Hammer, and
  {Mei{\ss}ner}}}]{Belushkin:2007dn}
\bibinfo{author}{\bibfnamefont{M.~A.} \bibnamefont{Belushkin}},
  \bibinfo{author}{\bibfnamefont{H.~W.} \bibnamefont{Hammer}},
  \bibnamefont{and} \bibinfo{author}{\bibfnamefont{U.~G.}
  \bibnamefont{{Mei{\ss}ner}}}, \bibinfo{journal}{Phys. Rev.}
  \textbf{\bibinfo{volume}{C75}}, \bibinfo{pages}{035202}
  (\bibinfo{year}{2007}{\natexlab{b}}), \eprint{hep-ph/0608337}.

\bibitem[{\citenamefont{Gasser et~al.}(1988)\citenamefont{Gasser, Sainio, and
  Svarc}}]{Gasser:1987rb}
\bibinfo{author}{\bibfnamefont{J.}~\bibnamefont{Gasser}},
  \bibinfo{author}{\bibfnamefont{M.~E.} \bibnamefont{Sainio}},
  \bibnamefont{and} \bibinfo{author}{\bibfnamefont{A.}~\bibnamefont{Svarc}},
  \bibinfo{journal}{Nucl. Phys.} \textbf{\bibinfo{volume}{B307}},
  \bibinfo{pages}{779} (\bibinfo{year}{1988}).

\bibitem[{\citenamefont{Gail and Hemmert}()}]{Hemmert:CBChPT}
\bibinfo{author}{\bibfnamefont{T.~A.} \bibnamefont{Gail}} \bibnamefont{and}
  \bibinfo{author}{\bibfnamefont{T.~R.} \bibnamefont{Hemmert}},
  \bibinfo{note}{forthcoming}.

\bibitem[{\citenamefont{Bernard et~al.}(1997)\citenamefont{Bernard, Kaiser, and
  {Mei{\ss}ner}}}]{Bernard:1996gq}
\bibinfo{author}{\bibfnamefont{V.}~\bibnamefont{Bernard}},
  \bibinfo{author}{\bibfnamefont{N.}~\bibnamefont{Kaiser}}, \bibnamefont{and}
  \bibinfo{author}{\bibfnamefont{U.-G.} \bibnamefont{{Mei{\ss}ner}}},
  \bibinfo{journal}{Nucl. Phys.} \textbf{\bibinfo{volume}{A615}},
  \bibinfo{pages}{483} (\bibinfo{year}{1997}), \eprint{hep-ph/9611253}.

\bibitem[{\citenamefont{Fettes et~al.}(1998)\citenamefont{Fettes,
  {Mei{\ss}ner}, and Steininger}}]{Fettes:1998ud}
\bibinfo{author}{\bibfnamefont{N.}~\bibnamefont{Fettes}},
  \bibinfo{author}{\bibfnamefont{U.-G.} \bibnamefont{{Mei{\ss}ner}}},
  \bibnamefont{and}
  \bibinfo{author}{\bibfnamefont{S.}~\bibnamefont{Steininger}},
  \bibinfo{journal}{Nucl. Phys.} \textbf{\bibinfo{volume}{A640}},
  \bibinfo{pages}{199} (\bibinfo{year}{1998}), \eprint{hep-ph/9803266}.

\bibitem[{\citenamefont{Entem and Machleidt}(2002)}]{Entem:2002sf}
\bibinfo{author}{\bibfnamefont{D.~R.} \bibnamefont{Entem}} \bibnamefont{and}
  \bibinfo{author}{\bibfnamefont{R.}~\bibnamefont{Machleidt}},
  \bibinfo{journal}{Phys. Rev.} \textbf{\bibinfo{volume}{C66}},
  \bibinfo{pages}{014002} (\bibinfo{year}{2002}), \eprint{nucl-th/0202039}.

\bibitem[{\citenamefont{Procura et~al.}(2006)\citenamefont{Procura, Musch,
  Wollenweber, Hemmert, and Weise}}]{Procura:2006bj}
\bibinfo{author}{\bibfnamefont{M.}~\bibnamefont{Procura}},
  \bibinfo{author}{\bibfnamefont{B.~U.} \bibnamefont{Musch}},
  \bibinfo{author}{\bibfnamefont{T.}~\bibnamefont{Wollenweber}},
  \bibinfo{author}{\bibfnamefont{T.~R.} \bibnamefont{Hemmert}},
  \bibnamefont{and} \bibinfo{author}{\bibfnamefont{W.}~\bibnamefont{Weise}},
  \bibinfo{journal}{Phys. Rev.} \textbf{\bibinfo{volume}{D73}},
  \bibinfo{pages}{114510} (\bibinfo{year}{2006}), \eprint{hep-lat/0603001}.

\bibitem[{\citenamefont{Ali~Khan et~al.}(2004)}]{Khan:2003cu}
\bibinfo{author}{\bibfnamefont{A.}~\bibnamefont{Ali~Khan}} \bibnamefont{et~al.}
  (\bibinfo{collaboration}{QCDSF-UKQCD}), \bibinfo{journal}{Nucl. Phys.}
  \textbf{\bibinfo{volume}{B689}}, \bibinfo{pages}{175} (\bibinfo{year}{2004}),
  \eprint{hep-lat/0312030}.

\bibitem[{\citenamefont{Syritsyn and Negele}(2007)}]{Syritsyn:2007mp}
\bibinfo{author}{\bibfnamefont{S.}~\bibnamefont{Syritsyn}} \bibnamefont{and}
  \bibinfo{author}{\bibfnamefont{J.~W.} \bibnamefont{Negele}},
  \bibinfo{journal}{PoS} \textbf{\bibinfo{volume}{LAT2007}},
  \bibinfo{pages}{078} (\bibinfo{year}{2007}), \eprint{0710.0425}.

\bibitem[{\citenamefont{Greil et~al.}()\citenamefont{Greil, Hemmert, and
  Sch{\"a}fer}}]{Greil:2009-in-prep}
\bibinfo{author}{\bibfnamefont{L.}~\bibnamefont{Greil}},
  \bibinfo{author}{\bibfnamefont{T.~R.} \bibnamefont{Hemmert}},
  \bibnamefont{and}
  \bibinfo{author}{\bibfnamefont{A.}~\bibnamefont{Sch{\"a}fer}},
  \bibinfo{note}{in preparation}.

\bibitem[{\citenamefont{Beane and Savage}(2004)}]{Beane:2004rf}
\bibinfo{author}{\bibfnamefont{S.~R.} \bibnamefont{Beane}} \bibnamefont{and}
  \bibinfo{author}{\bibfnamefont{M.~J.} \bibnamefont{Savage}},
  \bibinfo{journal}{Phys. Rev.} \textbf{\bibinfo{volume}{D70}},
  \bibinfo{pages}{074029} (\bibinfo{year}{2004}), \eprint{hep-ph/0404131}.

\bibitem[{\citenamefont{Detmold and Lin}(2005)}]{Detmold:2005pt}
\bibinfo{author}{\bibfnamefont{W.}~\bibnamefont{Detmold}} \bibnamefont{and}
  \bibinfo{author}{\bibfnamefont{C.~J.~D.} \bibnamefont{Lin}},
  \bibinfo{journal}{Phys. Rev.} \textbf{\bibinfo{volume}{D71}},
  \bibinfo{pages}{054510} (\bibinfo{year}{2005}), \eprint{hep-lat/0501007}.

\bibitem[{\citenamefont{Edwards and Joo}(2005)}]{Edwards:2004sx}
\bibinfo{author}{\bibfnamefont{R.~G.} \bibnamefont{Edwards}} \bibnamefont{and}
  \bibinfo{author}{\bibfnamefont{B.}~\bibnamefont{Joo}}
  (\bibinfo{collaboration}{SciDAC}), \bibinfo{journal}{Nucl. Phys. Proc.
  Suppl.} \textbf{\bibinfo{volume}{140}}, \bibinfo{pages}{832}
  (\bibinfo{year}{2005}), \eprint{hep-lat/0409003}.

\bibitem[{\citenamefont{Dolgov et~al.}(2002)}]{Dolgov:2002zm}
\bibinfo{author}{\bibfnamefont{D.}~\bibnamefont{Dolgov}} \bibnamefont{et~al.}
  (\bibinfo{collaboration}{LHPC}), \bibinfo{journal}{Phys. Rev.}
  \textbf{\bibinfo{volume}{D66}}, \bibinfo{pages}{034506}
  (\bibinfo{year}{2002}), \eprint{hep-lat/0201021}.

\bibitem[{\citenamefont{{L\"uscher}}(1977)}]{Luscher:1976ms}
\bibinfo{author}{\bibfnamefont{M.}~\bibnamefont{{L\"uscher}}},
  \bibinfo{journal}{Commun. Math. Phys.} \textbf{\bibinfo{volume}{54}},
  \bibinfo{pages}{283} (\bibinfo{year}{1977}).

\bibitem[{\citenamefont{Bratt and Negele}(2008)}]{Bratt:2008ur}
\bibinfo{author}{\bibfnamefont{J.}~\bibnamefont{Bratt}} \bibnamefont{and}
  \bibinfo{author}{\bibfnamefont{J.~W.} \bibnamefont{Negele}}
  (\bibinfo{year}{2008}), \eprint{0810.1954}.

\end{thebibliography}
